\PassOptionsToPackage{nonamebreak}{natbib}
\documentclass{aastex701}

\journalinfo{}
\makeatletter
\def\frontmatter@title@above{}
\makeatother

\shorttitle{IRIS: Deciphering the Galactic Center with Machine Learning}
\shortauthors{B. L. DuBois et al.}

\usepackage{enumerate}
\usepackage[table]{xcolor}
\usepackage{pifont}
\makeatletter
\let\aastex@tablewidth\tablewidth
\def\save@natural@width{
  \ifnum\@d@t@@flag=0
    \setlength{\@d@t@a}{0pt}
    \let\@d@t@b=\LT@entry
    \def\LT@entry##1##2{
      \addtolength{\@d@t@a}{##2}
    }
    \expandafter\csname LT@\romannumeral\c@deluxe@table@num\endcsname
    \setlength{\@d@t@a}{-\@d@t@a}
    \aastex@tablewidth{\the\@d@t@a}
    \def\LT@entry{\@d@t@b}
  \fi
}
\let\tablewidth\relax
\makeatother
\usepackage{tabularray}
\UseTblrLibrary{booktabs}
\usepackage{xspace}
\usepackage{amsmath}
\numberwithin{equation}{subsection}
\usepackage{amssymb}
\usepackage{esint}
\usepackage{siunitx}
\usepackage{physics}
\usepackage{tikz-cd}
\usetikzlibrary{arrows.meta,shapes.geometric}

\let\oldautoref\autoref
\renewcommand{\autoref}[1]{\mbox{\oldautoref{#1}}}

\DeclareSIUnit{\parsec}{pc}
\DeclareSIUnit{\Jansky}{Jy}
\DeclareSIUnit{\year}{yr}
\DeclareSIUnit{\gauss}{G}
\newcommand{\SgrAStar}{{\text{Sgr A}^\ast}}
\newcommand{\HTwo}{{\text{H}_2}}
\newcommand{\HPlus}{{\text{H}^+}}
\newcommand{\TwelveC}{{^{12}\text{C}}}
\newcommand{\TwelveCO}{{^{12}\text{CO}}}

\newcommand{\ThirteenC}{{^{13}\text{C}}}
\newcommand{\ThirteenCO}{{^{13}\text{CO}}}

\newcommand{\ThirteenCOTwoOne}{{^{13}\text{CO} \; (\text{2--1})}}
\newcommand{\abundance}[1]{\chi_{\raisebox{-3pt}{$\scriptstyle{#1}$}}}

\newcommand{\total}{{[\text{T}]}}
\newcommand{\cont}{{[\text{C}]}}
\newcommand{\dust}{{[\text{D}]}}
\newcommand{\sline}{{[\text{L}]}}
\newcommand{\back}{{[\text{B}]}}
\newcommand{\any}{{[\cdot]}}
\newcommand{\N}{\mathbb{N}}
\newcommand{\Z}{\mathbb{Z}}
\newcommand{\Q}{\mathbb{Q}}
\newcommand{\R}{\mathbb{R}}
\newcommand{\C}{\mathbb{C}}
\DeclareMathOperator{\arcsinh}{\text{arcsinh}}
\DeclareMathOperator{\mean}{\text{mean}}
\DeclareMathOperator{\meanb}{\text{mean}_b}

\DeclareMathOperator*{\E}{\mathbb{E}}
\DeclareMathOperator{\TSRE}{\text{TSRE}}
\newcommand{\TSREcube}{{\overline{\TSRE}_{\ell b v}}}
\newcommand{\TSRElb}{{\overline{\TSRE}_{\ell b}}}
\newcommand{\TSRElv}{{\overline{\TSRE}_{\ell v}}}
\DeclareMathOperator{\TAM}{\text{TAM}}
\DeclareMathOperator{\MASE}{\text{MASE}}

\def\Msun{\hbox{M$_{\odot}$}\xspace}

\newcommand{\citeLipman}{\text{Lipman et al. (in prep)}\xspace}
\newcommand{\citationLipman}{\text{Lipman et al., in prep}\xspace}
\newcommand{\citepLipman}{\text{(\citationLipman)}\xspace}

\NewPageAfterKeywords

\begin{document}

\title{IRIS: Deciphering Spectral-Line Imagery of the Galactic Center by Machine-Learning on Simulations}

\author[orcid=0009-0006-1435-2439, sname='B.L. DuBois']{B.L. DuBois}
\affiliation{University of Connecticut, Department of Physics, 196A Hillside Road, Unit 3046
Storrs, CT 06269-3046, USA}
\email{brendan@bldubois.com}

\author[orcid=0000-0002-6073-9320,sname='Cara Battersby']{Cara Battersby}
\affiliation{University of Connecticut, Department of Physics, 196A Hillside Road, Unit 3046
Storrs, CT 06269-3046, USA}
\email{cara.battersby@uconn.edu}

\author[orcid=0009-0009-5056-6938, sname='Jonah C. Baade']{Jonah C. Baade}
\affiliation{University of Connecticut, Department of Physics, 196A Hillside Road, Unit 3046
Storrs, CT 06269-3046, USA}
\email{igg24002@uconn.edu}

\author[0000-0002-5776-9473]{Dani R. Lipman}
\affiliation{University of Connecticut, Department of Physics, 196A Hillside Road, Unit 3046
Storrs, CT 06269-3046, USA}
\email{dani.lipman@uconn.edu}

\author[0000-0003-0946-4365]{H Perry Hatchfield}
\affiliation{University of Connecticut, Department of Physics, 196A Hillside Road, Unit 3046
Storrs, CT 06269-3046, USA}
\email{h.hatchfield@uconn.edu}

\author[0000-0002-2782-1082]{Jack Sullivan}
\affiliation{University of Connecticut, Department of Physics, 196A Hillside Road, Unit 3046
Storrs, CT 06269-3046, USA}
\email{john.2.sullivan@uconn.edu}

\author[orcid=0009-0008-1198-526X,sname=‘Russell Bentley’]{Russell Bentley}
\affiliation{Stony Brook University, Department of Computer Science, Stony Brook, NY 11794-2424 , USA}
\email{rbentley@cs.stonybrook.edu}

\author[0000-0001-5222-9139]{Stefan Reissl}
\affiliation{Universit\"{a}t Heidelberg, Zentrum f\"{u}r Astronomie, Institut f\"{u}r Theoretische Astrophysik, Albert-Ueberle-Straße 2, D-69120 Heidelberg, Germany}
\email{reissl@uni-heidelberg.de}

\author[orcid=0000-0002-0560-3172]{Ralf S.\ Klessen}
\affiliation{Universit\"{a}t Heidelberg, Zentrum f\"{u}r Astronomie, Institut f\"{u}r Theoretische Astrophysik, Albert-Ueberle-Str.\ 2, 69120 Heidelberg, Germany}
\affiliation{Universit\"{a}t Heidelberg, Interdisziplin\"{a}res Zentrum f\"{u}r Wissenschaftliches Rechnen, Im Neuenheimer Feld 225, 69120 Heidelberg, Germany}
\email{klessen@uni-heidelberg.de}

\author[orcid=0000-0002-0294-799X, sname='Victor F. Ksoll']{Victor F. Ksoll}
\affiliation{Universit\"{a}t Heidelberg, Zentrum f\"{u}r Astronomie, Institut f\"{u}r Theoretische Astrophysik, Albert-Ueberle-Stra{\ss}e 2, D-69120 Heidelberg, Germany}
\email{v.ksoll@uni-heidelberg.de}

\author[0000-0001-6113-6241]{Mattia~C.~Sormani}
\affiliation{Como Lake centre for AstroPhysics (CLAP), DiSAT, Universit{\`a} dell’Insubria, via Valleggio 11, 22100 Como, Italy}
\email{mattiacarlo.sormani@uninsubria.it}

\author[orcid=0009-0004-0121-1560,sname=‘Zi-Xuan Feng']{Zi-Xuan Feng}
\affiliation{Como Lake centre for AstroPhysics (CLAP), DiSAT, Universit{\`a} dell’Insubria, via Valleggio 11, 22100 Como, Italy}
\email{zixuan.feng@uninsubria.it}

\author[0000-0001-6431-9633]{Adam Ginsburg}
\affiliation{Department of Astronomy, University of Florida, Gainesville, FL 32611 USA}
\email{adamginsburg@ufl.edu}

\author[orcid=0000-0002-9483-7164,sname=‘Robin G. Tress’]{Robin Tress}
\affiliation{Institute of Physics, Laboratory for Galaxy Evolution and Spectral Modelling, EPFL, Observatoire de Sauverny, Chemin Pegasi 51, 1290 Versoix, Switzerland}
\email{robin.tress@epfl.ch}

\begin{abstract}
    \par In understanding the 3D structure of the Milky Way's Central Molecular Zone (CMZ), we are limited by our edge-on perspective. Towards addressing this problem, we introduce \textit{\textbf{\underline{I}}magery \textbf{\underline{R}}eversion \textbf{\underline{I}}nformed by \textbf{\underline{S}}imulation (\textbf{IRIS})}. IRIS is a novel machine-learning code base featuring a deep convolutional neural network (CNN), which we have designed to translate edge-on observations of our Milky Way Galaxy into top-down images by training on data generated from AREPO galaxy simulations and synthetic observations of those simulations. 
    \par We develop a large custom dataset on which we train our bespoke model, and then test the trained model on synthetic data to probe the potential of this machine-learning method, which we call \textit{supervised reversion}. We then apply our trained model to real observations from the SEDIGISM $\ThirteenCOTwoOne$ survey, yielding new top-down views of our CMZ. Though our SEDIGISM reversions are not fully consistent across model training runs, we posit that this lack of convergence can be alleviated by expansion of the training dataset. We argue that these results represent a strong proof-of-concept for the use of supervised reversion to decipher our CMZ's 3D structure. 
    \par Crucial in generating our training dataset's $\sim 100\text{k}$ synthetic observations, we introduce IRIS Synthetic Observation (IRIS-SO), a new GPU-accelerated and fully differentiable code implemented in PyTorch for the non-LTE synthetic observation of spectral lines and dust. We find that IRIS-SO provides up to $10{,}000\times$ speedups in comparison to the synthetic-observation code RADMC-3D. We release all the IRIS code open-source at \url{https://github.com/bldubois/IRIS}.
\end{abstract}

\tableofcontents
\vspace{1cm}
\twocolumngrid

\section{Introduction} \label{sec:Introduction}
    \begin{figure*}[t]
        \centering
        \includegraphics[width=1\linewidth]{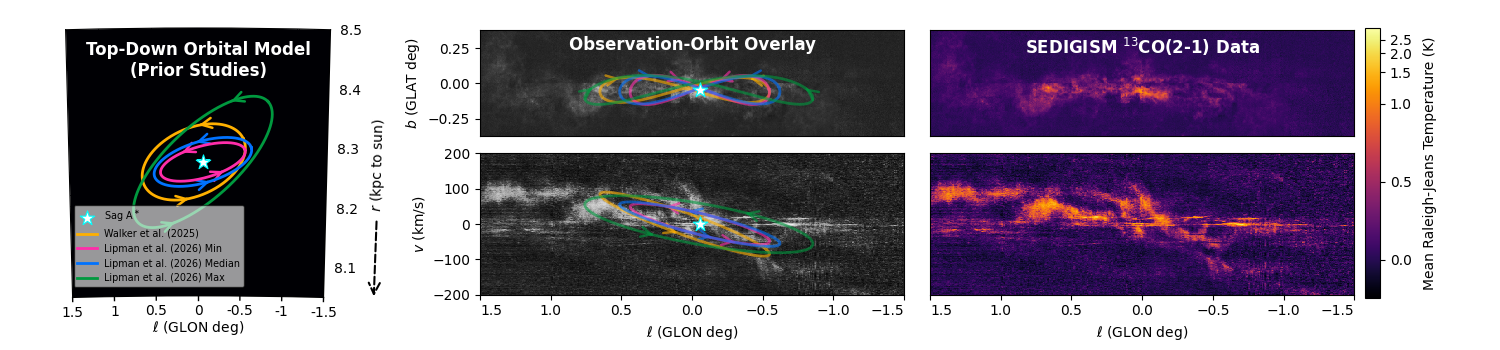}
        \caption{\textbf{CMZ Overview:} A visualization of the CMZ structure. Depicted in the left panel are scaled versions of the $x_2$ orbital fittings determined in the 3D CMZ paper series by \citet{Walker2025} and \citet{Lipman2026}. In the right panel are $\ell, b$ and $\ell, v$ mean projections of the continuum-subtracted Raleigh-Jeans brightness temperature (see \autoref{subsec:Continuum Subtraction} and \autoref{subsec:Intensity Versus Temperature}) recorded in the SEDIGISM $\ThirteenCOTwoOne$ data. The color bar is scaled via an $\arcsinh$ nonlinearity. In the center panel, the orbital fittings are overlaid onto the SEDIGISM data. Note that orbital sizes have been scaled via an angle-preserving transformation in order to adapt the solar distance to $\SgrAStar$ that is originally assumed in these studies ($R_0 = \SI{8100}{\parsec}$) to our alternate assumption for this radius \citep[$R_0 = \SI{8277}{\parsec}$ from][]{GRAVITY2021, GRAVITY2022}. This transformation was applied to orbital size only, however, and did not scale the velocities in the $\ell, v$ plots.}
        \label{fig:CMZ Overview}
    \end{figure*}

    \par Our edge-on vantage poses a unique challenge in observing our galaxy's Central Molecular Zone (CMZ). These inner $\sim \SI{300}{\parsec}$ of the Milky Way harbor an interstellar medium (ISM) characterized by dense molecular gas with a rich chemical makeup, areas of intense star formation, and a complex orbital structure. While understanding this anomalous region of the galaxy is a critical effort not just in characterizing the Milky Way but in advancing our science of stars and galaxies in general, our ability to refine our understanding of the CMZ is limited by the observational challenges produced by its complex structure. We are hindered, ultimately, by the fact that we lack a top-down perspective of our own galaxy.
    \par Our first tool in circumventing this observational challenge is theoretical context. As a barred spiral galaxy \citep{Dwek1995}, the Milky Way is theorized to exhibit a barred gravitational potential giving rise to families of approximately elliptical $x_1$ and $x_2$ orbits concentric with the disk's rotational center \citep[e.g.][]{Contopoulos1989, Athanassoula1992}. Consistent with this theory, models of the CMZ's orbital structure informed by observation are generally a variation on some form of ellipse. This elliptical form, however, is complicated by additional out-of-plane oscillations and/or open-ended inflows and outflows, with proposed models including a twisted elliptical ring of constant velocity \citep{Molinari2011}, inner spiral arms \citep{Sofue1995, Sofue2022, Sofue2025}, a pretzel-like orbit \citep{kdl2015}, and a closed ellipse with constant momentum \citep{Binney1991, Sormani2015, Tress2020}. These models are further complicated by an observed asymmetry of total mass in the CMZ between positive and negative galactic longitudes, suggesting that material is not evenly distributed across these orbits \citep{Bally1987, Sormani2018}.
    \par The 3D CMZ paper series \citep[][see \autoref{fig:CMZ Overview}]{Battersby2025a, Battersby2025b, Walker2025, Lipman2025, Lipman2026} introduces new methods to quantitatively differentiate molecular clouds on the near versus far side of the galactic center and combines all existing observational constraints to present the first unified 3D model of the CMZ. This comprehensive synthesis of observation and theory suggests that gas flows inwards along the Milky Way bar in dust lanes that approximately follow $x_1$ orbits \citep[see also][]{Sormani2015}, and that the CMZ is likely the culmination of a family of $x_2$ orbits in the inner 100-$\SI{200}{\parsec}$ of the galaxy. But locations of the injection points of material where these $x_1$ and $x_2$ orbits meet, as well as the size, orientation, and exact shape of the CMZ orbits, and the trajectories of material inflowing towards the Circumnuclear Disk (CND) remain topics of debate \citep[see, e.g.][]{Henshaw2023}.
    \par In further refining these CMZ models, we rely on the combination of theory, observation, and, increasingly, simulation. Our ability to observe the CMZ is complicated not only by our edge-on vantage but by the high density of gas and dust in this region \citep[e.g.][]{Gusten1983, Ginsburg2016, Mills2018}, which absorbs short-wavelength light, obscuring many features. Surveys of the CMZ therefore cover this region at long wavelengths ranging from the radio \citep[e.g.][]{Stil2006, Beuther2016, Schuller2021}, to the submillimeter \citep[e.g.][]{Longmore2026, Schuller2009, Ginsburg2013}, to the far- and mid-infrared \citep[e.g.][]{Butterfield2024, Molinari2011, Benjamin2003}, and into the near-infrared \citep[e.g.][]{Nogueras-Lara2021}. These observations include both continuum surveys \citep{Stil2006, Beuther2016, Longmore2026, Schuller2009, Ginsburg2013, Butterfield2024, Molinari2011, Benjamin2003, Nogueras-Lara2021}, which, in particular, provide information regarding the distribution of dust, stars, and proto-stars, and spectral-line surveys \citep{Beuther2016, Schuller2021, Longmore2026, Schuller2009}, which are more useful in characterizing the distribution of molecular gas.
    \par Spectral-line surveys are of particular use because they provide line-of-sight velocity information at each image coordinate via analysis of the line Doppler shift. These 3D observations---spectral or position-position-velocity (PPV) cubes, expressed as fields of radiative intensity or equivalent brightness temperature over dimensions of galactic longitude $\ell$, galactic latitude $b$, and velocity $v$---thus provide a proxy for the true 3D spatial description of the CMZ. The ability to recover 3D spatial information from PPV cubes is clearly imperfect, as there is ambiguity regarding line-of-sight distances and non-radial velocities, and, in the optically thick case, masking obscures far-field features. But analysis of optically thin lines combined with the application of physics-based constraints improves the amount of information that is recoverable. When projected into $\ell, v$ space, for example, spectral emission of gas uniformly distributed along a circular orbit forms a linear feature that defies alternative physical explanation.
    \par A rich source of physics-based constraints, and a parallel source of insight regarding CMZ structure, is that of astrophysical simulations, which provide increasing resolution and account for an ever-broadening collection of physical effects. Recent numerical simulations \citep[e.g.][]{Armillotta2019, Salas2020, Moon2021, Sormani2020, Tress2020, Hatchfield2021} perform 3D hydrodynamic (HD) simulations of the CMZ and incorporate a variety of physics, including gas self-gravity, realistic gravitational potential, star formation, and supernova feedback. The newest numerical simulations \citep[e.g.][\citationLipman]{Tress2024, Tress2025} include all these features with the addition of magnetic fields and self-consistent gas inflow from a galactic disk, and begin to resolve down to the scale of star formation. The resulting simulations have high degrees of complexity and now reproduce many observational features of the real CMZ.
    \par We conjecture that present methods of understanding the CMZ have not made complete use of the sum of all available information. We argue that new insight may be gleaned not just by acquiring new data to analyze, but by analyzing more efficiently the data we already have---in particular, with regards to the comparison of observation and simulation. Many current methods of comparing observations and simulations incorporate side-by-side visualizations or comparison of overall properties (such as star formation rate, gas fraction, etc.). But the outputs of such comparisons are intrinsically limited in scope, often revealing only qualitative similarities or contrasting properties of bulk regions. We believe that these challenges demand an automated solution that is capable of analyzing a breadth of data with a level of complexity beyond what human-driven or ``by-eye'' methods are able to provide. In moving towards such advanced, automated approaches, we hope to pave the way not merely towards corroborating or falsifying existing models of large-scale CMZ geometry, but towards unlocking details of the cloud-scale structure beyond the scope of current predictions.
    \par In this research, we set out to prove that such an automated method can be learned by a machine system provided a combination of observational and simulation-based data. We focus on one specific use case---inferring the top-down structure of the CMZ---but the methodology could be applied to a wide range of problems. Specifically, we take galactic simulations computed in the AREPO moving-mesh framework \citepLipman and perform synthetic observations of these simulations in the $\ThirteenCOTwoOne$ spectral line. We then train a neural network to transform mean-brightness-temperature images in $\ell, v$ space into mean-$\HTwo$-density images in a ``top-down'', longitude-line-of-sight ($\ell, r$) space. In a simulated context on a constrained training and validation dataset generated from a single simulation run, we find that our trained model is very effective at performing these transformations, which we term our \textit{image reversions}. 
    \par We release alongside this paper a fully documented and open-source Python package, which we name \textit{\textbf{\underline{I}}magery \textbf{\underline{R}}eversion \textbf{\underline{I}}nformed by \textbf{\underline{S}}imulation} (\textit{\textbf{IRIS}}, \url{https://github.com/bldubois/IRIS}). While IRIS is not a general-use API, we provide this release not just in the hopes of providing transparency and inviting collaboration in future research, but because we hope the release of this documented code provides other astrophysicists interested in machine learning with one more workable example. Broadly, IRIS consists of three primary code modules---a simulation-processing module that converts snapshots of CMZ simulations computed in AREPO into a usable training dataset using a scalably parallelized set of CPU and GPU workers; a GPU-parallelized and fully differentiable synthetic-observation module that performs spectral-line imaging of the training dataset with up to $10{,}000\times$ speedups against CPU-based equivalents; and a deep, convolutional neural network (CNN) with attention that learns, via a distributed training setup, to perform reversions on the synthetically observed dataset. The IRIS code is developed exclusively in Python, with implementation in PyTorch \citep{Paszke2019} of the machine-learning elements and synthetic-observation engine.    
    \par The structure of this paper closely follows that of the IRIS code. In \autoref{sec:Simulation Processing and Data Production}, we discuss processing of AREPO snapshots and production of IRIS training data. In \autoref{sec:Synthetic Observation Overview and Theory}, we provide a theoretically-focused overview of radiative-transfer computation as pertains to synthetic observation. We then provide an algorithmically-focused description of the synthetic-observation process in \autoref{sec:Synthetic Observation Implementation and Algorithms}. Our discussion of synthetic observation culminates in \autoref{sec:Synthetic Observation Testing and Verification} with a verification of the IRIS synthetic-observation code and preprocessing pipeline via a series of tests and visualizations. These tests include comparison of IRIS against the proven synthetic-observation codes RADMC-3D \citep{Dullemond2012} and POLARIS \citep{Reissl2016, Brauer2017}, as well as a battery of speed comparisons against RADMC-3D. In \autoref{sec:Machine-Learning Methods and Reversion}, we outline the theory of our machine-learning approach as well as describe the architecture and training regimen of the IRIS neural network. We additionally discuss some alternative machine-learning approaches that we also explored that may still prove relevant to future work. In \autoref{sec:Results and Discussion}, we analyze results. We show a variety of figures demonstrating reversions of both synthetic data and real observational data from the SEDIGISM $\ThirteenCOTwoOne$ survey \citep[see \autoref{fig:CMZ Overview},][]{Schuller2021}. We then characterize current shortcomings and failure modes. We summarize in \autoref{sec:Summary and Conclusion} and conclude with remarks regarding possible future avenues of expansion to this research.

\section{Simulation Processing and Data Production} \label{sec:Simulation Processing and Data Production}
    \begin{deluxetable*}{lccc}
        \tablecaption{\textbf{Training-Dataset Parameters:} Summary of dataset-construction parameters for the training, foreground, and full-cone test datasets. During training and validation of our reversion model (see \autoref{subsec:Supervised Reversion: Training Theory} and \autoref{subsec:Implementation of Reversion: Training Hyperparameters, Overfitting, and Regularization}), we combined our training dataset with random samples from our foreground dataset (see \autoref{subsec:Foreground and Background}). We used the full-cone test dataset for generating the full-cone visualizations in \autoref{fig:Synthetic Reversions}. See \autoref{table:Synthetic Observation Parameters} for the synthetic-observation parameters we used over all these datasets. \textsuperscript{a}~All compute was executed on the UConn Storrs High Performance Cluster (HPC). \textsuperscript{b}~Total job-time included time spent by each job to download each 5--13GB snapshot over a network connection from a remote server to a local machine with limited storage space. \textsuperscript{c}~Each physical tensor generated per snapshot represents a different observer position (see \autoref{subsec:Physical Tensor Interpolation}). \textsuperscript{d}~See \autoref{subsec:Resolution Convergence} for discussion regarding the sensitivity of synthetic observation to the radial resolution of physical tensors, as well as for notes regarding the alternate radial resolutions adopted for some of the figures and diagnostics presented in this study. \textsuperscript{e}~We choose $r_\text{observer}$ to be the approximate solar distance $R_0 \approx \SI{8277}{\parsec}$ to $\SgrAStar$ given in \citet{GRAVITY2022} and \citet{GRAVITY2021}. \label{table:Dataset Parameters}}
        \tablehead
        {
            & \colhead{Training Dataset} & \colhead{Foreground Dataset} & \colhead{Full-Cone Test Dataset}
        }
        \startdata
        Number of compute jobs & 80 & 8 & 1 \\
        CPU allocation per job & 6 AMD EPYC 7713 cores \textsuperscript{a} & 6 AMD EPYC 7713 cores \textsuperscript{a} & 3 AMD EPYC 7713 cores \textsuperscript{a} \\
        GPU allocation per job & 2 40GB NVIDIA A100 \textsuperscript{a} & 2 40GB NVIDIA A100 \textsuperscript{a} & 1 40GB NVIDIA A100 \textsuperscript{a} \\
        Memory allocation per job & 80GB & 270GB & 256GB \\
        Simulation runs in database & 1 & 1 & 1 \\
        Snapshots in database & 710 & 710 & 710 \\
        First/last snapshot timestamps & $\SI{237}{\mega\year}$, $\SI{244}{\mega\year}$ & $\SI{237}{\mega\year}$, $\SI{244}{\mega\year}$ & $\SI{237}{\mega\year}$, $\SI{244}{\mega\year}$ \\
        Snapshots sampled per job & 20 \textsuperscript{c} & 10 \textsuperscript{c} & 4 \textsuperscript{c} \\
        Physical tensors per snapshot & 64 & 32 & 1 \\
        Total number of data points & 102{,}400 & 2560 & 4 \\
        Total compute time & $\sim 350$ job-hours \textsuperscript{b} & $\sim 40$ job-hours \textsuperscript{b} & $\sim 6$ job-hours \textsuperscript{b} \\
        Physical tensors stored on disk? & NO & NO & NO \\
        Top-down images stored on disk? & YES & YES & YES \\
        $\ell, v$ observations stored on disk? & YES & YES & YES \\
        Total disk usage & 210 GB & 13 GB & 8MB \\
        Distance perturbation scaling & $\alpha \sim U[0.25, 1)$ & $0.5$ & $\alpha \sim U[0.4, 0.6)$ \\
        Density perturbation scaling & None applied & None applied & None applied \\
        Grid resolution $(r, \ell, b, v)$ & $(512, 512, 128, 512)$ \textsuperscript{d} & $(2048, 512, 128, 512)$ \textsuperscript{d} & $(16384, 512, 128, 512)$ \textsuperscript{d} \\
        $r_{\text{observer}}$ & $\SI{8277}{\parsec}$ \textsuperscript{e} & $\SI{8277}{\parsec}$ \textsuperscript{e} & $\SI{8277}{\parsec}$ \textsuperscript{e} \\
        $r_{\text{min}}$ & $\SI{8052}{\parsec}$ & $\SI{6000}{\parsec}$ & $\SI{1302}{\parsec}$ \\
        $r_{\text{max}}$ & $\SI{8502}{\parsec}$ & $\SI{7500}{\parsec}$ & $\SI{15702}{\parsec}$ \\
        $\ell_{\text{min}}$ & $\SI{-1.5}{\degree}$ & $\SI{-1.5}{\degree}$ & $\SI{-1.5}{\degree}$ \\
        $\ell_{\text{max}}$ & $\SI{1.5}{\degree}$ & $\SI{1.5}{\degree}$ & $\SI{1.5}{\degree}$ \\
        $b_{\text{min}}$ & $\SI{-0.375}{\degree}$ & $\SI{-0.375}{\degree}$ & $\SI{-0.375}{\degree}$ \\
        $b_{\text{max}}$ & $\SI{0.375}{\degree}$ & $\SI{0.375}{\degree}$ & $\SI{0.375}{\degree}$ \\
        $v_{\text{min}}$ & $\SI{-200}{\km\s^{-1}}$ & $\SI{-200}{\km\s^{-1}}$ & $\SI{-200}{\km\s^{-1}}$ \\
        $v_{\text{max}}$ & $\SI{200}{\km\s^{-1}}$ & $\SI{200}{\km\s^{-1}}$ & $\SI{200}{\km\s^{-1}}$ \\
        \enddata
    \end{deluxetable*}

    \par Simulation represents the universe from which our reversion model learns. Modern galactic simulations, which are beginning to closely approximate observed reality in the CMZ \citep[e.g.][\citationLipman]{Tress2025, Tress2024, Tress2020, Petkova2023, Sormani2020}, provide a ``ground truth'' from which the model can learn physical constraints. This ground truth includes gas density, velocity, and temperature, as well as chemical abundances and dust temperature. Paired with synthetic observations of simulation snapshots (see \autoref{sec:Synthetic Observation Overview and Theory} and \autoref{sec:Synthetic Observation Implementation and Algorithms}), we are able to construct a dataset of processed simulations (or \textit{physical tensors}, see \autoref{subsec:Physical Tensors}) over which to perform supervised training of the reversion model (see \autoref{subsec:Supervised Reversion: Training Theory} and \autoref{subsec:Implementation of Reversion: Training Hyperparameters, Overfitting, and Regularization} for training details). All parameters for our final dataset construction are listed in \autoref{table:Dataset Parameters}.
    \par From the simulation dataset, the reversion model learns not only simple correlations, but more subtle necessities of prediction. In our case, for example, a synthetic observation in the $\ThirteenCOTwoOne$ line only provides direct information about the density distribution of CO. Since we wish, however, to predict the density distribution of $\HTwo$, which is more diffuse, the network must learn to ``color outside the lines'', i.e. assign plausible $\HTwo$-density predictions outside the boundaries of observable $\ThirteenCO$. Additionally, the network must learn to differentiate both noise and foreground/background features from meaningful emission originating within the top-down window over which we are seeking to infer a density image. Since all this intelligence must be learned organically from the simulations via the neural network training process, it is important that the simulations used in this research be sophisticated and physically representative. To meet this requirement, we use AREPO simulations with a zoom-in refinement on the CMZ, as developed by \citeLipman and described in \autoref{subsec:AREPO Zoom Simulations}.

\subsection{AREPO Zoom Simulations} \label{subsec:AREPO Zoom Simulations}
    \begin{figure*}[t]
        \centering
        \includegraphics[width=1\linewidth]{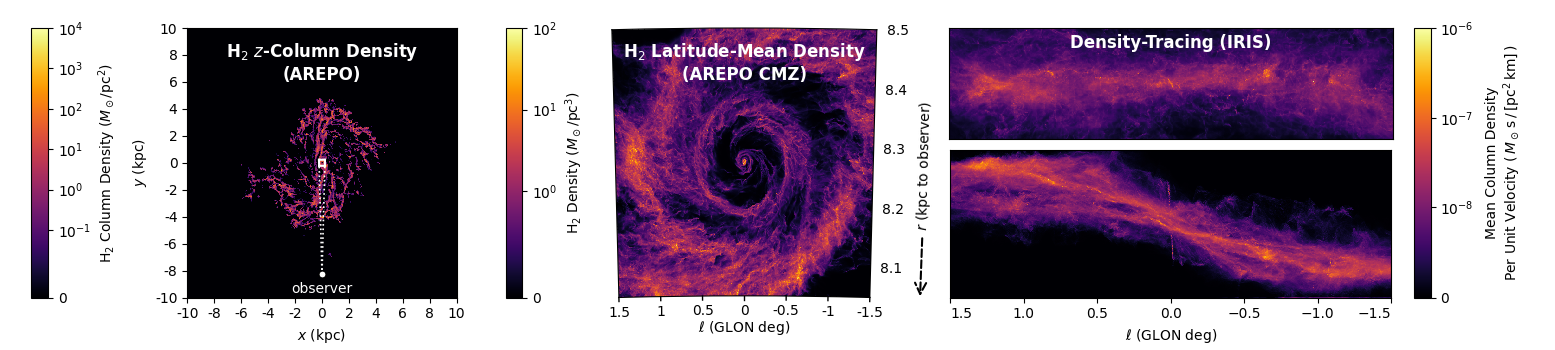}
        \caption{\textbf{Simulation Overview:} A visualization of the AREPO simulations due to \citeLipman, as described in \autoref{subsec:AREPO Zoom Simulations}, that we use in generating our training dataset (see \autoref{table:Dataset Parameters} for dataset parameters). In the left panel, we see the full top-down projection of $\HTwo$ density, integrated over the entire reduced $z$-axis to yield a column density. In the center panel, we see a zoomed-in plot of $\HTwo$ density over just the high-resolution CMZ region, meaned over the latitude extent specified for our training dataset in \autoref{table:Dataset Parameters}. In the right panel, we see a density-tracing observation, as described in \autoref{subsec:Synthetic Observation Versus Density-Tracing}, of this same CMZ region, in mean $\ell, b$ and $\ell, v$ reduction. The outline of the observed section, which is featured in top-down projection in the center panel, is overlaid onto the wide top-down projection in the left panel, as is the observer location. The observer distance is set according to the same value of $R_0 = \SI{8277}{\parsec}$ \citep[from][]{GRAVITY2022, GRAVITY2021} that we use in all synthetic observations in our training dataset (see, again, \autoref{table:Dataset Parameters}). All color bars are scaled with an $\arcsinh$ nonlinearity.}
        \label{fig:Sims Overview}
    \end{figure*}

    \par To construct the training dataset for our reversion model, we use ideal magneto-hydrodynamic (MHD) simulations produced by \citeLipman using the moving mesh code AREPO \citep{Springel2010,Pakmor2011,Pakmor2013,Pakmor2016,Weinberger2020}. The simulations of \citeLipman, for which we provide a visual overview in \autoref{fig:Sims Overview}, model a Milky Way-like galaxy while providing increased resolution in the CMZ region. These simulations build upon the robust frameworks from \citet{Tress2020}, \citet{Sormani2020}, \citet{Tress2024}, and \citet{Tress2025}, and incorporate ideal MHD, gas self-gravity, star formation, supernova feedback, and a rudimentary chemical network. Star formation is modeled by a combination of both static-mass star particles with supernova feedback \citep{Goller2025, Tress2025}, and sink particles that model mass accretion at greater computational expense \citep{Federrath2010, Tress2020}.
    \par \citeLipman initialize a disk of $\SI{5}{\kilo\parsec}$ radius, in an external Milky Way-like gravitational potential \citep{Hunter2024}, with a seed magnetic field of initially uniform and poloidal intensity given by $\mathbf{B} = \SI{0.02}{\micro\gauss} \, \hat{\mathbf{z}}$, and star particles to generate turbulence in the disk. This initial disk is run for $\sim \SI{100}{\mega\year}$ at a global resolution of $500\Msun$, where $\Msun$ is the solar mass unit. After the disk reaches a stable state, the galactic bar potential is slowly turned on between 100--$\SI{240}{\mega\year}$ to avoid transient artifacts.
    \par The sub-grid zoom-in geometry is activated at $\SI{200}{\mega\year}$, and consists of three resolution regions:
    \begin{enumerate}[(i)]
        \item an $R = \SI{500}{\parsec}$ region around the CMZ at the target resolution;
        \item an elliptical region with axial lengths of $R_{\mathrm{major}} = \SI{3}{\kilo\parsec}$ and $R_{\mathrm{minor}} = \SI{0.6}{\kilo\parsec}$ surrounding and co-rotating with the galactic bar at a resolution of 100 times the target resolution; and
        \item the rest of the galactic disk at 500 times the target resolution.
    \end{enumerate}
    The inclusion of the intermediate-resolution bar region mitigates potential artifacts from overshooting gas, which may enter and exit the CMZ multiple times on short dynamical timescales. Additionally, to aid in computational expense, material that deviates from the galactic plane is de-refined to 100 times the target resolution for scale heights $R_z > \SI{200}{\parsec}$, or 500 times for $R_z > \SI{500}{\parsec}$. The zooms are initially run at a target resolution of $100\Msun$ from $\sim 200$--$\SI{240}{\mega\year}$. The final target resolution of $1\Msun$ is activated between $\sim 240$--$\SI{250}{\mega\year}$, at which point new star-particle formation is halted in the spherical CMZ region, and accreting sink particles are allowed to form within the CMZ.
    \par \citeLipman use the implementation of star particles as described in \citet{Goller2025} and \citet{Tress2025}, in which stars are formed if the gas cell is unresolved compared to the Jeans' Mass and converging at a local minimum of the gravitational potential. Resolved gas is converted into stars over the free-fall time, with star-formation efficiency of $\epsilon = 0.1$, and formation is otherwise determined stochastically \citep{Springel2003}. Stellar masses are determined by sampling the high-end of a Kroupa initial-mass function \citep[IMF,][]{Kroupa2001}, and subsequent stellar lifetimes are based on Table 25.6 from \citet{Maeder2008}. The star particles represent stellar associations with masses between $10$--$1000\Msun$.
    \par Sink particles are only allowed to form in the CMZ region during the highest-resolution zooms, and are interpreted as small stellar clusters and pre-stellar cores. Following the prescription from \citet{Federrath2010} and \citet{Tress2020}, a sink particle may form if gas is within an accretion radius above a given density threshold, which is set in these simulations to $\rho_{\rm{crit}} = \SI{1e6}{\cm^{-3}}$. Similar to star particles, the gas flow must be converging on a gravitationally bound, local minimum of potential. Additionally, candidates are required to not be located inside or near the accretion radius of another sink. A star-formation efficiency of $\epsilon_{\rm{sink}} = 1$ is assumed, which immediately converts all accreted gas into stellar mass and avoids trapping non-stellar gas inside the sink. Both stars and sinks are allowed to freely enter and exit the CMZ after formation, and may undergo supernovae at any location.
    \par In addition to gas and star-formation physics, the simulations incorporate the NL97 chemical network from \citet{NL97_GloverClark_2012}, which tracks the chemical abundance of hydrogen \citep[specifically H, $\HTwo$, and $\HPlus$,][]{Glover2007a,Glover2007b} and CO \citep{Nelson_Langer1997}. In modeling the chemical reaction rates necessary for tracking abundances, the network follows radiative heating and cooling of the gas based on a detailed atomic and molecular cooling function \citep[for more detailed explanation please refer to][]{NL97_GloverClark_2012,Mackey2019}. This cooling function also incorporates dust cooling, from which the simulation computes a variable dust temperature we incorporate in physical-tensor construction (\autoref{subsec:Physical Tensors}, \autoref{subsec:Tensor Variables and Gas Temperature}) and then utilize for synthetic observation (\autoref{subsec:Dust Emission and Absorption}).
    \par The accuracy of this chemical network is somewhat limited by a variety of simplifications introduced by \citet{NL97_GloverClark_2012} as well as by the simulation resolution. In particular, the high sensitivity to local turbulence of $\HTwo$ and CO abundance leads to a necessary resolution for attaining abundance convergence that is substantially higher than even the peak resolution of these simulations \citep{Joshi2019}. Note, however, that the measure of convergence investigated by \citet{Joshi2019} is with respect to computationally negligible error in the abundances themselves. \citeLipman find that this error in chemical abundances has negligible impact on the overall structure and thermodynamical behavior of the simulated gas and ISM. We also find that our own error tolerance in computing synthetic observations derived from these abundances is far higher than this computational negligibility measure. For those reasons, we find that the NL97 chemical network in these simulations is suitable for our purposes.
    \par In constructing the training dataset for our reversion model (see \autoref{table:Dataset Parameters} for dataset parameters), we were limited by availability of simulation snapshots. The zoom-simulations of \citeLipman, in particular, demand not only massive computational expense to run but substantial storage to save snapshots. At full resolution, a single snapshot of this simulation is $\sim 13$GB. For these reasons, we were limited in access to 710 snapshots from a single zoom-simulation run with time-stamps ranging over the interval $[\SI{237}{\mega\year}, \SI{244}{\mega\year}]$ ($\sim \SI{7}{\mega\year}$ total time evolution with snapshots every $\sim \SI{10}{\kilo\year}$). For this proof-of-concept, we prioritized proving that our reversion model could learn the most broadly varying synthetic dataset we could construct, as opposed to prioritizing maximal physical accuracy for best generalization to the SEDIGISM data \citep{Schuller2021}. We therefore sampled randomly from this 710-snapshot repository in constructing all our datasets for this publication, even though the gravitational bar potential and zoom resolution are only fully turned on after $\SI{240}{\mega\year}$. For more mature iterations of this research in which we focus in greater depth on generalization to real observations, we aim to create a massively multi-run dataset in which snapshots are only sampled from portions of the simulation where refinement is at full resolution and all physics is turned on (see \autoref{sec:Summary and Conclusion} for more commentary on future research directions).

\subsection{Physical Tensors} \label{subsec:Physical Tensors}
    \par AREPO simulations output HDF5 snapshots that describe a Voronoi mesh of gas cells \citep{Springel2010}. Each cell is ascribed a multitude of properties---position, velocity, density, chemical abundances, etc. This native format is not convenient to the purposes of producing synthetic observations and training a reversion model. Instead, we require a data product that is readily accessible in PyTorch \citep{Paszke2019} and that encodes the minimal necessary information. IRIS achieves this objective first by processing a set of AREPO snapshots into a dataset of objects we term \textit{physical tensors}. In the current implementation of IRIS, A physical tensor is a PyTorch data tensor composed of seven channels---line-of-sight velocity, density, gas temperature, $\HTwo$ abundance, CO abundance, and dust temperature---each sampled from AREPO cell particles over a spherical-coordinate grid. The choice of spherical as opposed to Cartesian coordinates is made to simplify synthetic observation of spectral-line cubes in $\ell, b, v$ coordinates, as described in \autoref{sec:Synthetic Observation Overview and Theory}, \autoref{sec:Synthetic Observation Implementation and Algorithms}, and \autoref{sec:Synthetic Observation Testing and Verification}.
    \par The origin of the spherical-coordinate system is an observer point in the galactic plane at a distance $r_\text{observer}$ to the galactic center. The lattice points are sampled at regular intervals over radius $r$, galactic longitude $\ell$, and galactic latitude $b$, where galactic longitude and latitude are defined per the standard convention with respect to the observer point and the $(r, \ell, b) = (r_\text{observer}, 0, 0)$ vector pointing from the observer to the galactic center. The upper and lower limits of $r$ are the radial limits of the synthetic-observation process. For our training dataset, these radial limits also correspond to the bounds of the top-down window over which the reversion model predicts a density image (see \autoref{subsec:Supervised Reversion: Objective}). For our full-cone dataset, which we use for model testing and visualization, these radial bounds are taken as the fore and rear ends of the entire observational cone, which is substantially longer (see \autoref{subsec:Foreground and Background} and \autoref{table:Dataset Parameters} for comparison of the training and full-cone datasets). The upper and lower limits of $\ell, b$ are taken to match the size of the real observation desired for comparison, which for this study we choose as a cropped SEDIGISM cube \citep{Schuller2021}. These coordinate constants are all configurable in the IRIS code and values for the training data generated for this paper are listed in \autoref{table:Dataset Parameters}.

\subsection{Tensor Variables and Gas Temperature} \label{subsec:Tensor Variables and Gas Temperature}
    \par Velocity, density, $\HTwo$ and CO abundance, and dust temperature are directly available from AREPO cell variables. The abundances are expressed as they are in AREPO---as fractional abundances relative to the total number density of all H atoms. Dust temperature is computed by the NL97 chemical network based on the radiative cooling of dust in thermal equilibrium \citep{NL97_GloverClark_2012}. Gas temperature, on the other hand, must be computed from each cell's internal energy density per unit mass, $u$. In the AREPO simulations of \citeLipman the internal energy of each cell is modeled as incorporating translational kinetic energy only. Rotational and vibrational kinetic energy are not modeled as separate internal energy compartments. This simplification arises because the hydrodynamics of these simulations are evolved based upon the ideal-gas equation of state, $P = (\gamma - 1){\rho}u$, where $P$ is the gas pressure, $\rho$ is the gas density, and the adiabatic index is taken to be $\gamma = 5/3$ \citep[e.g.][]{Tress2020}. By the Equipartition Theorem, the accessible degrees of thermodynamic freedom of the simulated gas are related to this adiabatic index by $f = 2/(\gamma - 1) = 3$ \citep{Landau1980}. Hence, only the three translational degrees of freedom are relevant to the simulation hydrodynamics and $u$---even for $\HTwo$, which may be rotationally active within the temperature regimes spanned by the simulation.
    \par Continuing under the assumption that the gas within each cell is ideal, we then have again via the Equipartition Theorem that
    \begin{equation}
        u\mu = \frac{3}{2}kT_\text{kin} \text{ ,}
    \end{equation}
    where $\mu$ is the mean mass per gas molecule, $k$ is the Boltzmann constant, and $T_\text{kin}$ is the exact kinetic temperature, i.e. the temperature corresponding to translational degrees of freedom only. Note that we do not assume local thermal equilibrium (LTE). Since only translational freedom is present ($f = 3$), the equation for $T_\text{kin}$ is by definition rather than by thermal equilibrium (as would be the case for some other value of $\gamma$). We then define the IRIS gas temperature to be $T_\text{kin}$. In terms of the cell mass $M$ and the total number of gas molecules $N$---each expressed in terms of the masses $m_\text{H}, m_\text{He}$ of H and He, respectively, and the abundances $\abundance{\text{H}}, \abundance{\HPlus}, \abundance{\text{e}^-}, \abundance{\HTwo}, \abundance{\text{He}}$ per total H atoms $N_{\text{H TOT}}$---we compute $\mu$ as
    \begin{align}
        \mu &= \frac{M}{N} \\
        &= \frac{N_{\text{H TOT}}(m_H + m_{He}\abundance{\text{He}})}{N_{\text{H TOT}}(\abundance{\text{H}} + \abundance{\HPlus} + \abundance{\text{e}^-} + \abundance{\HTwo} + \abundance{\text{He}})} \text{ .} \notag
    \end{align}
    The $N_{\text{H TOT}}$ dependence cancels, and the negligible contribution of low-abundance CO is ignored.
    \par The downstream uses of the gas-temperature channel of a physical tensor are three-fold, all within the synthetic-observation pipeline: (i) temperature is used in looking up collision rates for a given tracer molecule as pertains to solving the tracer level populations (see \autoref{subsec:Level Populations: Computational Challenges}, \autoref{subsec:Level Populations: The Optically Thin Assumption}, and \autoref{subsec:Level Populations: Mathematical Solution}); (ii) temperature is used in determining thermal broadening of the spectral line (see \autoref{subsec:The Radiative Transfer Equation}); and (iii) temperature is one of the variables over which the IRIS synthetic-observation module permits specification of a derived tracer abundance if observing a complex tracer not modeled in the AREPO chemical network (i.e. other than CO, see \autoref{subsec:Spectral-Line Configuration}). To the first two applications, the kinetic temperature computed above is the exact variable required. Regarding the third application, the significance of this temperature variable is dependent upon the definition of the abundance function, which is a downstream consideration. For the $\ThirteenCOTwoOne$ test-case explored in this paper, we use a constant $\ThirteenCO$ abundance function that is temperature-independent (see \autoref{subsec:Spectral-Line Configuration}).

\subsection{Physical Tensor Interpolation} \label{subsec:Physical Tensor Interpolation}
    \par From each single snapshot, IRIS produces a configurable number $n_\text{points}$ of separate physical tensors, with individual observer points arranged as the vertices of a regular $n_\text{points}$-gon about the galactic center. That is, from each snapshot, we produce $n_\text{points}$ different physical tensors corresponding to $n_\text{points}$ different observational perspectives. For the IRIS training dataset, we set $n_\text{points} = 64$ (see \autoref{table:Dataset Parameters}). The most computationally intensive step in this physical-tensor generation process is then the interpolation of the cell values over the spherical-coordinate lattice. IRIS uses the nearest-neighbor interpolation algorithm, which it can compute either on CPU via the SciPy library \citep{Virtanen2020} or on GPU via the CuPy library \citep{Nishino2017}. Nearest-neighbor interpolation prioritizes computational efficiency at the expense of sharp edges produced at cell boundaries. At the resolution of our publication training dataset, by virtue of the zoom-in refinement employed by the simulations on which this dataset was generated (\citationLipman, see \autoref{subsec:AREPO Zoom Simulations}), the AREPO cells are comparable in size to physical-tensor voxels, and so these sharp boundaries are minimized. When using lower-resolution simulations, however, these boundaries can become highly noticeable in both top-down density projection and synthetic observation.
    \par The sharp boundaries generated by the nearest-neighbor interpolation scheme have the most substantial effect over the physical-tensor velocity channel. The wide regions of constant velocity separated by discontinuous boundaries appear, in the $\ell, v$ projection of a synthetic observation, as flat and disconnected streaks of intensity. In order to yield a smoother synthetic observation, which we conjecture provides more reliable data in training the reversion model to generalize onto real observations that lack these streak artifacts, we apply a Gaussian blurring convolution to each physical tensor over the velocity channel only. This blurring convolution, which is implemented as a fast PyTorch operation \citep{Paszke2019}, eliminates sharp velocity boundaries and the resulting intensity streaks in synthetic observations. We opt not to apply a blur over the physical-tensor channels as even small changes that are locally mean-preserving can yield substantial fluctuations in synthetically observed intensities that are not strictly necessary in terms of producing clean training data.

\subsection{Physical Tensor Perturbations} \label{subsec:Physical Tensor Perturbations}
    \par One issue we encountered in generating a large training dataset of sufficient variation for training of our reversion model (see \autoref{subsec:Supervised Reversion: Training Theory} and \autoref{subsec:Implementation of Reversion: Training Hyperparameters, Overfitting, and Regularization}) was the limited number of AREPO snapshots output by independent simulation runs. These simulations are highly computationally intensive and it is not straightforward to generate the large suites of varied simulations needed for more complete training. Practically, for this study, we aimed for a proof-of-concept using snapshots from only a single simulation run. In order to artificially enhance the variation of training data given this constraint on simulation availability, we experimented with different types of random perturbations made to each physical tensor. The primary perturbation type with which we experimented was that of distance perturbations. Distance perturbations uniformly alter the simulation size by scaling the unit of length in both the positions and velocities of all AREPO cells by a constant factor. These distance perturbations are density-conserving and thus not mass-conserving.
    \par Distance perturbations are particularly important because the AREPO simulations from \citep[][\citationLipman]{Sormani2020, Tress2020, Tress2024, Tress2025} tend to yield CMZ-like regions that are larger than the observed Milky-Way CMZ by a factor of roughly two. We refer to \citet{Henshaw2023}, \citet{Sormani2024}, \citet{Tress2025}, and \citeLipman for more detailed discussion on this scale difference, but note that its meaning, and even the interpretation of the real CMZ's size, are subjects of active discussion. For our publication training dataset, we implemented distance perturbations with scaling constants randomly sampled per physical tensor from a uniform distribution over the interval $[0.25, 1)$ (See \autoref{table:Dataset Parameters}). We applied similar random distance perturbations with scaling constants sampled over the interval $[0.4, 0.6)$ to each physical tensor shown in the visualization of our trained model's performance on synthetic data (\autoref{fig:Synthetic Reversions}). In all other figures, no distance perturbations were applied. We also experimented with perturbations that uniformly scale the densities of all AREPO cells by a constant, but did not implement these perturbations into the publication training dataset.

\subsection{Foreground and Background} \label{subsec:Foreground and Background}
    \par Another issue we encountered in producing a large training dataset for our reversion model was in the substantial compute requirements for generating ``full-cone'' physical tensors and synthetic observations. By ``full-cone'', we mean the full observational cone, with lines of sight extending across the entire galactic disk, as opposed to standard physical tensors and synthetic observations made over the constrained radial window of the CMZ region only, which leaves out the foreground and background. While, for our chosen simulations (\autoref{subsec:AREPO Zoom Simulations}) and tensor resolution (\autoref{table:Dataset Parameters}), construction and synthetic observation of a physical tensor over the CMZ region can be computed by a single CPU-plus-GPU process in approximately 10 seconds, construction and synthetic observation of a full-cone physical tensor introduces orders of magnitude more computational complexity.
    \par This added complexity has multiple sources. First, a full-cone physical tensor occupies substantially more space in memory. Physical-tensor interpolation, when performed on GPU via the fast CuPy option (see \autoref{subsec:Physical Tensor Interpolation}), is radially chunked into smaller interpolation problems that fit within GPU memory. This chunking yields time complexity that scales linearly with total radial resolution even before considering the time scaling of the nearest-neighbor tree search employed by the interpolation algorithm itself. But more substantially, synthetic-observation time complexity scales quadratically with radial resolution, since the transfer solution must not only be iterated over more radial steps but must be chunked into smaller ray batches to prevent GPU memory overflow. (See \autoref{sec:Synthetic Observation Implementation and Algorithms} for details on synthetic-observation implementation.) We find that full-cone physical-tensor construction and synthetic observation at our desired data resolution requires approximately 1 hour in compute time for a single CPU and GPU process.
    \par Full-cone observations differ from synthetic observations constrained to the CMZ region in the presence of foreground and background features. These foreground and background features---in particular, the Milky Way's spiral arms, which manifest in $\ell, v$ images as low-velocity, horizontal bands of intensity---are prominently visible in real observations of the CMZ, and specifically the SEDIGISM $\ThirteenCOTwoOne$ data \citep{Schuller2021} used as the test case for this paper. It is therefore important that the reversion model sees these foreground and background features during training so that it learns how to intelligently ignore these features when inferring a top-down density image.
    \par Due to the computational time and complexity, we opted not to generate an entire training dataset of full-cone observations. We instead generated a smaller, separate dataset of foreground/background features only, which we added randomly, during training, to the primary dataset of observations constructed over the radial window of the CMZ only. This simplification, of course, supposes that the ideal behavior of the reversion model with respect to foreground/background features is only in ignoring these features, i.e. that there is no information in these features useful in inferring a top-down image of the CMZ region only. We conjecture that this simplification is a reasonable first-order approximation, though it may in theory be possible that foreground/background features encode global information regarding galaxy structure that provides meaningful insight into the local structure of the CMZ region.
    \par We find that training via a random foreground dataset yields a reversion model that produces good results when tested on actual full-cone observations. (See \autoref{subsec:Supervised Reversion: Training Theory}, \autoref{subsec:Implementation of Reversion: Choice of Loss Function}, and \autoref{subsec:Implementation of Reversion: Training Hyperparameters, Overfitting, and Regularization} for training details.) We feature full-cone observations in the visualization of our trained model's performance on synthetic data (\autoref{fig:Synthetic Reversions}), but show CMZ-only synthetic observations in all other figures. None of the visualizations for this publication are of foreground-added observations.

\subsection{Computational Considerations} \label{subsec:Computational Considerations}
    \par Because of the substantial compute requirements in dataset generation, parallelization is a crucial optimization. We built IRIS to employ a scalable approach of mixed CPU and GPU parallelization via synthesis of the PyTorch backend \citep{Paszke2019}, which enables streamlined GPU tasking, with CPU-parallel task-management using the MPI for Python (\texttt{mpi4py}) package \citep{Dalcin2021}. Individual CPU workers each work independently to process a single physical tensor at a time, coordinating for shared GPU access at key computational bottlenecks. This approach scales cleanly to an unstructured CPU/GPU cluster of any size and topology, including across multiple networked compute nodes, and thus provides the scaffolding for future efforts in massively distributed data generation.
    \par The most substantial limitation for our compute resources is dataset size, which translates to both storage space on disk and load latency when streaming data from the disk during training. For our large datasets, it was more practical to synthetically observe the physical tensors during data generation, and then store, to the disk, only pairs of synthetically observed $\ell, v$ images and top-down density images, which are each only 2D data products. This setup prevents having to store the entire 3D physical tensor on disk, and removes the necessity of redundant synthetic observation during training, while still enabling the supervised training approach (see \autoref{subsec:Supervised Reversion: Training Theory} for details on this training scheme). This approach thus improves disk-usage efficiency and load latencies by orders of magnitude compared with storage of the full physical tensors. Nonetheless, this shortcut is incompatible with the future concepts we hope to explore regarding 3D-to-3D reversion (see \autoref{subsec:Supervised Reversion: Higher Dimensionality} and \autoref{sec:Summary and Conclusion}), from which we conclude that such future efforts may require orders of magnitude more storage space. We retained full physical tensors as needed for generating some of the figures in this publication.
    \par One last computational consideration in physical-tensor construction and synthetic observation is the choice of units and numerical stability. The process of synthetic observation, in particular, requires an enormous dynamic range in order to balance small quantities---such as observed intensities, molecular masses, and the Planck and Boltzmann constants---with large quantities such as distances and molecular number densities over astronomical scales. Despite the breadth of this required range, IRIS achieves numerical stability in single-precision, floating-point arithmetic across the entire data-construction and training pipeline. To achieve single-precision stability, we employed multiple separate units systems, which we describe in further detail in the IRIS code documentation. An especially useful trick was the introduction of a large number-unit for handling number densities.
    \par Following data generation, we renormalized our publication training dataset into a new set of units. These units were chosen to achieve unit standard deviation, over the whole dataset, of Raleigh-Jeans temperature in $\ell, v$ observations (see \autoref{subsec:Intensity Versus Temperature}) and of density in top-down images. Normalizing these statistics in the input and output spaces of the reversion model promote more stable gradients for better training. To ensure accurate estimation of the population standard deviation from a random sample of physical tensors, we applied Bessel's correction.

\section{Synthetic Observation Overview and Theory} \label{sec:Synthetic Observation Overview and Theory}
    \definecolor{spectral}{RGB}{255,248,200}
    \definecolor{continuum}{RGB}{220,245,220}
    \definecolor{postproc}{RGB}{220,235,255}
    \definecolor{darkgreen}{RGB}{0,150,100}
    \newcommand{\yes}{\textcolor{darkgreen}{\ding{51}}}
    \newcommand{\no}{\textcolor{red}{\ding{55}}}
    \begin{table*}[t]
        \centering
        \small
        \begin{tblr}
        {
            width=.875\linewidth,
            colspec={|Q[c,m,wd=6.5mm]|Q[l,m,wd=0.69\textwidth]|Q[c,m,wd=0.25\textwidth]|},
            rows={ht=4.5mm},
            row{1}={font=\bfseries, ht=6.2mm},
            column{2}={leftsep=2pt,rightsep=2pt},
            column{3}={leftsep=2pt,rightsep=2pt},
            vlines,
            hlines,
            cell{2}{1}  = {r=11}{bg=spectral,   valign=m, halign=c},
            cell{13}{1} = {r=18}{bg=continuum,  valign=m, halign=c},
            cell{31}{1} = {r=6}{bg=postproc,    valign=m, halign=c},
        }
            & Effect & Modeled in IRIS-SO? \\

            \rotatebox[origin=c]{90}{\bfseries\textcolor{orange!80!black}{Spectral-Line Effects}}
            & \SetCell{bg=spectral} Single and/or multiple spectral lines & \SetCell{bg=spectral} \yes \\
            & \SetCell{bg=spectral} All chemical data automatically computed from LAMDA files & \SetCell{bg=spectral} \yes \\
            & \SetCell{bg=spectral} Spontaneous emission & \SetCell{bg=spectral} \yes \\
            & \SetCell{bg=spectral} Stimulated emission & \SetCell{bg=spectral} \yes \\
            & \SetCell{bg=spectral} Absorption & \SetCell{bg=spectral} \yes \\
            & \SetCell{bg=spectral} Line scattering & \SetCell{bg=spectral} \no \\
            & \SetCell{bg=spectral} Tracer level balance (local thermal equilibrium) & \SetCell{bg=spectral} \no \\
            & \SetCell{bg=spectral} Tracer level balance (optically thin) & \SetCell{bg=spectral} \yes \\
            & \SetCell{bg=spectral} Tracer level balance (nonzero escape probability) & \SetCell{bg=spectral} \no \\
            & \SetCell{bg=spectral} Zeeman splitting (magnetic field effects) & \SetCell{bg=spectral} \no \\
            & \SetCell{bg=spectral} Line polarization & \SetCell{bg=spectral} \no \\

            \rotatebox[origin=c]{90}{\bfseries\textcolor{green!50!black}{Continuum Effects}}
            & \SetCell{bg=continuum} Cosmic microwave background (CMB) & \SetCell{bg=continuum} \yes \\
            & \SetCell{bg=continuum} Thermal dust emission & \SetCell{bg=continuum} \yes \\
            & \SetCell{bg=continuum} Dust extinction & \SetCell{bg=continuum} \yes \\
            & \SetCell{bg=continuum} Dust scattering & \SetCell{bg=continuum} \no \\
            & \SetCell{bg=continuum} Single-species dust & \SetCell{bg=continuum} \yes \\
            & \SetCell{bg=continuum} Multi-species dust & \SetCell{bg=continuum} \no \\
            & \SetCell{bg=continuum} Dust opacity constant over single-line spectral extent & \SetCell{bg=continuum} \yes \\
            & \SetCell{bg=continuum} Dust opacity varying between separate lines & \SetCell{bg=continuum} \yes \\
            & \SetCell{bg=continuum} Dust opacity varying inside single-line spectral extent & \SetCell{bg=continuum} \no \\
            & \SetCell{bg=continuum} Dust density distribution coupled to gas density & \SetCell{bg=continuum} \yes \\
            & \SetCell{bg=continuum} Dust density independent of gas density & \SetCell{bg=continuum} \no \\
            & \SetCell{bg=continuum} Dust temperature independent of gas temperature & \SetCell{bg=continuum} \yes \\
            & \SetCell{bg=continuum} Complex/multi-scale dust-grain geometry & \SetCell{bg=continuum} \no \\
            & \SetCell{bg=continuum} Dust polarization & \SetCell{bg=continuum} \no \\
            & \SetCell{bg=continuum} Spinning and magnetic dust emission \citep{Erickson1957, Draine2003} \textsuperscript{\textdagger} & \SetCell{bg=continuum} \no \\
            & \SetCell{bg=continuum} Bremsstrahlung (free-free) emission\textsuperscript{\textdagger} & \SetCell{bg=continuum} \no \\
            & \SetCell{bg=continuum} Synchrotron emission\textsuperscript{\textdagger} & \SetCell{bg=continuum} \no \\
            & \SetCell{bg=continuum} Stellar radiation\textsuperscript{\textdagger} & \SetCell{bg=continuum} \no \\

            \rotatebox[origin=c]{90}{\bfseries\textcolor{blue!70!black}{Post-processing Effects}}
            & \SetCell{bg=postproc} Post-transfer continuum subtraction & \SetCell{bg=postproc} \yes \\
            & \SetCell{bg=postproc} Configurable antenna resolution (Gaussian beam convolution) & \SetCell{bg=postproc} \yes \\
            & \SetCell{bg=postproc} Coarse noise simulation (Gaussian speckle) & \SetCell{bg=postproc} \yes \\
            & \SetCell{bg=postproc} Real noise simulation (atmospheric effects) & \SetCell{bg=postproc} \no \\
            & \SetCell{bg=postproc} Real noise simulation (antenna electronics) & \SetCell{bg=postproc} \no \\
            & \SetCell{bg=postproc} Real noise simulation (interferometric artifacts) & \SetCell{bg=postproc} \no \\
        \end{tblr}
        \caption{\textbf{Effects Modeled in IRIS-SO:} Summary of the effects modeled and not modeled in the IRIS synthetic-observation code (IRIS-SO). IRIS-SO provides reasonable modeling of one or multiple spectral lines that are optically thin in the omnidirectional average, but not necessarily along all observed lines of sight. See \autoref{subsec:Level Populations: The Optically Thin Assumption} for more details on why an optically thin level balance is a weaker assumption than that of total non-absorption and non-stimulation of the line. \textsuperscript{\textdagger}~Contributors to the continuum that are believed to be weak throughout most regions of the CMZ in the $\SI{1}{\mm}$ band \citep{Battersby2020}. \label{table:IRIS-SO Modeling Effects}}
    \end{table*}
    
    \par Synthetic observations are the bridge between simulation and observable reality. The synthetic observations we produce for this publication are brightness-temperature images in the $\ThirteenCOTwoOne$ spectral line, parameterized over dimensions of galactic longitude, galactic latitude, and velocity. These PPV-cube images are intended to be directly comparable to those produced by the SEDIGISM survey \citep{Schuller2021}. Many codes exist for computing such synthetic line images, but, due to both the high volume of synthetic observations we require for this study, as well as the need for granular control over every element of our data pipeline, the IRIS synthetic-observation module (IRIS-SO) is implemented from scratch with minimal external dependencies.
    \par In particular, by virtue of GPU acceleration in PyTorch \citep{Paszke2019}, IRIS-SO achieves orders of magnitude of speedup (up to $10,000\times$, see \autoref{subsec:Speed Testing}) in comparison to CPU-based synthetic-observation codes, which is essential in producing the $\sim 100\text{k}$ synthetic observations needed for the IRIS training dataset. (See \autoref{sec:Synthetic Observation Implementation and Algorithms} for a more detailed description of the advantages of IRIS-SO over traditional synthetic-observation codes, including its end-to-end differentiability, as well as comparisons with comparable codes.) In order to optimize speed, however, we need to make some simplifying assumptions, which are summarized in \autoref{table:IRIS-SO Modeling Effects}. The aim of these assumptions is to reasonably model one or multiple spectral lines that are optically thin in the omnidirectional average, but not necessarily along all observed lines of sight. Dust is also coarsely modeled, as is antenna resolution and noise. These assumptions form a reasonable approximation of the true behavior of the $\ThirteenCOTwoOne$ line in the CMZ and the observed characteristics of the SEDIGISM survey.
    \par For technical completeness, and because some aspects of the IRIS-SO implementation represent unique contributions, we provide a detailed exposition of the IRIS-SO design over the next two sections. Specifically, in this section, we focus on the physical theory of radiative transfer enabling synthetic observation. In \autoref{sec:Synthetic Observation Implementation and Algorithms}, we describe the algorithmic aspects of the IRIS-SO design. Since the theory and implementation of synthetic observation has been explored thoroughly in the literature, however, some readers may wish to skip ahead to \autoref{sec:Synthetic Observation Testing and Verification}, in which we analyze the products of our synthetic-observation pipeline. In particular, we provide a detailed side-by-side verification of IRIS-SO against both RADMC-3D \citep[][see \autoref{fig:RADMC-3D vs IRIS}]{Dullemond2012} and POLARIS \citep[][see \autoref{fig:POLARIS vs IRIS}]{Reissl2016, Brauer2017}---a pair of more established synthetic-observation codes with a higher level of general utility but not equally optimized to the unique needs of this research. We also provide a detailed series of diagnostics and visual comparisons justifying various assumptions in our synthetic-observation code, as well as a detailed speed comparison against RADMC-3D (\autoref{fig:Speed Test}).

\subsection{The Radiative Transfer Equation} \label{subsec:The Radiative Transfer Equation}
    \par In a non-scattering medium, the change in radiative intensity $I_\nu$ per unit frequency at some fixed frequency $\nu$, along an observational ray parameterized by distance $s$, is related to the coefficients $j_\nu, \alpha_\nu$ of emission and absorption, respectively, by the radiative-transfer equation of the form
    \begin{equation} \label{eqn:Radiative Transfer Equation}
        \frac{dI_\nu}{ds} = j_\nu - \alpha_{\nu}I_\nu
    \end{equation}
    \citep{Rybicki1986}. For the purpose of synthetic observation, it is also useful to consider separately the total intensity $I_\nu^\total$ and the idealized continuum baseline $I_\nu^\cont$, which the total intensity approaches outside the spectral neighborhood of the line, where the line contribution is negligible. As a product of the simplifying assumptions outlined in \autoref{table:IRIS-SO Modeling Effects}, we treat dust as the only source of continuum emission and absorption, although the Cosmic Microwave Background (CMB) is modeled as a background radiation source. It is then also useful to consider separately the line emission and absorption coefficients $j_\nu^\sline, \alpha_\nu^\sline$ and the dust emission and absorption coefficients $j_\nu^\dust, \alpha_\nu^\dust$. We thus have
    \begin{equation} \label{eqn:Total Radiative Transfer}
        \frac{dI_\nu^\total}{ds} = \left(j_\nu^\sline + j_\nu^\dust\right) - \left(\alpha_\nu^\sline + \alpha_\nu^\dust\right)I_\nu^\total \text{ .}
    \end{equation}
    and
    \begin{equation} \label{eqn:Continuum Radiative Transfer}
        \frac{dI_\nu^\cont}{ds} = j_\nu^\dust - \alpha_\nu^{\dust}I_\nu^\cont \text{ .}
    \end{equation}

\subsection{Dust Emission and Absorption} \label{subsec:Dust Emission and Absorption}
    \par Dust emission and absorption are complex phenomena. In reality, the average dust emission and absorption over any volume of the ISM is the product of a multitude of dust grains of differing scales, geometries, polarizations, and chemical compositions. Moreover, dust emission is the product of a number of independent mechanisms, including thermal, spinning, and magnetic dust emission \citep{Draine2003}. These complexities prevent true dust emission from adhering strictly to a blackbody source function, and additionally prevent dust absorption from assuming a frequency-independent form over a large spectral range.
    \par As a result, many approaches have been explored in the literature to model the dust spectrum, most of which simplify some degree of this complexity. The most widely cited models describe observed intensity via modified-emissivity graybody spectra with scaling constants dependent on a power law of frequency \citep{Odegard2016}. These models are consistent with the approximation of the dust spectrum via an absorption coefficient that varies over frequency according to some emissivity factor, but that retains a blackbody source function
    \begin{equation} \label{eqn:Dust Source Equation}
        S_\nu\left(T\right) = \frac{j_\nu^\dust}{\alpha_\nu^\dust} = \frac{2h\nu^3}{c^2\left[e^{h\nu/\left(kT\right)} - 1\right]} \text{ ,}
    \end{equation}
    where $T$ is the dust temperature, $h$ is the Planck constant, $c$ is the vacuum speed of light, and $k$ is the Boltzmann constant. The immutability of this source function is also consistent with Kirchoff's Law of Thermal Radiation, which holds under the assumption that dust emission is purely thermal and that the dust is in local thermodynamic equilibrium (LTE) \citep{Rybicki1986}.
    \par Since dust emission and absorption are only minority effects in our synthetic observations (which we prove in \autoref{subsec:Optical Depth and Dust Effects}), we construct only a very coarse model using a combination of the most simplifying assumptions. We assume that dust has a single-species population and adheres to the blackbody source function. Note that we thus assume dust LTE, but that this assumption is not equivalent to that of gas LTE, which we do not assume (see \autoref{subsec:Level Populations: Computational Challenges}). We also assume that absorption is proportional to dust density via an opacity constant $\kappa$, which holds approximately in the limit where the dust radius is much smaller than the radiation wavelength, and in which the effects of any anisotropic distribution of dust-grain geometry and alignment are ignored \citep{Rybicki1986}.
    \par We further assume that $\kappa$ is constant over the small frequency extent of a single-line PPV cube. (We do allow the specification of different $\kappa$ values for different spectral lines, but this is a feature we reserve for future IRIS-SO uses, which we do not use for the $\ThirteenCOTwoOne$ synthetic observations computed for this publication.) We then assume that the distribution of dust density is coupled to the distribution of gas density via a standard gas-to-dust mass ratio. We take this mass ratio to be implicit within the constant $\kappa$, which we specify per unit mass of gas only. In terms of the gas density $\rho_\text{gas}$, the dust absorption coefficient is then given independently of frequency by
    \begin{equation}
        \alpha^\dust \equiv \alpha_\nu^\dust = \kappa\rho_\text{gas} \text{ ,}
    \end{equation}
    and the dust emission coefficient is determined via \autoref{eqn:Dust Source Equation}.
    \par We adopt values for $\kappa$ from \citet{Ossenkopf1994}. In the neighborhood of the $\ThirteenCOTwoOne$ line \citep[$\lambda \approx \SI{1.36}{\mm}$][]{Schoier2005}, multiplying by a conversion factor of $\SI{1e-2}{}$ to account for the standard assumption of a $100 : 1$ gas-to-dust mass ratio, we interpolate a rough value $\kappa \approx \SI{1e-2}{\cm^2\g^{-1}}$ from the values computed for the Mathis-Rumpl-Nordsieck \citep[MRN][]{Mathis1977} distribution with thin ice mantles at $\SI{1e5}{\year}$ coagulation and gas number-density of $\SI{1e6}{\cm^{-3}}$. We note that these values were determined from models of dust in protostellar cores. As such, they may overestimate average dust grain size in comparison to the average grain size distribution over the entire CMZ. Coarsely, we can treat average opacity as independent from grain-size distribution (see \autoref{subsec:POLARIS Dust Treatment} for a derivation), but we note this independence relies on idealized assumptions of spherical grains with radii much smaller than the observed wavelength. In reality, even noting that the value $\kappa$ is specified per unit mass, the inaccuracy in grain size distribution may yield a systemic bias in dust emission and absorption. As a first-order approximation, however, we take these values as sufficient since we find, in the literature, no more suitable study of average dust opacity values throughout the ISM over the CMZ specifically.

\subsection{Line Emission and Absorption} \label{subsec:Line Emission and Absorption}
    \par We determine the line emission and absorption coefficients $j_\nu^{\sline}, \alpha_\nu^{\sline}$ in terms of the level populations $n_u, n_l$. We define these populations as the number densities of tracer molecules in the upper and lower energy levels, respectively, of the line transition, and leave their estimation as the topic of \autoref{subsec:Level Populations: Computational Challenges}, \autoref{subsec:Level Populations: The Optically Thin Assumption}, and \autoref{subsec:Level Populations: Mathematical Solution}. The term $j_\nu^{\sline}$ accounts for spontaneous emission only, and is given in terms of the transition frequency $\nu_{ul}$, the Einstein coefficient $A_{ul}$, and a line profile function $\varphi(\nu)$ as
    \begin{equation} \label{eqn:Line Emission}
        j_\nu^{\sline} = \frac{h\nu_{ul}}{4\pi}n_uA_{ul}\varphi(\nu) \equiv j^{\sline}\varphi(\nu)
    \end{equation}
    \citep{Rybicki1986}. We define, here, the frequency-independent emission coefficient $j^{\sline}$, which holds a computational convenience we explain in \autoref{subsec:Grid Precomputation}. The term $\alpha_\nu^{\sline}$ accounts for both absorption and---as a negative contribution---stimulated emission, and is given in terms of the Einstein coefficients $B_{lu}$ and $B_{ul}$ as
    \begin{equation}  \label{eqn:Line Absorption}
        \alpha_\nu^{\sline} = \frac{h\nu_{ul}}{4\pi}(n_lB_{lu} - n_uB_{ul})\varphi(\nu) \equiv \alpha^{\sline}\varphi(\nu)
    \end{equation}
    \citep{Rybicki1986}. We likewise define the frequency-independent absorption coefficient $\alpha^{\sline}$.
    \par The line profile $\varphi(\nu)$ determines the spread of the spectral line in frequency space. While, in the most naive context, $\varphi(\nu)$ may be treated as a Dirac delta, multiple processes contribute to the broadening of the line in reality. Natural broadening, which stems from the Uncertainty Principle, yields the Lorentz profile, which is dominant at the line ``wings''. Near the line center, however, the dominant broadening process is the Doppler shift stemming from the thermal kinetic energy of the tracer molecules \citep{Rybicki1986}. Taken independently of the natural broadening, this thermal broadening produces the Gaussian profile
    \begin{equation} \label{eqn:Thermal Broadening Profile}
        \varphi(\nu) = \frac{e^{-(\nu - \nu_{ul})^2/\Delta\nu_{\scriptscriptstyle{D}}^2}}{\Delta\nu_{\scriptscriptstyle{D}}\sqrt{\pi}}
    \end{equation}
    where
    \begin{equation}
        \Delta\nu_{\scriptscriptstyle{D}} = \frac{\nu_{ul}}{c}\sqrt{\frac{2kT}{m_a}} \text{ ,}
    \end{equation}
    and $m_a$ is the molecular mass of the tracer \citep{Rybicki1986}. This Doppler profile can easily be adjusted to account for microturbulent kinetic energy in addition to thermal kinetic energy, and the Lorentz profile and Doppler profile can be convolved into a more general Voigt profile, which accounts for both broadening processes. For simplicity, however, IRIS uses only the purely thermal Doppler broadening as described by \autoref{eqn:Thermal Broadening Profile}.

\subsection{Level Populations: Computational Challenges} \label{subsec:Level Populations: Computational Challenges}
    \par The tracer level populations $n_u, n_l$, while requisite in determining the line emission and absorption coefficients, are themselves functions of the radiation field. In particular, provided the tracer gas is sufficiently optically thick to absorb or be stimulated by its own line radiation, the tracer levels are then coupled specifically to the line component of this radiation field. Moreover, the function of the radiation field that is germane to the task of estimating the level populations is not just the intensity $I_\nu$ along the observed line of sight, but the intensity averaged over all omnidirectional rays, $J_\nu$, at each point along the observed line of sight \citep{Rybicki1986, vanderTak2007}.
    \par This coupling poses a massive computational challenge to synthetic line observation. Naively, we would need to wrap the numerical solution method to the radiative-transfer equation over the entire simulated volume and all omnidirectional rays in a nonlinear solver for the level populations. This naive approach would require iterating the nonlinear solver over the iterated differential equation solver, which is computationally infeasible. In practice, we must decouple the level populations from the radiation field via some simplifying approximation \citep{vanderTak2007}.
    \par The most general such approximation involves estimation of an escape/reabsorption probability of a photon emitted by the tracer molecule under the assumption of a large velocity gradient (LVG), which ensures that line self-absorption/stimulation is strictly local \citep{Sobolev1960, vanderTak2007}. IRIS does not use this method for the following reasons: First, it would require estimation of the omnidirectional mean-intensity field at all points within the transfer medium, which is far more intensive than computation of the radiative transfer along observed rays only. Second, this mean intensity would need to be incorporated into the emission/absorption-coefficient grids that IRIS precomputes as a critical computational optimization (see \autoref{subsec:Grid Precomputation}). This additional grid dimension would be highly expensive in GPU memory.
    \par Another more simple approximation for decoupling the level populations from the radiation field is that of assuming local thermodynamic equilibrium (LTE), which supposes that the level populations are determined by temperature and density according to Maxwell-Boltzmann statistics \citep{Rybicki1986}. While LTE is the most simple means of estimating the level populations, IRIS likewise does not assume LTE, as LTE is often not attained within the low-density ISM, in which substantial level-degeneracy due to the dominance of spontaneous emission routinely arises. We demonstrate this failing definitively in \autoref{subsec:Validity of the Optically Thin Level Balance}, in which we show that the LTE approximation yields substantially different results than the LVG approximation under our observed conditions of the $\ThirteenCOTwoOne$ spectral line over our AREPO simulations (see \autoref{fig:OT vs LVG Balance} and \autoref{fig:OT vs LTE Balance}). Note, however, that we do assume dust LTE, which is not equivalent to gas LTE (see \autoref{subsec:Dust Emission and Absorption}).

\subsection{Level Populations: The Optically Thin Assumption} \label{subsec:Level Populations: The Optically Thin Assumption}
    \par Rather than the LVG or LTE assumptions, IRIS uses an optically thin (OT) assumption. In the simplest version of the OT approximation, it is assumed that the tracer gas cannot absorb or be stimulated by radiation, and so the level balance is a function of spontaneous emission and collisions only, which is a good approximation in low optical depth \citep{vanderTak2007}. Two nuances arise regarding the OT level balance, however. The first of these nuances is that this assumption is not strictly equivalent to the assumption that the line transfer itself is everywhere optically thin. Indeed (depending on the enabled transfer mode, see \autoref{subsec:Transfer Solution: Overview}) IRIS does compute both absorption and stimulated emission of the line. We find in \autoref{subsec:Optical Depth and Dust Effects} that for the case of $\ThirteenCOTwoOne$ on our test simulations, results differ substantially from emission-only computations (see \autoref{fig:Optically Thin vs Thick} for a visual comparison of synthetic observation with and without absorption and stimulation).
    \par Such optically thick behavior, however, occurs only on the minority of dense lines of sight within our edge-on vantage of the galactic disk along which there is material to observe, as the sum behavior of the entire observed ray. At any point along such optically thick lines of sight, the OT level balance will still be a reasonable approximation if the medium is locally optically thin in the omnidirectional average over all (primarily out-of-disk) lines of sight. Ignoring absorption and stimulated emission in the OT level balance is thus a much weaker assumption than ignoring all absorption and stimulated emission in the transfer itself, which would necessarily induce an additional source of error. For these reasons, we argue that the OT level balance is still a reasonable approximation in many scenarios in which the line transfer demonstrates some optically thick behavior. We return to this argument in \autoref{subsec:Validity of the Optically Thin Level Balance}, in which we analyze in depth the validity of the OT assumption over our test simulations, and show definitively that the OT assumption yields results comparable to the LVG assumption in our observational regime.
    \par The second nuance regarding the OT level balance is that the initial pure assumption that the line may not absorb or be stimulated by any radiation may be refined to the assumption that the line may not self-interact but may still absorb or be stimulated by a constant external radiation background. In IRIS, rather than the pure OT assumption, we use a version of this refined OT assumption in which we account for an external blackbody radiation profile at a fixed temperature $\overline{T}$. If we set $\overline{T} = \SI{0}{\K}$, then the assumption simplifies back to the pure case. As a best estimate, we set $\overline{T} = T_\text{CMB} \approx \SI{2.73}{\K}$. Some ambiguity arises, however, as to whether this choice of $\overline{T}$ is appropriate, as the total radiation field is, in our simple observational model, additionally fed by both spectral-line emission and dust emission. In \autoref{subsec:Validity of the Optically Thin Level Balance}, we return to this question also, and show definitively that $\overline{T} = T_\text{CMB}$ is the most appropriate choice. In the IRIS code, however, in order to provide the most robust experimental framework, we keep separate the parameters $\overline{T}$---which sets the radiation background in the OT level balance---and $T_\text{CMB}$, which sets the background source for radiative transfer.

\subsection{Level Populations: Mathematical Solution} \label{subsec:Level Populations: Mathematical Solution}
    \par In general, determination of the level populations requires solution of an equilibrium system involving particle collisions, spontaneous emission, stimulated emission, absorption, and the radiation field. That is, assuming that the level populations are not evolving in time, we then suppose that the populating mechanisms of each tracer energy level are in equilibrium with the depopulating mechanisms. The populating and depopulating mechanisms are:
    \begin{enumerate}[(i)]
        \item collisions exciting or de-exciting tracer molecules from energy level $i$ to $j$, governed by the all-partner collision frequency $C_{ij}$ of the tracer (measured per tracer number-density) associated with the $i$--$j$ transition;
        \item spontaneous emission from higher energy levels $i$ down to lower energy levels $j$, governed by the Einstein $A_{ij}$ coefficient (measured per tracer number density);
        \item stimulated emission from higher energy levels $i$ down to lower energy levels $j$, governed by the Einstein $B_{ij}$ coefficient (measured per tracer number density, per unit intensity $J_{ij}$ of the omnidirectionally averaged radiation field at the transition frequency $\nu_{ij}$); and
        \item absorption from lower energy levels $j$ up to higher energy levels $i$, governed by the Einstein $B_{ji}$ coefficient (measured per tracer number density, per unit intensity $J_{ij}$) \citep{Rybicki1986, vanderTak2007}.
    \end{enumerate}
    \par These constraints yield a level balance equation for each energy level $i$, which, considering only the finite set of energy levels up to some maximal $0 \leq i \leq k$, together form a linear system of rank $k$ for the $k + 1$ level-population variables $n_i$. We then substitute, for the ground-state ($i = 0$) equation, a conservation equation constraining the sum of all level populations to be equal to the total tracer population, which yields a well-determined system. Under the notational convention $A_{ij} \equiv 0$ for $i < j$ and $J_{ij} \equiv J_{ji}$, this system for the level populations $n_i$ can be expressed as
    \begin{align} \label{eqn:OT Levels System}
        0 &= \sum_{j \neq i} \Big[n_j(C_{ji} + A_{ji} + B_{ji}J_{ji}) - n_i(C_{ij} + A_{ij} + B_{ij}J_{ij})\Big] \text{ ,} \notag \\
        N &= \sum_i n_i \text{ .}
    \end{align}
    \par Under the OT assumption, we reduce the radiation field term $J_{ij}$ to a blackbody background with a constant temperature $\overline{T}$ via Planck's law:
    \begin{equation}
        J_{ij} \equiv J_{ji} \equiv J(\nu_{ij}) = \frac{2h\nu_{ij}^3}{c^2\left[e^{h\nu_{ij}/\left(kT\right)} - 1\right]} \qquad i > j \text{ ,}
    \end{equation}
    where $h$ is the Planck constant, $c$ is the vacuum speed of light, and $k$ is the Boltzmann constant. Lastly, we estimate the all-partner collision rates $C_{ij}$ via standard single-partner rates $K_{ij}$ (measured per tracer number density, per partner number density) over all relevant partner species. That is,
    \begin{equation}
        C_{ij} = \sum_{s \in S} K_{ij}^{[s]}N_s \text{ ,}
    \end{equation}
    where $S$ is the set of all considered partner species, $K_{ij}^{[s]}$ is the single-partner collision rate for some partner $s$, and $N_s$ is the partner number density. (See \autoref{subsec:Spectral-Line Configuration} and \autoref{subsec:Grid Precomputation} for details on which species are considered.)

\subsection{Einstein Coefficients, Collisions, and the Detailed Balance} \label{subsec:Einstein Coefficients, Collisions, and the Detailed Balance}
    \par We obtain the Einstein $A$ coefficients and single-partner collision rates $K_{ij}$ from the standard molecular databases LAMDA \citep{Schoier2005} and BASECOL \citep{Dubernet2024} (see \autoref{subsec:Spectral-Line Configuration}). The Einstein $B$ coefficients are related to the $A$ coefficient via the relations
    \begin{align}
        B_{ul} &= \frac{c^2}{2h\nu_{ul}^3}A_{ul} &\text{and} \\
        B_{lu} &= \frac{g_u}{g_l}B_{ul} \text{ ,} \notag
    \end{align}
    where $g_u$ and $g_l$ are the statistical weights of the upper and lower levels, respectively, also available from standard databases \citep{Rybicki1986}.
    \par One remaining technical point is that such standard molecular databases often give only the values of the collisional de-excitation rates $K_{ij}, i > j$. To determine the excitation rates $K_{ji}$, $i > j$, we use the detailed balance
    \begin{equation}
        K_{ji} = K_{ij}\frac{g_i}{g_j}e^{(\varepsilon_j - \varepsilon_i)/[kT]} \quad (i > j) \text{ ,}
    \end{equation}
    where $T$ here refers to the gas kinetic temperature \citep{vanderTak2007}. This detailed balance relation is derived directly from the application of Maxwell-Boltzmann statistics, which hold only in LTE. Since collision rates, however, are microscopic properties dependent only upon the average geometries and quantum mechanical properties of the tracer and partner molecules, the detailed balance is a more general relation that holds outside of LTE also.

\section{Synthetic Observation Implementation and Algorithms} \label{sec:Synthetic Observation Implementation and Algorithms}
    \par We now turn to the computational implementation of IRIS-SO. While we leave to the IRIS documentation issues of code organization and internal functionality, we describe the high-level algorithmic design in this section. In part, we believe these algorithmic details are of independent academic interest and that we have contributed some novel design choices. A robust description of these design choices in publication may be of utility to other researchers designing novel synthetic-observation codes. Additionally, with the aim in mind of enabling future research, we have designed IRIS-SO with substantial functionality and configurability, and we leave open the possibility of a future independent release of IRIS-SO as a standalone Python package for the general radioastronomy user base. We intend this section to serve as a primary reference in the event of such a release. As such, we take some care to discuss the theoretical and algorithmic aspects of a variety of IRIS-SO features, even those we have not used in this study specifically.
    \par The entire IRIS-SO code is implemented in PyTorch \citep{Paszke2019}, which provides a number of advantages. First, it streamlines the data-generation process, allowing flexible and granular control over all details within the data-generation routine itself, and eliminating painful interfacing with external code platforms or executables. Second, it provides automatic GPU-parallelization of the synthetic-observation process within the familiar and flexible schema of the PyTorch ecosystem, which yields orders of magnitude of speedup (up to $10{,}000\times$, see \autoref{subsec:Speed Testing}) against traditional, CPU-based transfer codes. And third, the entire synthetic-observation pipeline is designed to be differentiable, which opens up possibilities for a number of future use-cases. IRIS is one of the first line radiative transfer codes implemented in PyTorch, and that provides GPU-acceleration and end-to-end differentiability.
    \par Some comparable synthetic-observation codes preceding the IRIS release include Magritte-torch, which is a GPU-accelerated and fully differentiable PyTorch adaptation of the code Magritte \citep{Ceulemans2024} by the same developers, RadJAX \citep{Levis2025}, which is a GPU-accelerated and differentiable synthetic-observation code implemented in JAX, and THOR \citep{Byrohl2025}, which is a non-differentiable but GPU-accelerated synthetic-observation code implemented in C++. (See also \autoref{subsec:Neural Fields: General Approach} for more discussion regarding the conceptual intersection between the IRIS and RadJAX projects, in particular.) We highlight, however, that IRIS was conceived and developed concurrently with and independently from each of these projects, without collaboration or mutual knowledge. We believe, rather than introducing over-redundancy, the recent and nearly concurrent emergence of a few such options instead represents an encouraging diversity of work and is indicative of a strong demand for GPU-accelerated and/or differentiable codes in enabling promising avenues of future research.
    \par Though not ultimately used in this publication, we take a moment to highlight the automatic-differentiability of IRIS-SO within the PyTorch ecosystem, which may be of substantial utility in future machine-learning applications. While disabled by default in order to improve computational efficiency, IRIS-SO allows differentiability to be enabled in one of two modes. The first of these modes is end-to-end differentiability, which allows backpropagation of gradients through the entire synthetic-observation process to any proceeding computations, or even to a differentiable physical tensor as a neural field. Such gradient backpropagation to a neural physical tensor was a core aim of some of our earlier explorations in this research project, and was part of the original impetus for developing IRIS-SO. A related approach to a similar problem in the reverse-imaging of protoplanetary disks (as opposed to the CMZ) was also employed successfully in \citet{Levis2025}, suggesting the potential future utility of such neural-fields approaches (see \autoref{subsec:Neural Fields: General Approach}). The second differentiability mode provides backpropagation to the tracer abundance function only, at a substantial compute savings in comparison to end-to-end differentiability. This mode provides optionality for the solution of complex tracer functions based upon some observational prior.
    \par The IRIS synthetic-observation process is broken into a few separate stages. The first stage is \textit{spectral-line configuration} (\autoref{subsec:Spectral-Line Configuration}), in which all data necessary to make a spectral-line observation is specified to the IRIS code. This stage occurs pre-startup. In the next stage, \textit{grid precomputation} (\autoref{subsec:Grid Precomputation}), IRIS computes emission and absorption coefficients over a large parameter grid, which improves the efficiency of emission/absorption determination at runtime via fast linear interpolation. This stage occurs once at startup only. The remaining observation stages all occur at runtime. The first runtime stage is \textit{observability determination} (\autoref{subsec:Observability Determination}), in which emission and absorption coefficients are computed over the spatial and frequency/velocity dimensions. Next, during \textit{transfer solution} (\autoref{subsec:Transfer Solution: Overview}, \autoref{subsec:Transfer Solution: Optically Thick Transfer}, \autoref{subsec:Transfer Solution: Selectively Thin Transfer}, \autoref{subsec:Transfer Solution: Optically Thin Transfer}, and \autoref{subsec:Transfer Solution: Formal and Smooth Integration of Optically Thick and Selectively Thin Transfer}), the radiative-transfer equation is solved over each ray for both the total cube and continuum-baseline cube. In the following stage, \textit{continuum subtraction} (\autoref{subsec:Continuum Subtraction}), a spectral-line observation is generated by subtracting the baseline from the total cube. During \textit{beam convolution} (\autoref{subsec:Beam Convolution}), this raw observation is convolved with a beam (point-spread) function to model the nonzero angular resolution of a real antenna. Finally, during \textit{noise addition} (\autoref{subsec:Noise Addition}), simulated antenna noise is added to yield a realistic synthetic observation.

\subsection{Spectral-Line Configuration} \label{subsec:Spectral-Line Configuration}
    \par IRIS-SO is designed to compute continuum-subtracted synthetic observations over one or multiple spectral lines in parallel. To do so, IRIS requires only two user-specified elements. The first configurable element required is a molecular data file, from which IRIS pulls transition frequencies, emission rates, and collision rates. IRIS employs the \texttt{.dat} file convention used by the Leiden Atomic and Molecular Database \citep[LAMDA][]{Schoier2005}, but a file from any source can be utilized provided it is formatted according to this convention. By default, IRIS pulls H, para-$\HTwo$, ortho-$\HTwo$, and He collision rates from the LAMDA file. Combined $\HTwo$ collision rates are pulled if no separate rates for para-$\HTwo$ and ortho-$\HTwo$ are listed, and H and He collisions are ignored if not present in the data file. For this study, IRIS uses the official $\ThirteenCO$ data file from LAMDA \citep{Schoier2005}, modified by the addition of H collision rates due to \citet{Walker2015} from BASECOL \citep{Dubernet2024}.
    \par The second of these configurable elements is an abundance function to determine number density of the molecular tracer from the densities, gas kinetic temperatures, and base abundances of $\HTwo$ and CO recorded per grid point in each physical tensor. For this study, we determine the abundance of the isotopologue $\ThirteenCO$ as a constant multiple of the total CO abundance:
    \begin{equation}
        \abundance{\ThirteenCO} \approx \left(\SI{4e-2}{}\right)\abundance{\text{CO}} \text{ .}
    \end{equation}
    \par We derive this abundance from studies of the $\TwelveC/\ThirteenC$ abundance ratio in the ISM, assuming this ratio is equal to the $\TwelveCO/\ThirteenCO$ ratio, which is likely a good approximation, though not necessarily true throughout the Milky-Way ISM in general \citep[see][]{Wilson1994, Sheffer2002}. A study by \citet{Langer1990} cites the value of the $\TwelveC/\ThirteenC$ ratio within the CMZ as 24, while \citet{Riquelme2010} cites a $\TwelveC/\ThirteenC$ ratio in the CMZ of 20--25, which are consistent with our choice of $\abundance{\ThirteenCO}$.
    \par We do note, however, that the $\TwelveC/\ThirteenC$ ratio does not appear to be constant throughout the Milky Way, with \citet{Wilson1994}, \citet{Sheffer2002}, and \citet{Langer1990} citing values around 70 towards the outer edges of the galactic disk. There thus appears to be a higher $\ThirteenCO$ abundance in the galactic center, with \citet{Langer1990} noting a gradient of intermediate abundances between the CMZ and the outer galaxy. This deviation from a true constant $\ThirteenCO$ abundance may yield overestimation in the brightness of foreground/background features in our full-cone observations (see \autoref{subsec:Foreground and Background}). We leave open the possibility of implementing more complex, spatially-dependent abundance functions in future IRIS versions. We also leave open the possibility of more definitive solutions involving augmentation of the simulation chemical network at runtime or in postprocessing (see \autoref{sec:Summary and Conclusion}). But we do not attempt such complexities in this proof-of-concept research.
    \par While the current IRIS version is limited, by the simplified H-CO chemical network implemented in our current AREPO simulations (see \autoref{subsec:AREPO Zoom Simulations}), to the base physical-tensor abundances of $\HTwo$ and CO, we do also note that IRIS allows for synthetic observation in spectral tracers other than CO, provided a tracer abundance function is specified in terms of the physical-tensor variables. In case an abundance function cannot be determined a priori, IRIS provides the ability to learn an abundance function based on some observational prior, using either the end-to-end or abundance-only differentiability modes discussed in the \autoref{sec:Synthetic Observation Implementation and Algorithms} introduction. In the simplest example, IRIS can learn a constant abundance, as a fraction of total H, $\HTwo$, or CO number density, that reproduces, over a dataset of physical tensors, a mean intensity computed over a given real cube observation.

\subsection{Grid Precomputation} \label{subsec:Grid Precomputation}
    \par During grid precomputation, the frequency-independent line emission and absorption coefficients $j^\sline, \alpha^\sline$ defined via \autoref{eqn:Line Emission} and \autoref{eqn:Line Absorption} are solved over a 3D grid of dimensions gas density, $\HTwo$ abundance, and gas kinetic temperature, using collision rates from the LAMDA file that are interpolated in temperature. The purpose of this grid precomputation is that the OT level balance detailed in \autoref{subsec:Level Populations: The Optically Thin Assumption} and \autoref{subsec:Level Populations: Mathematical Solution} requires solution to a large linear system (\autoref{eqn:OT Levels System}). Solution of this system over each separate physical-tensor grid point, at runtime, is computationally prohibitive. Instead, the observability grid is computed just once at startup and is saved in memory for reference at runtime, eliminating the necessity of computing the level balance over all physical-tensor grid points, and front-loading redundant computations to a startup-only phase. IRIS uses the GPU-accelerated PyTorch linear solver \citep{Paszke2019} to perform this grid computation. Choosing, for this study, grid dimensions of $n_\rho, n_\chi, n_T = (128, 64, 64)$, we find that grid precomputation takes about $\sim 3$ seconds.
    \par Since significant tracer abundance (for this paper, $\ThirteenCO$ abundance) occurs only in the areas of a simulation with low $\HPlus$ abundance, we assume here that all H is either neutral atomic H or $\HTwo$. It is this assumption, combined with an assumption of a constant He abundance $\abundance{\text{He}} \approx 0.1$ per total H number density, and with the assumption of a constant ortho-$\HTwo$-to-para-$\HTwo$ ratio (configurable in IRIS, but set in this study as $3 \, : \, 1$), that allows specification of the level system detailed in \autoref{subsec:Level Populations: Mathematical Solution} in terms of just the three dimensions of gas density, kinetic temperature, and $\HTwo$ abundance. This reduction to a minimal three grid dimensions is, in turn, a critical memory optimization. In addition to the grids themselves, the grid gradients are also computed to allow the runtime computation of emission and absorption coefficients via a fast linear interpolation. By virtue of this interpolation scheme, observability determination itself remains differentiable, and yields emission and absorption that exhibit linear behavior, rather than merely constant behavior, outside observability grid bounds and between grid points.
    \par We note, here, that these grids and grid gradients are both solved per tracer abundance, where tracer abundance is expressed as a fraction of total H number density. This parameterization allows the abundance function to be applied as the final step of observability determination, which shortens the gradient path in abundance-only differentiability mode (see \autoref{sec:Synthetic Observation Implementation and Algorithms}), yielding improved computational efficiency. We denote these per-abundance coefficients as $\widetilde{\jmath}^\sline, \widetilde{\alpha}^\sline$, i.e.
    \begin{equation}
        j^\sline = \abundance{}\widetilde{\jmath}^\sline \qquad \text{and} \qquad \alpha^\sline = \abundance{}\widetilde{\alpha}^\sline \text{ .}
    \end{equation}
    This per-abundance computation is possible since both the emission/absorption coefficients and the OT level system itself (\autoref{eqn:OT Levels System}) are linear with respect to tracer abundance.

\subsection{Observability Determination} \label{subsec:Observability Determination}
    \par Observability determination is the first runtime stage of IRIS-SO. To begin, the per-tracer-abundance, frequency-independent line coefficients $\widetilde{\jmath}^\sline, \widetilde{\alpha}^\sline$ are computed over each point in the physical tensor via linear interpolation of the precomputed grid. We also set an optional temperature threshold $T_\infty$, above which no emission or absorption are computed, as the tracer is assumed to have thermally decomposed above this point. Assuming some inefficiency and error in the simplified AREPO chemical network (see \autoref{subsec:AREPO Zoom Simulations}), or if using derived tracer abundances not computed in this chemical network (see \autoref{subsec:Spectral-Line Configuration}), this threshold serves to additionally mitigate extreme emission from cells of anomalously high temperature. For this study, while we find that the AREPO CO network is already efficient at eliminating CO abundance in such anomalous cells by itself, we impose $T_\infty = \SI{5e4}{\K}$ (see \autoref{table:Synthetic Observation Parameters}).
    \par Next, the line $\widetilde{\jmath}^\sline, \widetilde{\alpha}^\sline$ tensors are expanded to incorporate a velocity/frequency dimension, and the Doppler-shifted line profile $\varphi(\nu)$ is computed and applied via multiplication to the coefficient tensors. IRIS uses either the standard profile or the integrated profile, depending upon the transfer mode (see \autoref{subsec:Transfer Solution: Overview}). Here, we use the classical Doppler shift:
    \begin{equation}
        \nu' = \left(1 + \frac{v}{c}\right)\nu \text{ .}
    \end{equation}
    Finally, tracer abundance is computed over all physical-tensor grid points and applied via multiplication to the per-abundance, frequency-dependent coefficients $\widetilde{\jmath}_\nu^\sline, \widetilde{\alpha}_\nu^\sline$. Again, delayed-abundance application enables more efficient abundance-only differentiability (see the \autoref{sec:Synthetic Observation Implementation and Algorithms} introduction). This yields true emission/absorption coefficients $j_\nu^\sline, \alpha_\nu^\sline$ over the velocity/frequency-expanded physical-tensor grid.
    \par The other component of observability determination is the computation of the dust emission and absorption coefficients $j_\nu^\dust, \alpha_\nu^\dust$. IRIS treats the opacity $\kappa$ of the simple, single-species dust, distributed according to a standard gas mass ratio, as constant over the small frequency window associated with each separate line cube, but varying over the large frequency gaps between separate lines. IRIS uses a constant (user-defined) opacity $\kappa$ for each separate line to compute the dust absorption coefficients $\alpha^\dust$, which are, in turn, constant in frequency over each separate line. The emission coefficients $j_\nu^\dust$ are then computed over all points on the velocity/frequency-expanded physical-tensor grid according to the Doppler-shifted Planck source function given in \autoref{eqn:Dust Source Equation}. Note that the massive size of the observability tensors $j_\nu^\sline, j_\nu^\dust, \alpha_\nu^\sline, \alpha^\dust$, and thus the memory required on GPU, is the primary compute bottleneck for the synthetic-observation process. To prevent GPU memory overflow, IRIS performs observability determination and transfer solution iteratively over smaller batches of observed rays.

\subsection{Transfer Solution: Overview} \label{subsec:Transfer Solution: Overview}
    \par The next stage of synthetic observation, transfer solution, involves computation of an observed PPV cube by solution of the radiative-transfer equation along each observed ray. We experimented with a number of different integration schemes and approximative modes, and implemented several of these modes separately into IRIS-SO for comparison. We class these schemes as a matrix of three alternate transfer modes---\textit{optically thick}, \textit{selectively thin}, and \textit{optically thin}---and two separate integration modes---\textit{formal} and \textit{smooth}. With the exception of two diagnostic figures we include in \autoref{sec:Synthetic Observation Testing and Verification} detailing our testing and verification of IRIS-SO (\autoref{fig:Optically Thin vs Thick} and \autoref{fig:Formal vs Smooth}), all of our synthetic observations for this study were made with optically thick transfer and smooth integration, which we found to provide the best combination of accuracy and speed. For completeness in describing IRIS-SO for future reference, however, we illustrate the mathematics of all modes in the following sections.
    \par The three separate transfer modes provide different solutions to a substantial complication in the transfer-solution process. Namely, while the radiative transfer is given by \autoref{eqn:Radiative Transfer Equation} in terms of the instantaneous, per-frequency intensity $I_\nu$, what we want to compute in a synthetic observation is the channel-averaged intensity. Specifically, letting
    \begin{equation}
        \fint f(\nu) \; \equiv \Delta{\nu}_{\scriptscriptstyle{W}}^{-1}\int_{\nu_\text{lo}}^{\nu_{\text{hi}}} f(\nu) \, d\nu
    \end{equation}
    denote the channel-averaging operator, where $\nu_\text{hi}, \nu_\text{lo}$ are the frequency-channel edges and $\Delta{\nu}_{\scriptscriptstyle{W}} = \nu_\text{hi} - \nu_\text{lo}$ is the channel width, we want to compute
    \begin{equation}
        \overline{I}_\nu \equiv \fint I_\nu \text{ .}
    \end{equation}
    Such channel-averaging is a valid approximation of real telescope behavior, provided the spectrometer response function associated with the telescope can be approximated as a rectangular (bandpass) convolution, which is reasonable in most cases \citep{Wilson2013}. For dust emission, which has no fine-frequency features, we have the simple approximation
    \begin{equation}
        \overline{\jmath}_\nu^\dust \equiv \fint j_\nu^\dust \approx j_\nu^\dust \text{ .}
    \end{equation}
    For dust absorption $\alpha^\dust$, which we model as constant over the entire frequency dimension of a single line cube, the issue is moot. Thus, for the continuum $\overline{I}_\nu^\cont$, we have the simple approximation
    \begin{equation}
        \overline{I}_\nu^\cont \approx I_\nu^\cont \text{ ,}
    \end{equation}
    which we use in all transfer modes.
    \par Depending on the chosen velocity resolution, however, the line profile may be much narrower than the frequency-channel width itself. Thus, application of this same simplification to the total intensity, via solution of the total radiative-transfer equation only at the frequency-channel centers, can yield spotty observations where substantial emission or absorption within each velocity channel is missed due to minute misalignment of the channel centers with line-profile contributions from Doppler-shifted physical-tensor cells. We discuss this failure mode in greater detail in \autoref{subsec:Resolution Convergence}. In optically thick transfer, we solve the problem by performing a secondary numerical integration over frequency/velocity space. This is the most accurate mode and is our preferred mode for all synthetic observations in this study. In selectively thin and optically thin transfer, we make some simplifying assumptions that allow analytic frequency integration of the radiative-transfer equation.
    \par The two separate integration modes provide different means of solving the transfer equation itself. In all transfer and integration modes, we must compute the solutions of the total-intensity and continuum-baseline-intensity cubes separately, over each ray, by applying a stepwise numerical method beginning from the background intensity as an initial condition. We denote this background intensity as $I_\nu^\back$, which is a configurable field in IRIS, but which we take for all synthetic observations in this study simply to be the pure blackbody profile of the CMB. As with the continuum intensity, we again assume that
    \begin{equation}
        \overline{I}_\nu^\back \equiv \fint I_\nu^\back \approx I_\nu^\back \text{ .}
    \end{equation}
    \par Formal integration uses a step computation derived from the analytic solution to the radiative-transfer equation assuming emission and absorption coefficients are constant within each single radial step. Smooth integration treats emission and absorption coefficients as step-center values, and integrates a smooth interpolation between these values. Formal integration is theoretically more stable at a variety of radial step sizes. Smooth integration is substantially faster than formal integration, by the elimination of expensive transcendental operations in optically thick and selectively thin modes (more than $1.5 \times$ speedup) and a factor-of-2 reduction of the sum order in optically thin mode ($\sim 1.3 \times$ speedup). We find smooth integration to achieve negligibly different results from formal integration in standard regimes (see \autoref{subsec:Comparison of Formal and Smooth Integration} and \autoref{fig:Formal vs Smooth}), and so prefer it as our standard integration for all synthetic observations in this study.

\subsection{Transfer Solution: Optically Thick Transfer} \label{subsec:Transfer Solution: Optically Thick Transfer}
    \par In optically thick transfer, which was used for all the synthetic observations in this study (excepting \autoref{fig:Optically Thin vs Thick}), the full radiative-transfer equation is solved for both the total per-frequency intensity and the continuum-baseline per-frequency intensity, given by \autoref{eqn:Total Radiative Transfer} and \autoref{eqn:Continuum Radiative Transfer}, reproduced below with the replacement of the theoretical $\alpha_\nu^\dust$ with the implementationally frequency-constant $\alpha^\dust$:
    \begin{equation*}
        \frac{dI_\nu^\total}{ds} = \left(j_\nu^\sline + j_\nu^\dust\right) - \left(\alpha_\nu^\sline + \alpha^\dust\right)I_\nu^\total \text{ ,}
    \end{equation*}
    and
    \begin{equation*}
        \frac{dI_\nu^\cont}{ds} = j_\nu^\dust - \alpha^{\dust}I_\nu^\cont \text{ .}
    \end{equation*}
    We postpone discussion of the numerical solution of these equations over an observed ray to \autoref{subsec:Transfer Solution: Formal and Smooth Integration of Optically Thick and Selectively Thin Transfer}.
    \par In order to eliminate the frequency under-sampling problem, however, we must compute the total intensity $I_\nu^\total$, as well as the line emission and absorption coefficients, over a fine frequency/velocity grid. The fine grid is therefore introduced previously during observability determination (\autoref{subsec:Observability Determination}). The total intensity is then numerically integrated in frequency post-transfer in order to yield the frequency-averaged cubes. IRIS conducts frequency integration via a fast, vectorized implementation of Simpson's Rule \citep{Stoer1980}. Letting $\nu_\text{lo}, \nu_\text{mid}, \nu_\text{hi}$ be a sequence of adjacent, fine-frequency steps, such that $\nu_\text{mid}$ is the midpoint between $\nu_\text{lo}$ and $\nu_\text{hi}$, with $\Delta{\nu}_{\scriptscriptstyle{W}} = \nu_\text{hi} - \nu_\text{lo}$ as before, a single integration step is computed as
    \begin{equation}
        \int_{\nu_\text{lo}}^{\nu_{\text{hi}}} I_\nu \, d\nu \approx \frac{\Delta{\nu}_{\scriptscriptstyle{W}}}{6}\left[I_{\nu_\text{lo}} + 4I_{\nu_\text{mid}} + I_{\nu_\text{hi}}\right] \text{ .}
    \end{equation}
    \par Incorporation of a fine frequency grid and post-transfer frequency integration each add time and memory complexity. This added complexity is the primary drawback of optically thick transfer. The benefit is that this mode accurately models all line-continuum interaction and line self-interaction, provided a suitable velocity subsampling resolution is chosen, and is thus the best option to model optically thick regimes. We characterize this subsampling resolution by a \textit{velocity-subsampling factor}, which we define as the number of added midpoints between each pair of coarse velocity grid points. At the velocity resolution of our training dataset (see \autoref{table:Dataset Parameters}) we find a subsampling factor of 1 to be sufficient. We return to a more detailed discussion of sufficient resolution in \autoref{subsec:Resolution Convergence}.

\subsection{Transfer Solution: Selectively Thin Transfer} \label{subsec:Transfer Solution: Selectively Thin Transfer}
    \par In selectively thin mode, the total cube is solved over the coarse frequency grid via an approximate analytic integral. In particular,
    \begin{align}
        \frac{d\overline{I}_\nu^{\total}}{ds} &= \frac{d}{ds} \fint I_\nu^{\total} \\
        &= \fint \frac{dI_\nu^{\total}}{ds} \notag \\
        &= \fint \left[j_\nu^{\sline} + j_\nu^{\dust} - \left(\alpha_\nu^{\sline} + \alpha^{\dust}\right)I_\nu^{\total}\right] \notag \\
        &= \fint \bigg[j^{\sline}\varphi(\nu) + j_\nu^{\dust} - \left(\alpha^{\sline}\varphi(\nu) + \alpha^{\dust}\right)I_\nu^{\total}\bigg] \notag \\
        &= j^{\sline} \fint \varphi(\nu) + \overline{\jmath}_\nu^{\dust} - \alpha^{\sline}\fint \varphi(\nu)I_\nu^{\total} - \alpha^{\dust} \fint I_\nu^{\total} \notag \\
        &\approx j^{\sline} \fint \varphi(\nu) + \overline{\jmath}_\nu^{\dust} - \alpha^{\sline}\fint \varphi(\nu) \fint I_\nu^{\cont} - \alpha^{\dust} \fint I_\nu^{\total} \notag \\
        &= j^{\sline} \fint \varphi(\nu) + \overline{\jmath}_\nu^{\dust} - \alpha^{\sline}\left(\fint \varphi(\nu)\right)\overline{I}_\nu^{\cont} - \alpha^{\dust}\overline{I}_\nu^{\total} \text{ .} \notag
    \end{align}
    \par Here, we make the critical approximation
    \begin{equation}
        \fint \varphi(\nu)I_\nu^{\total} \approx \fint \varphi(\nu) \fint I_\nu^{\cont} \text{ .}
    \end{equation}
    That is, the channel average of the total intensity, weighted by the line profile (a probability density function), is just the continuum average (approximately constant over the channel width) multiplied by the probability mass of the line profile contained in the channel. In other words, the line may absorb or be stimulated by the continuum, and dust may absorb the line, but the line is entirely non-self-interacting. This is a good approximation in the scenario that the line is both locally optically thin everywhere and globally non-self-interacting by virtue of a large velocity gradient. We then have an analytic solution for the integrated line profile, up to the standard error function erf:
    \begin{align}
        \overline{\varphi}(\nu) &\equiv \fint \varphi(\nu) \\
        &= \fint \frac{e^{-(\nu - \nu_{ul})^2/\Delta\nu_{\scriptscriptstyle{D}}^2}}{\Delta\nu_{\scriptscriptstyle{D}}\sqrt{\pi}} \notag \\
        &= \Delta{\nu}_{\scriptscriptstyle{W}}^{-1}\frac{1}{2}\text{erf}\left(\frac{\nu - \nu_{ul}}{\Delta\nu_{\scriptscriptstyle{D}}}\right) \text{ .} \notag
    \end{align}
    In this transfer mode, this integrated line profile is applied during observability determination instead of the standard profile, yielding
    \begin{equation}
        \overline{\jmath}_\nu^\sline \equiv j^\sline \, \overline{\varphi}(\nu) \qquad \text{and} \qquad \overline{\alpha}_\nu^\sline \equiv \alpha^\sline \, \overline{\varphi}(\nu) \text{ .}
    \end{equation}
    This approximation entirely eliminates the need for frequency subsampling and numerical integration, instead achieving the same integration analytically, which saves substantial compute. As in optically thick transfer, the continuum baseline requires no frequency integration at all, since it has no fine-frequency features. We again postpone discussion of the numerical solution of these equations over an observed ray to \autoref{subsec:Transfer Solution: Formal and Smooth Integration of Optically Thick and Selectively Thin Transfer}.

\subsection{Transfer Solution: Optically Thin Transfer} \label{subsec:Transfer Solution: Optically Thin Transfer}
    \par In optically thin transfer mode, absorption and stimulation are ignored entirely, yielding the most coarse but also most computationally efficient approximation. IRIS-SO does, however, still compute dust emission and a continuum-baseline cube in this mode in order to account for the Planck nonlinearity in continuum subtraction, if conducted in brightness-temperature space (as opposed to in intensity space or equivalent subtraction in Raleigh-Jeans temperature space, see \autoref{subsec:Continuum Subtraction}). In this case, the radiative-transfer equation is approximated as
    \begin{align}
        &\frac{d\overline{I}_\nu^{\total}}{ds} = \frac{d}{ds} \fint I_\nu^{\total} \\
        &= \fint \frac{dI_\nu^{\total}}{ds} \notag \\
        &= \fint \left[j_\nu^{\sline} + j_\nu^{\dust} - \left(\alpha_\nu^{\sline} + \alpha^{\dust}\right)I_\nu^{\total}\right] \notag \\
        &\approx \fint \left[j_\nu^{\sline} + j_\nu^{\dust}\right] \notag \\
        &= \fint \left[j^{\sline}\varphi(\nu) + j_\nu^{\dust}\right] \notag \\
        &= j^{\sline}\fint \varphi(\nu) + \overline{\jmath}_\nu^{\dust} \text{ .} \notag
    \end{align}
    As in selectively thin mode, the integrated line profile is again applied during the observability determination step via
    \begin{equation}
        \overline{\jmath}_\nu^\sline \equiv j^\sline \, \overline{\varphi}(\nu) \text{ ,}
    \end{equation}
    which again eliminates the need for frequency subsampling and numerical integration.
    \par Unlike in the optically thick and selectively thin transfer modes, however, these transfer equations both map to direct integral solutions:
    \begin{equation}
        \overline{I}_\nu^\any = I_\nu^\back + \int_\text{ray} f^{\any}(s) \, ds \text{ ,}
    \end{equation}
    where
    \begin{equation}
        f^{\total}(s) = \frac{dI^{\total}}{ds} = \overline{\jmath}_\nu^{\sline} + \overline{\jmath}_\nu^{\dust} \text{ ,}
    \end{equation}
    and
    \begin{equation}
        f^{\cont}(s) = \frac{dI^{\cont}}{ds} = \overline{\jmath}_\nu^{\dust} \text{ .}
    \end{equation}
    \par In formal integration mode, we simply compute the stepwise-linear integral, which is evaluated via a fast, vectorized operation. Taking $s_\text{far}, s_\text{near}$ to be the bounds of a single radial step with midpoint $s_\text{mid}$ and width $h = s_\text{far} - s_\text{near}$, and assuming $f^\any$ to be linear over the interval $(s_\text{far}, s_\text{near})$,
    \begin{equation}
        \int_{s_\text{far}}^{s_{\text{near}}} \overline{I}_\nu^\any \, ds = hf^\any(s_\text{mid}) \text{ .}
    \end{equation}
    Here, every physical tensor step is taken as a midpoint $s_\text{mid}$, yielding a sum of order equal to the total number $R$ of radial steps in the observed physical tensor.
    \par In smooth integration mode, rather than treating emission and absorption coefficients as midpoint values of a stepwise linear function, we treat them as midpoint values connected by a smooth function that is locally cubic on the scale of the step size. The integrals are then evaluated via a fast, vectorized Simpson's Rule \citep{Stoer1980}:
    \begin{equation}
        \int_{s_\text{far}}^{s_{\text{near}}} \overline{I}_\nu^\any \, ds \approx \frac{h}{6}\left[f^\any(s_\text{far}) + 4f^\any(s_\text{mid}) + f^\any(s_\text{near})\right] \text{ .}
    \end{equation}
    Here, physical tensor steps must be taken as both midpoints $s_\text{mid}$ and edges $s_\text{far}, s_\text{near}$, yielding a sum of order $\lfloor(R - 1) / 2\rfloor$. Step resolution is reduced by an approximate factor of 2, but at the gain of a higher order of accuracy (exact for cubic functions, as opposed to the midpoint method, which is only exact for linear functions). Smooth integration is thus generally preferable, not only since it provides superior accuracy assuming smooth as opposed to constant emission and absorption coefficients, but since it provides on average a 1.3 times speedup due to the reduction of the sum order.

\subsection{Transfer Solution: Formal and Smooth Integration of Optically Thick and Selectively Thin Transfer} \label{subsec:Transfer Solution: Formal and Smooth Integration of Optically Thick and Selectively Thin Transfer}
    \par In optically thick and selectively thin transfer, the transfer solution is not, as in optically thin transfer, a direct integral. This complication is due to the $I_\nu$ dependence of the radiative-transfer equation with nonzero absorption. Therefore, we must use a stepwise solution method for ordinary differential equations (ODEs). Care must be taken here because the radiative-transfer equation is a classic stiff ODE, as are all linear ODEs of the form
    \begin{equation}
        y' = \gamma - {\alpha}y \text{ .}
    \end{equation}
    More precisely, for high coefficients $\alpha$ that fall outside a particular solution method's domain of stability (or, in application to the radiative-transfer equation, high absorption coefficients), explicit stepwise solution methods such as 4th-order Runge-Kutta (RK4) are prone to producing unpredictable divergence if the $s$ step size is chosen too large \citep{Hairer1996}. Applied to IRIS-SO, these divergences manifested during our early experiments, in which RK4 was used, as anomalous points of extreme intensity in a synthetic observation.
    \par In formal integration mode, we use the analytic, exponential-form solution to each radiative-transfer equation assuming emission and absorption coefficients that are radially stepwise-constant, which provides unconditional stability. Besides failing to apply any stepwise smoothing, however, which may negatively impact accuracy under some conditions, this analytically tailored numerical method is computationally expensive because it requires the evaluation of the exponentiation operator over massive tensors. Like all transcendental operations, exponentiation requires evaluation of a high-degree polynomial approximation under the hood, in the PyTorch/CUDA-GPU backend \citep{Paszke2019}. In smooth integration mode, we find speedups in excess of $1.5 \times$ by instead treating the transfer equations as general stiff ODEs and applying an appropriate numerical method with a lower polynomial degree. We find smooth integration mode for optically thick and selectively thin transfer to yield results negligibly different from formal integration (see \autoref{subsec:Comparison of Formal and Smooth Integration} and \autoref{fig:Formal vs Smooth}). Additionally, these methods also treat emission and absorption coefficients as smoothly varying over the line of sight as opposed to stepwise-constant, which may even improve accuracy under some conditions.
    \par For stiff ODEs, the gold standard for numerical methods are implicit, A-stable methods. These methods ensure that the failure mode for all positive $\alpha$ values (in the case of the radiative-transfer equation, absorption coefficients) is regression to zero \citep{Hairer1996}. Divergence may still occur for negative $\alpha$ values, but while negative line absorption coefficients are possible due to stimulated emission (indicating high pumping/masing) they are extraordinarily unlikely for most spectral line tracers, and, in particular, $\ThirteenCO$. The likelihood of negative total absorption coefficients is thus even lower, considering the addition of dust absorption, which is strictly positive. We use the 2nd-order Backwards Differentiation Formula (BDF2), along with the single-step Trapezoidal Rule (TR) for the initial step computation. Both are implicit and A-stable \citep{Hairer1996}.
    \par As such, the step computation is not given by an explicit step formula, but as the implicit solution to a step relation. For BDF2, the step relation for the initial value problem
    \begin{equation}
        y' = f(x, y) \qquad y(x_0) = y_0
    \end{equation}
    is
    \begin{equation}
        y_{n + 2} - \frac{4}{3}y_{n + 1} + \frac{1}{3}y_n = \frac{2}{3}hf(x_{n + 2}, y_{n + 2}) \text{ ,}
    \end{equation}
    where $h = x_{n + 1} - x_n$ is the step size. For TR, the step relation for the same initial-value problem and step size $h$ is
    \begin{equation}
        y_{n + 1} = y_n + \frac{1}{2}h\left[f(x_x, y_n) + f(x_{n + 1}, y_{n + 1})\right]
    \end{equation}
    \citep{Hairer1996}. If the ODE is nonlinear, this step relation is also nonlinear and so generally requires an iterative solution at each individual step, which entails a high computational cost. Critically, however, the radiative-transfer equation is linear in intensity. Therefore, the BDF2 and TR step relations both have explicit algebraic solutions, and so each step can still be computed in constant time as with using an explicit method or the formal integration.
    \par In optically thick transfer mode, eliminating $\nu$ subscripts for legibility, the formally solved steps are given as
    \begin{equation}
        I_{n + 1}^\any = I_n^{\any}e^{-\tau^\any} + S^\any(1 - e^{-\tau^\any}) \text{ ,}
    \end{equation}
    where the source function $S^\any$ and optical depth $\tau^\any$ \citep{Rybicki1986} are given by
    \begin{align}
        S^{\total} &= \frac{j^{\sline} + j^{\dust}}{\alpha^{\sline} + \alpha^{\dust}} \text{ ,} \\
        S^{\cont} &= \frac{j^{\dust}}{\alpha^{\dust}} \notag
    \end{align}
    and
    \begin{align}
        \tau^{\total} &= h(\alpha^{\sline} + \alpha^{\dust}) \text{ ,} \\
        \tau^{\cont} &= h\alpha^{\dust} \text{ .} \notag
    \end{align}
    \par For smooth integration, letting
    \begin{align}
        j_\nu^\total &\equiv j^\sline + j^\dust \text{ ,} \\
        j_\nu^\cont &\equiv j^\dust \text{ ,} \notag \\
        \alpha_\nu^\total &\equiv \alpha^\sline + \alpha^\dust \text{ ,} &\text{and} \notag \\
        \alpha^\cont &\equiv \alpha^\dust \text{ ,} \notag
    \end{align}
    the BDF2 step relations take the form
    \begin{equation}
        I_{n + 2}^\any - \frac{4}{3}I_{n + 1}^\any + \frac{1}{3}I_n^\any = \frac{2}{3}h\left[j_{n + 2}^\any - \alpha_{n + 2}^{\any}I_{n + 2}^\any\right] \text{ ,}
    \end{equation}
    yielding a step solution
    \begin{equation}
        I_{n + 2}^\any = \frac{2hj_{n + 2}^\any + 4I_{n + 1}^\any - I_n^\any}{3 + 2h\alpha_{n + 2}^\any} \text{ .}
    \end{equation}
    \par We also apply a ReLU nonlinearity on the computed $I_{n + 2}$ (i.e. clamp to nonnegative values), in order to prevent physically impossible negative true (non-subtracted) intensities, which may be produced by solver inaccuracy and numerical instability, from producing divergent, negative-intensity absorption. The TR step relations used for the first step take the form
    \begin{equation}
        I_{n + 1}^\any = I_n^\any + \frac{h}{2}\left[\left(j_n^\any - \alpha_n^{\any}I_n^\any\right) + \left(j_{n + 1}^\any - \alpha_{n + 1}^{\any}I_{n + 1}^\any\right)\right] \text{ ,}
    \end{equation}
    yielding a step solution
    \begin{equation}
        I_{n + 1}^\any = \frac{h\left(j_n^\any + j_{n + 1}^\any\right) + \left(2 - h\alpha_n^\any\right)I_n^\any}{2 + h\alpha_{n + 1}^\any} \text{ .}
    \end{equation}
    We again also apply a ReLU to $I_{n + 1}$.
    \par In selectively thin transfer mode, the formal and smooth step solutions are all the same as in optically thick mode, except in the substitutions
    \begin{align}
        I_\nu^\any &\mapsto \overline{I}_\nu^\any \text{ ,} \\
        j_\nu^\any &\mapsto \overline{\jmath}_\nu^\any \text{ ,} &\text{and} \notag \\
        \alpha_\nu^\any &\mapsto \overline{\alpha}_\nu^\any \text{ ,} \notag
    \end{align}
    and in defining
    \begin{equation}
        \overline{\jmath}_\nu^{\total} \equiv \overline{\jmath}_\nu^\sline + \overline{\jmath}_\nu^\dust -\overline{\alpha}^{\sline} \, \overline{I}_\nu^{\cont}
    \end{equation}
    and
    \begin{equation}
        \overline{\alpha}^{\total} \equiv \alpha^{\dust} \text{ .}
    \end{equation}
    Since $\overline{I}_\nu^{\cont}$ is independent of $\overline{I}_\nu^{\total}$, we simply solve for the continuum step first, and then use the continuum intensity to solve for the total-intensity step. As in optically thick transfer mode, a ReLU is again applied to each step computation to prevent impossible negative intensities, produced by numerical imprecision, from inducing divergent effects.

\subsection{Intensity Versus Temperature} \label{subsec:Intensity Versus Temperature}
    \par An ambiguity that arises following transfer solution is that the same observational cubes may be equivalently represented in a variety of units. The most common units are those of radiative intensity, brightness temperature, and antenna temperature. Radiative intensity $I_\nu$, commonly expressed in units of $\SI{}{\Jansky/\steradian}$, is the quantity in terms of which the radiative-transfer equations are expressed and synthetically solved. Intensity may be converted into the brightness temperature $T_\text{B}$ of a blackbody source of equivalent intensity by application of Planck's Law \citep{Wilson2013}:
    \begin{equation} \label{eqn:Brightness Temperature}
        I_\nu = \frac{2h\nu^3}{c^2\left[e^{h\nu/(kT_\text{B})} - 1\right]} \text{ .}
    \end{equation}
    Alternatively, an approximate brightness temperature $T_\text{RJ}$ may be computed via the linear Raleigh-Jeans Approximation to Planck's Law \citep{Wilson2013}:
    \begin{equation} \label{eqn:Raleigh-Jeans Temperature}
        I_\nu = \frac{2k\nu^2T_\text{RJ}}{c^2} \text{ .}
    \end{equation}
    \par In low-temperature and high-frequency regimes, the Raleigh-Jeans Approximation diverges substantially from Planck's Law. Since, however, the brightness temperature of a continuum-subtracted line observation, which may be negative, has no intrinsic physical meaning beyond being a convenient representational unit, the Raleigh-Jeans Approximation may in some contexts be taken as definitional. To alleviate this ambiguity, we use the terms \textit{brightness temperature} or \textit{Planck brightness temperature} to refer to the temperature $T_\text{B}$ computed via Planck's Law, and \textit{Raleigh-Jeans temperature} or \textit{Raleigh-Jeans brightness temperature} to refer to the temperature $T_\text{RJ}$ computed via the Raleigh-Jeans Approximation, although these terms may be used interchangeably by other authors.
    \par In addition to the brightness and Raleigh-Jeans temperatures, an observed cube may be expressed in terms of antenna temperature $T_\text{A}$, which is the weighted average of the Raleigh-Jeans temperature over the antenna beam profile, i.e. the Raleigh-Jeans temperature computed based on the actual detected (as opposed to ideal) intensity. Practically, the ideal Raleigh-Jeans temperature may be estimated from the antenna temperature via a constant beam efficiency factor $\eta$:
    \begin{equation}
        T_\text{A} \approx {\eta}T_\text{RJ} \text{ .}
    \end{equation}
    While IRIS-SO provides native options for producing intensity, brightness-temperature, and Raleigh-Jeans-temperature cubes, we work exclusively in this study with Raleigh-Jeans temperature (except as noted in \autoref{subsec:Validity of the Optically Thin Level Balance} and \autoref{fig:Continuum Temperature}), since these are the units in which the SEDIGISM $\ThirteenCO$ data is published \citep{Schuller2021}.

\subsection{Continuum Subtraction} \label{subsec:Continuum Subtraction}
    \par In all modes, having solved the entire total-intensity and continuum-baseline cubes via transfer over the full radial depth, the next step is continuum subtraction. When a real radio-telescope observation is refined into a spectral-line observation/PPV cube, an estimate of the continuum baseline is first subtracted from the raw observed cube \citep{Wilson2013}. The final observed cube will be zero where the raw cube matches the baseline estimate, positive where line emission is strong, and even negative where line absorption against the continuum dominates. In IRIS, we simulate this baseline-estimation and continuum-subtraction process by subtracting the true continuum-only cube from the total observed cube. Due to two distinct nonlinearities, it is important that the final, line synthetic observation be generated via this post-transfer subtraction, as opposed to short-cutting the subtraction process by simulating the pure line cube only, in the absence of any continuum contributions.
    \par The first of these nonlinearities involves the space in which continuum subtraction is performed within the real observational-processing pipeline. Since the radiative-transfer equation (\autoref{eqn:Radiative Transfer Equation}) is linear in intensity, continuum subtraction in intensity space will yield a subtracted cube that is identical to a pure-line synthetic observation, in which no continuum is simulated. Continuum subtraction in Raleigh-Jeans temperature space will also yield equivalent results to subtraction in intensity space, since the Raleigh-Jeans temperature is also linear in intensity (\autoref{eqn:Raleigh-Jeans Temperature}). If, however, subtraction is performed in brightness-temperature space, the results will be different due to the nonlinearity of Planck's Law. IRIS-SO therefore implements two separate continuum subtraction options---intensity/Raleigh-Jeans temperature and brightness temperature. The choice of which subtraction mode to use depends entirely upon the real observational-processing pipeline being modeled. For the SEDIGISM data analyzed in this article, which was baseline-subtracted in Raleigh-Jeans temperature \citep{Schuller2021}, we use intensity-space subtraction. It should also be noted that this nonlinearity persists even if the line and continuum are entirely optically thin. Thus, IRIS still computes a full continuum cube based on dust emission in optically thin transfer mode.
    \par The second nonlinearity is that of the radiative intensity derivative. Note that while the radiative-transfer equation is a linear ODE with respect to intensity, the intensity derivative is only bilinear, i.e. nonlinear, with respect to the absorption-intensity pair. For example, given the general form of the radiative-transfer equation reproduced from \autoref{eqn:Radiative Transfer Equation}
    \begin{equation*}
        \frac{dI_\nu}{ds} = j_\nu - {\alpha_\nu}I_\nu \text{ ,}
    \end{equation*}
    it is in theory possible that if $I$ accumulates sufficiently via contribution of
    continuum emission alone, a spectral line that would be optically thin in the absence
    of continuum emission could become optically thick. This effect only manifests in this extreme if the line is weak in comparison to the continuum, but will still have a graded effect even at higher line strength. Due to the combination of both these nonlinearities, IRIS performs an explicit post-transfer subtraction that directly models the real observational process. Once subtraction is complete, IRIS also provides the option to output the final observed cube in units of intensity, brightness temperature, or Raleigh-Jeans temperature.

\subsection{Beam Convolution} \label{subsec:Beam Convolution}
    \par In the next step of the synthetic-observation process, IRIS-SO applies a Gaussian beam convolution to simulate the nonzero angular resolution of a true antenna. Before this step, the continuum-subtracted PPV cube represents an ideal observation with infinitely fine resolution. In reality, however, a true antenna produces a spectral-line image by analysis of a receiver beam of nonzero angular width. This receiver beam adds a blurring effect that destroys detail below the beam scale. Typically, this true-antenna distortion is modeled during the synthetic-observation process by application of a \textit{beam convolution} or \textit{point-spread function} \citep{Wilson2013}. IRIS uses a standard Gaussian beam convolution, specified in terms of its full-width-half-max (FWHM), which we set for this study at $30''$ (see \autoref{table:Synthetic Observation Parameters}), corresponding to the approximate resolution of the APEX telescope on which the SEDIGISM survey was performed \citep{Schuller2021}. This convolution is then applied via a fast PyTorch operation \citep{Paszke2019}, although IRIS provides the option of skipping this step in order to obtain an ideal observation.

\subsection{Noise Addition} \label{subsec:Noise Addition}
    \par At this stage, the synthetic observation is a realistic model of a true, continuum-subtracted line observation with realistic antenna resolution, up to the application of noise. Modeling real observational noise, however, is extremely complex. To begin, noise arises from a variety of differing sources. Receiver noise originates at both the electrical and radio levels, integrating errant radio signals and instrumental electrical noise as false contributions to observed intensity. Interferometric distortions, arising from the imperfect synthesis of interferometer antennae, may persist in non-negligible degree even after interferometric correction is applied \citep[e.g. via the CLEAN algorithm, ][]{Hogbom1974}. Atmospheric effects also distort the observational target before it is ever observed by the antenna itself, which can in turn produce calibration defects that may manifest as a checkering artifact over the patchwork of independent antenna pointings from which a full observation is stitched together \citep{Wilson2013}.
    \par The combination of all these differing effects yields observational noise on a variety of angular and spectral scales, according to a variety of distinct probability distributions. Moreover, the exact noise characteristics depend entirely on the real observational process being modeled. The data refinement pipeline, the antenna hardware itself, and even the weather and background radiation at the time of observation are all factors. As such, real noise modeling is extraordinarily complex, and impractical for our purposes. As pertains to the training of the IRIS reversion model, however, real noise modeling, while potentially beneficial, is not strictly necessary. Noise is applied to synthetic observations during the training process only to teach the reversion model how to intelligently ignore noise. For these purposes, any noise distribution within which the real noise distribution is not a statistical outlier is sufficient, provided the noise distribution is not so broad as to obscure meaningful features not obscured in the real observation.
    \par To this end, IRIS-SO employs an extremely first-order noise approximation that, while unsophisticated, provides at least a baseline degree of noise inoculation to the reversion model during the training process. Specifically, IRIS uses a pixelwise Gaussian distribution, which can be specified in either intensity or temperature (see \autoref{subsec:Intensity Versus Temperature}),
    \begin{equation}
        I_\text{noise}, T_\text{noise} \sim N(\gamma\mu, \gamma\sigma) \text{ ,}
    \end{equation}
    where $N$ is the normal distribution, $\mu$ and $\sigma$ are a configurable noise mean and standard deviation, respectively (in units of either intensity or temperature), and $\gamma \sim U[0, 1)$ is another random variable distributed according to the uniform distribution. Noise is generated independently over each PPV voxel. Note that this voxel-wise independent generation yields noise that is uniformly speckled at the scale of a single voxel, which may represent a physically arbitrary size. While such noise generation is admittedly extremely coarse and unsophisticated, we find that its application during training improves generalization to real observational data, and so we incorporate it into this proof-of-concept study, while allowing the possibility of more sophisticated noise types in future research. Note also that we randomly generate this noise during each training step, as opposed to incorporating it into the synthetic-observation pipeline proper, which further diversifies training data and provides a source of regularization (see \autoref{subsec:Implementation of Reversion: Training Hyperparameters, Overfitting, and Regularization}). For this study, we set $\mu_T = \SI{0}{\K}$ and $\sigma_T = \SI{1}{\K}$ (see \autoref{table:Synthetic Observation Parameters}).

\section{Synthetic Observation Testing and Verification} \label{sec:Synthetic Observation Testing and Verification}
    \begin{deluxetable*}{lc}
        \tablecaption{\textbf{Synthetic-Observation Parameters:} Summary of the synthetic-observation parameters for the datasets described in \autoref{table:Dataset Parameters} and for the figures produced in this publication. See the IRIS code documentation for more details. \textsuperscript{a}~See \autoref{subsec:Spectral-Line Configuration}. \textsuperscript{b}~Used as a background intensity during radiative-transfer solution (see \autoref{subsec:Transfer Solution: Overview}), set to the approximate temperature of the CMB. \textsuperscript{c}~Used as a background in computing the OT level balance (see \autoref{subsec:Level Populations: Mathematical Solution}). In our training data, we set to the approximate temperature of the CMB, which we consider the most accurate assumption (see \autoref{subsec:Level Populations: The Optically Thin Assumption} and \autoref{subsec:Validity of the Optically Thin Level Balance}). In our POLARIS and RADMC-3D side-by-side figures (\autoref{fig:RADMC-3D vs IRIS}, \autoref{fig:POLARIS vs IRIS}), we use the alternate CMB temperature $T_\text{CMB} = \SI{2.75}{\K}$, since POLARIS sets this value automatically. In \autoref{fig:OT-OK vs OT-CMB Balance} and \autoref{fig:OT-CMB vs OT-5K Balance}, we explore the effect of the alternate values $\SI{0}{\K}$ and $\SI{5}{\K}$ (see also \autoref{subsec:Validity of the Optically Thin Level Balance}). \textsuperscript{d}~In order to minimize the observed effect of thermal outlier cells, no emission or absorption are computed above $T_\infty$, regardless of CO abundance (see \autoref{subsec:Observability Determination}). \textsuperscript{e}~Adapted from \citet{Ossenkopf1994} using the MRN with thin ice mantles model at $\SI{1e5}{\year}$ coagulation and gas number-density of $\SI{1e6}{\cm^{-3}}$. See \autoref{subsec:Dust Emission and Absorption} for more details on the validity of this assumed dust opacity as a CMZ average. In \autoref{fig:POLARIS vs IRIS}, we provide POLARIS an equivalent dust refractive index (see \autoref{subsec:POLARIS Dust Treatment}). In \autoref{fig:No Dust vs Dust x100}, we probe the effect of dust at 100 times this standard opacity. \textsuperscript{f}~See \autoref{subsec:Transfer Solution: Overview}. \textsuperscript{g}~See \autoref{subsec:Transfer Solution: Formal and Smooth Integration of Optically Thick and Selectively Thin Transfer}. \textsuperscript{h}~See \autoref{subsec:Transfer Solution: Optically Thick Transfer}. \textsuperscript{i}~See \autoref{subsec:Resolution Convergence} for details on radial resolution. \textsuperscript{j}~See \autoref{subsec:Continuum Subtraction} for details on continuum subtraction. \textsuperscript{k}~See \autoref{subsec:Intensity Versus Temperature} for temperature definitions. \textsuperscript{l}~See \autoref{subsec:Beam Convolution} for details on angular resolution. We adopt the resolution $30''$ from angular resolution of the APEX telescope used for the SEDIGISM survey \citep{Schuller2021}. \textsuperscript{m}~Used for antenna noise simulation (see \autoref{subsec:Noise Addition}). \label{table:Synthetic Observation Parameters}}
        \tablehead
        {
           \colhead{Parameter} & \colhead{Value in Dataset}
        }
        \startdata
        Observed line transition & $\ThirteenCOTwoOne$ \\
        Molecular data from & LAMDA \citep[primary source;][]{Schoier2005} \\
        & H collision rates from \citet{Walker2015} \\
        & accessed via BASECOL \citet{Dubernet2024} \\
        $\nu_\text{transition}$ & $\SI{220.4}{\giga\hertz}$ \\
        ortho-$\HTwo$-para-$\HTwo$ ratio & $3 \, : \, 1$ \\
        He abundance & 0.1 per total H-atom abundance \\
        $\ThirteenCO$ abundance & $\SI{4e-2}{}$ per total CO abundance \textsuperscript{a} \\
        Transfer background temperature $T_\text{CMB}$ & $\SI{2.73}{\K}$ (training data, other figures) \\
        & $\SI{0}{\K}$ (\autoref{fig:RADMC-3D vs IRIS}, \autoref{fig:POLARIS vs IRIS}, \autoref{fig:Optically Thin vs Thick}) \textsuperscript{b} \\
        Level-balance temperature $\overline{T}$ & $\SI{2.73}{\K}$ (training data, other figures) \\
        & $\SI{2.75}{\K}$ (\autoref{fig:RADMC-3D vs IRIS}, \autoref{fig:POLARIS vs IRIS}) \\
        & $\SI{0}{\K}$ (\autoref{fig:OT-OK vs OT-CMB Balance}) \\
        & $\SI{5}{\K}$ (\autoref{fig:OT-CMB vs OT-5K Balance}) \textsuperscript{c} \\
        $T_\infty$ & $\SI{5e4}{\K}$ \textsuperscript{d} \\
        $\kappa_\text{dust}$ & $\SI{1e-2}{\cm^2\g^{-1}}$ (training data, other figures) \textsuperscript{e} \\
        & $\SI{1}{\cm^2\g^{-1}}$ (\autoref{fig:No Dust vs Dust x100}) \\
        Transfer mode & Optically thick (training data, other figures) \\
        & Optically thin (\autoref{fig:Optically Thin vs Thick}) \textsuperscript{f} \\
        Integration mode & Smooth (training data, other figures) \\
        & Formal (\autoref{fig:Formal vs Smooth}) \textsuperscript{g} \\
        Velocity-integration resolution & $\SI{0.391}{\km/\s}$ \textsuperscript{h} \\
        $v$-subsampling factor & 1 at $\Delta{v_\text{channel}} = \SI{.781}{\km/s}$ (training data, other figures) \\ 
        & 8 at $\Delta{v_\text{channel}} = \SI{6.25}{\km/s}$ (\autoref{fig:RADMC-3D vs IRIS}, \autoref{fig:POLARIS vs IRIS}) \textsuperscript{h} \\
        Radial resolution & $\SI{0.881}{\parsec}$ (training data, \autoref{fig:Synthetic Reversions}) \textsuperscript{i} \\
        & $\SI{4.71}{\parsec}$ (\autoref{fig:RADMC-3D vs IRIS}), $\SI{0.586}{\parsec}$ (\autoref{fig:POLARIS vs IRIS}) \\
        & $\SI{0.110}{\parsec}$ (\autoref{fig:Low-Res vs. High-Res}), $\SI{0.220}{\parsec}$ (\autoref{fig:Low-Res vs. High-Res}, other figures) \\
        Continuum-subtraction space & $I$ (mathematically equivalent to subtraction in $T_\text{RJ}$) \textsuperscript{j, k} \\
        Output space & $T_\text{RJ}$ \textsuperscript{k} \\
        Angular resolution & $30''$ (training data, \autoref{fig:Synthetic Reversions}) \textsuperscript{l} \\
        & $0''$ (other figures) \\
        $\mu_\text{noise}$ & $\SI{0}{\K}$ \textsuperscript{m} \\
        $\sigma_\text{noise}$ & $\SI{1}{\K}$ \textsuperscript{m} \\
        \enddata
    \end{deluxetable*}
    
    \par Many readers may rightfully be curious not about the detailed theoretical or implementational details of IRIS-SO, but about whether these implementational details can be justified via concrete tests, whether IRIS-SO can be verified against known synthetic-observation codes, and how the performance of IRIS-SO stacks up against these established codes. In this section, we present a comprehensive series of such tests. Our primary results are verification against the well-known synthetic-observation codes RADMC-3D \citep{Dullemond2012} and POLARIS \citep{Reissl2016, Brauer2017} via side-by-side visualizations that we find yield near-exact matches (see \autoref{subsec:Side-by-Side Verification}, \autoref{fig:RADMC-3D vs IRIS}, and \autoref{fig:POLARIS vs IRIS}), and a speed-comparison in which we find that IRIS-SO yields up-to $10{,}000 \times$ speedups over RADMC-3D in like conditions (see \autoref{subsec:Speed Testing} and \autoref{fig:Speed Test}).
    \par We preface these tests with a discussion in \autoref{subsec:Error Metrics for Synthetic Observation} of suitable error metrics for synthetic observations, which we employ throughout the remainder of this section for quantitative comparisons. We then present a brief study in \autoref{subsec:Resolution Convergence} of the convergence of synthetic observation with respect to radial and velocity resolution. Crucially, we find that convergence requires a high radial resolution, which challenges the compute budget of this study. We discuss our choice to adopt a somewhat insufficient radial resolution in generating our training data, as this compute compromise still supports the primary objective of this study that is producing a proof-of-concept of the supervised-reversion method on synthetic data. We then identify a target radial resolution for future work focused on generalization of the trained model to real observations, which requires maximum physicality in the training data. We compare synthetic observation at our training-data resolution to synthetic observation at this approximate convergent resolution in \autoref{fig:Low-Res vs. High-Res}.
    \par We then turn to our side-by-side verifications against RADMC-3D and POLARIS, which we differentiate by purpose, in \autoref{subsec:RADMC-3D Configurations}, \autoref{subsec:POLARIS Configurations}, \autoref{subsec:POLARIS Dust Treatment}, and \autoref{subsec:Side-by-Side Verification}. RADMC-3D provides the most direct verification of the IRIS synthetic-observation algorithm itself (see \autoref{fig:RADMC-3D vs IRIS}), as we use it to observe the same physical tensor (see \autoref{subsec:Physical Tensors}), in the same spherical-coordinate system. In order to verify the IRIS physical-tensor preprocessing steps---to include interpolation of the AREPO Voronoi grid and velocity-blurring (see \autoref{subsec:Physical Tensor Interpolation})---we use POLARIS, which provides a native functionality for the solution of each ray directly over the AREPO Voronoi grid. POLARIS thus enables synthetic observation of an AREPO snapshot in its original state prior to the IRIS physical-tensor interpolation. For both codes, we find excellent visual and quantitative alignment against IRIS.
    \par In the next parts of this section, we turn to justifying other assumptions in our implementation and usage of synthetic observation via a series of additional diagnostics and visualizations. In \autoref{subsec:Optical Depth and Dust Effects}, we analyze the optical depth in our synthetic observations, justifying the choices we make in modeling spectral-line absorption/stimulation and dust emission/extinction. We find that line self-absorption is substantial in our synthetic observations, while dust extinction is mostly negligible at our assumed dust density and opacity. We also note that optical depth is sensitive to radial resolution, as discussed in \autoref{subsec:Resolution Convergence}, but that our comparisons in this section use sufficient resolution for observational convergence.
    \par In \autoref{subsec:Validity of the Optically Thin Level Balance}, we investigate the validity of our OT level balance (\autoref{subsec:Level Populations: The Optically Thin Assumption}), further analyzing the impact of optical depth, and testing the susceptibility of the OT balance to different continuum background intensities. We then use RADMC-3D to compare a synthetic observation computed under the OT assumption against ones computed under the weaker LTE assumption and the more general LVG assumption (see \autoref{subsec:Level Populations: Computational Challenges}). We find only a minimal difference between the products of the OT and LVG level balances under our observed conditions, while we find a substantial difference between these synthetic observations and one computed via the LTE balance, justifying our choice of the OT balance for expedient non-LTE transfer in the production of our training dataset.
    \par In \autoref{subsec:Comparison of Formal and Smooth Integration}, we compare the more exact formal integration scheme for solution of the radiative-transfer equation and our faster and preferred smooth integration scheme (\autoref{subsec:Transfer Solution: Formal and Smooth Integration of Optically Thick and Selectively Thin Transfer}). At our chosen observational resolution, we find the results of smooth integration to differ negligibly from formal integration, offering an essentially free $\sim 1.5\times$ speedup at minimal cost to accuracy. Then, in \autoref{subsec:Synthetic Observation Versus Density-Tracing}, we compare the results of the synthetic-observation process against density-tracing. We find strong correlation between synthetic observations and density features in our original AREPO snapshots, justifying the hypothesis that density information can be recovered from real or synthetic observations through supervised reversion (see also \autoref{subsec:Supervised Reversion: Objective} for a discussion of this machine-learning scheme). We then conclude in \autoref{subsec:Speed Testing} with our speed comparisons.

\subsection{Error Metrics for Synthetic Observation} \label{subsec:Error Metrics for Synthetic Observation}
    \par In much of the analysis contained within the following subsections, we need to conduct a quantitative comparison of two synthetic observations. In enabling such comparisons, we define a set of appropriate error metrics. We begin by characterizing the voxel- or pixel-wise error between two like PPV cube observations, or reduced images, via a metric we term \textit{Threshold-Symmetric Relative Error (TSRE)}. Following these initial definitions, we also define global statistics for comparing dislike cubes.
    \par Given Raleigh-Jeans brightness temperatures $T_1, T_2$ (see \autoref{subsec:Intensity Versus Temperature}), let
    \begin{align}
        \TSRE(T_1, T_2) &= \begin{cases} 2 \cdot \frac{|T_1 - T_2|}{|T_1| + |T_2|} & \min\big(|T_1|, |T_2|\big) > T_\text{min} \\ 0 & \text{otherwise} \end{cases} \text{ ,} \\
        T_\text{min} &= \SI{5e-2}{\K} \text{ .} \notag
    \end{align}
    Then, letting $T^{[\ell b]}, T^{[\ell v]}$ denote the mean reductions of a PPV cube $T$ over the $v$ and $b$ dimensions, respectively, define the following mean-TSRE metrics via the elementwise application of TSRE:
    \begin{align}
        \TSREcube(T_1, T_2) &= \mean_{\ell, b, v} \, \TSRE(T_1, T_2) \text{ ,} \\
        \TSRElb(T_1, T_2) &= \mean_{\ell, b} \, \TSRE\left(T_1^{[\ell b]}, T_2^{[\ell b]}\right) \text{ ,} \notag \\
        \TSRElv(T_1, T_2) &= \mean_{\ell, v} \, \TSRE\left(T_1^{[\ell v]}, T_2^{[\ell b]}\right) \text{ .} \notag \\
    \end{align}
    \par In defining TSRE and mean-TSRE, we compare voxels and pixels only above the threshold $T_\text{min} = \SI{5e-2}{\K}$, since the relative error of two very small temperatures can become large even when the numerical difference between these temperatures is negligible. Moreover, since the majority of PPV voxels in a synthetic observation have near-zero temperature, these large errors of near-equal voxels/pixels of approximately zero temperature can inflate any mean relative error in a way that is unbounded given sufficient numerical precision. Therefore, we only compare the active voxels/pixels. We prefer a symmetric relative error so that the metric equivalently and accurately illuminates the equally problematic scenarios $T_1 \gg T_2$ and $T_1 \ll T_2$. 
    \par Under these definitions, $\TSREcube$ characterizes the error between two entire synthetic observations. The reduced metric $\TSRElv$ is also enlightening, because it characterizes the error between two synthetic $\ell, v$ images, which are the inputs of our IRIS reversion model (\autoref{subsec:Supervised Reversion: Objective}). Ultimately, $\TSRElv$ is the best measure of whether the error between two synthetic observations will impact the accuracy or generalizability of the supervised-reversion scheme. We cite TSRE and mean-TSRE values as percentages, so they may be intuitively conceptualized as standard relative-error percentages. We note, however, that TSRE is capped at $200\%$. We have $\TSRE(T_1, T_2) = 200\%$ if $T_1, T_2$ are of opposite sign (note that negative continuum-subtracted brightness temperatures are admissible) Also, fixing $T_1$ with $T_2$ of the same sign, $\TSRE(T_1, T_2) \to 200\%$ as $|T_2| \to \infty$ (and similarly, fixing $T_2$). Given the specificity in our definition of TSRE, we find that this metric is most useful as a relative value comparing the quality of the multiple different side-by-side comparisons we detail throughout the rest of this section, rather than as an absolute metric of whether an individual side-by-side comparison demonstrates reasonable error. For such quality judgment of an individual synthetically observed side-by-side, we instead defer to visual comparison. Nonetheless, we find as a general tolerance that $\TSREcube$ and $\TSRElv$ values below $\sim 35\%$ indicate minimal visual discrepancy.
    \par While our TSRE metrics provide useful comparisons of PPV cubes with comparable spatial features, TSRE is limited in that it is a pointwise error metric. It is therefore not useful in comparing cubes with even very similar spatial features that are nonetheless offset according to some nonlinear spatial distortion that is nontrivial or impossible to correct. As described in \autoref{subsec:Side-by-Side Verification}, this problem is particularly relevant in comparing our POLARIS and IRIS observations, since the parallel and spherical projections employed by these respective codes yield incompatible observational geometry. We therefore prefer, in this case, the comparison of global statistics.
    \par As a first analysis, we may compare the voxel-wise temperature maxima over each whole PPV cube. This comparison provides a weak indication of whether one synthetic-observation code is computing substantially higher emission, but fails to characterize many patterns of error since it is a function of only two voxels. Similarly, we may wish to compare the mean temperatures of each cube for a more reliable measure of error. But we again encounter the issue that most voxels are nearly zero in temperature, and so voxel-wise mean temperatures differ negligibly from zero. We therefore define a \textit{Threshold-Active Mean (TAM)}, which compares only the active voxels of a cube. Specifically, given a PPV cube $T$ of Raleigh-Jeans brightness temperature, we define the set of active voxels of $T$ as
    \begin{align}
        \mathcal{A}(T) &= \Big\{(\ell, b, v) \, : \, |T| > T_\text{min}\Big\} \text{ ,} \\
        T_\text{min} &= \SI{5e-2}{\K} \text{ ,} \notag
    \end{align}
    using the same temperature threshold as before. We then define
    \begin{equation}
        \TAM(T) = \mean_{\mathcal{A}(T)} \, T \text{ .}
    \end{equation}
    And we extend the definition of $\TAM$ to $\ell, b$ and $\ell, v$ images by considering the image mean over the threshold-active pixels rather than the cube mean over active voxels. 

\subsection{Resolution Convergence} \label{subsec:Resolution Convergence}
    \begin{deluxetable*}{lcccccc}
        \tablecaption{\textbf{Resolution-Convergence Test:} Parameters and results for the resolution-convergence test described in \autoref{subsec:Resolution Convergence} for synthetic observation. We conducted two test series, the first varying radial resolution of the synthetically observed physical tensor (\autoref{subsec:Physical Tensors}), and the second varying the velocity-subsampling factor for optically thick transfer (\autoref{subsec:Transfer Solution: Optically Thick Transfer}). For each trial, the table shows the radial resolution $r$, the radial cell size $\delta{r}$, the velocity resolution $v$, the velocity-subsampling factor $f$, the velocity-integration step size $\delta{v}$, and the $\TSRElv$ value for the synthetic observation compared to that of the preceding trial (\autoref{subsec:Error Metrics for Synthetic Observation}). The test snapshot is from \citeLipman, as described in \autoref{subsec:AREPO Zoom Simulations}. All other parameters are as in the training dataset, as given in \autoref{table:Dataset Parameters} and \autoref{table:Synthetic Observation Parameters}. \label{table:Convergence Test}}
        \tablehead
        {
           \colhead{Test Series} & \colhead{$r$} & \colhead{$\delta{r}$} & \colhead{$v$} & \colhead{$f$} & \colhead{$\delta{v}$} & \colhead{$\TSRElv$ (with preceding trial)}
        }
        \startdata
        Radial Resolution Test & 128 & $\SI{3.54}{\parsec}$ & 512 & 1 & $\SI{0.391}{\km/\s}$ & \\
        & 256 & $\SI{1.76}{\parsec}$ & 512 & 1 & $\SI{0.391}{\km/\s}$ & $32.8\%$ \\
        & 512 & $\SI{0.881}{\parsec}$ & 512 & 1 & $\SI{0.391}{\km/\s}$ & $26.6\%$ \\
        & 1024 & $\SI{0.440}{\parsec}$ & 512 & 1 & $\SI{0.391}{\km/\s}$ & $21.4\%$ \\
        & 2048 & $\SI{0.220}{\parsec}$ & 512 & 1 & $\SI{0.391}{\km/\s}$ & $16.0\%$ \\
        & 4096 & $\SI{0.110}{\parsec}$ & 512 & 1 & $\SI{0.391}{\km/\s}$ & $10.4\%$ \\
        & 8192 & $\SI{5.49e-2}{\parsec}$ & 512 & 1 & $\SI{0.391}{\km/\s}$ & $7.38\%$ \\
        & 16384 & $\SI{2.75e-2}{\parsec}$ & 512 & 1 & $\SI{0.391}{\km/\s}$ & $7.78\%$ \\
        & & & & & & \\
        Velocity Resolution Test & 512 & $\SI{0.881}{\parsec}$ & 128 & 0 & $\SI{3.13}{\km/\s}$ & \\
        & 512 & $\SI{0.881}{\parsec}$ & 128 & 1 & $\SI{1.56}{\km/\s}$ & $6.62\%$ \\
        & 512 & $\SI{0.881}{\parsec}$ & 128 & 2 & $\SI{0.781}{\km/\s}$ & $10.5\%$ \\
        & 512 & $\SI{0.881}{\parsec}$ & 128 & 3 & $\SI{0.521}{\km/\s}$ & $5.93\%$ \\
        & 512 & $\SI{0.881}{\parsec}$ & 128 & 4 & $\SI{0.391}{\km/\s}$ & $4.15\%$ \\
        & 512 & $\SI{0.881}{\parsec}$ & 128 & 5 & $\SI{0.313}{\km/\s}$ & $3.54\%$ \\
        & 512 & $\SI{0.881}{\parsec}$ & 128 & 6 & $\SI{0.260}{\km/\s}$ & $3.30\%$ \\
        & 512 & $\SI{0.881}{\parsec}$ & 128 & 7 & $\SI{0.223}{\km/\s}$ & $3.50\%$ \\
        \enddata
    \end{deluxetable*}
    \begin{figure*}[t]
        \centering
        \includegraphics[width=1\linewidth]{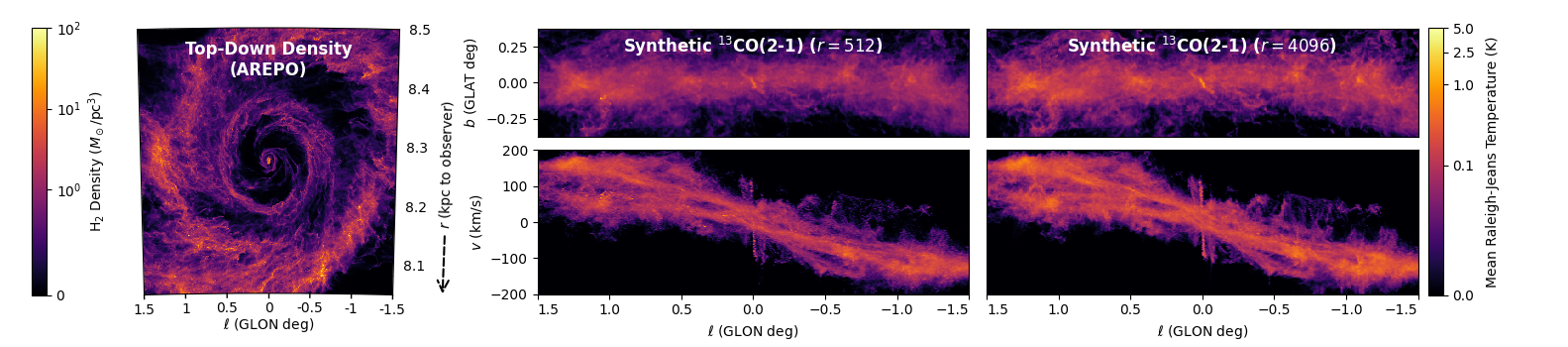}
        \caption{\textbf{Synthetic-Observation Radial-Resolution Comparison:} We compare IRIS-SO applied to a physical tensor (see \autoref{subsec:Physical Tensors}) generated at our training-data radial resolution of $r = 512$, corresponding to $\delta{r} = \SI{0.881}{\parsec}$, against IRIS-SO applied to a physical tensor generated over the same observational field but at an approximate convergent resolution of $r = 4096$ and $\delta{r} \leq \SI{0.110}{\parsec}$. The simulation featured is due to \citeLipman, as described in \autoref{subsec:AREPO Zoom Simulations}, which we feature in our training dataset. All color bars are scaled with an $\arcsinh$ nonlinearity. See \autoref{subsec:Resolution Convergence} for a detailed discussion of the figure. We find that our training-data resolution is somewhat insufficient for convergence, but that this computational simplification remains sufficient in providing a proof-of-concept of the supervised-reversion method on synthetic data, in which maximal training-data physicality for generalization to real observations is not the priority.}
        \label{fig:Low-Res vs. High-Res}
    \end{figure*}

    \par A key factor determining the accuracy of the synthetic-observation process is resolution. In any synthetic-observation code, the output only converges to a stable result in the limit towards infinite resolution in the observed field. Here, we distinguish angular, line-of-sight, and velocity resolution. Insufficient angular resolution yields pixelation in the synthetic observation that is visually obvious. Insufficient line-of-sight and velocity resolution can produce more subtle failure modes, and should be considered with greater care. In IRIS-SO, in optically thick transfer (\autoref{subsec:Transfer Solution: Optically Thick Transfer}), these failure modes associated with line-of-sight and velocity resolution are only avoided by setting the physical-tensor resolution (\autoref{subsec:Physical Tensors}) and velocity-subsampling factor (\autoref{subsec:Transfer Solution: Optically Thick Transfer}) sufficiently high. We take a moment to discuss our convergence testing in these resolutions, as well as to note the impacts of the differing resolutions we use across the verifications in this section and our training data (see \autoref{table:Dataset Parameters}).
    \par The velocity-resolution problem is discussed in \autoref{subsec:Transfer Solution: Overview}. While selectively thin (\autoref{subsec:Transfer Solution: Selectively Thin Transfer}) and optically thin (\autoref{subsec:Transfer Solution: Optically Thin Transfer}) transfer provide analytic solutions to velocity channel-integration that are resolution-independent, they rely on approximations that do not hold when line self-absorption is substantial. In the more robust optically thick transfer (\autoref{subsec:Transfer Solution: Optically Thick Transfer}), which we use in constructing our training dataset, each velocity channel must be numerically integrated by subsampling. Setting the velocity-subsampling factor too low can yield subsampling gaps that are larger than the line width itself. In this case, both line emission and absorption can be substantially underestimated. Whether such underestimation yields observed intensities that are too low or too high depends upon whether the observed source is bright or dark against the background. But in typical regimes of spectral-line observation, this failure mode presents as a shotgun appearance in the observation, where full intensity is only observed at a sparse subset of points separated by regions of low intensity.
    \par The most notable failure mode associated with line-of-sight resolution also concerns velocity. In IRIS, line-of-sight resolution is the radial resolution of the physical tensor. Since, in the transfer solution, velocity is treated as constant within each physical-tensor cell, a cell with a high radial depth will yield line self-absorption that is higher than in a series of radially smaller cells in which line peaks are separated by fine Doppler shifts. This effect will tend to produce a systematic overestimation of line self-absorption and underestimation of observed intensity. In IRIS, this failure mode is only alleviated by setting the radial resolution of the physical tensor sufficiently high to attain approximate convergence. In contrast to other synthetic-observation codes, IRIS employs no adaptive or recursive resolution methods. Fixing resolution enables tensorization, which is the critical efficiency that allows GPU acceleration leading to the large speedups we describe in \autoref{subsec:Speed Testing}. Static resolution, however, imposes greater import on choosing an appropriate resolution through convergence testing.
    \par We conducted two series of convergence tests, probing radial resolution and velocity-subsampling factor separately. For both series, we synthetically observed a test snapshot from \citeLipman, as described in \autoref{subsec:AREPO Zoom Simulations}, under the radial, angular, and velocity extents specified in \autoref{table:Dataset Parameters} for our training dataset. In the first series, we fixed our angular resolution, velocity resolution, and velocity-subsampling factor at the nominal values given in \autoref{table:Dataset Parameters} and \autoref{table:Synthetic Observation Parameters} and varied radial resolution. In the second series, we set angular and radial resolutions at the nominal values and, fixing velocity resolution at a low base value, varied the velocity-subsampling factor. We then judged convergence both visually in the synthetic observations and via $\TSRElv$ on consecutive pairs within each series. The parameters and $\TSRElv$ statistics for each test are summarized in \autoref{table:Convergence Test}.
    \par On our test snapshot, under our observational parameters at a velocity resolution of $v = 64$, we found approximate visual convergence above a subsampling factor of $f = 2$, corresponding to a velocity-integration step size of $\delta{v} \leq \SI{0.781}{\km/\s}$. Our primary visual findings are supported by the minimal $\TSRElv$ values shown in \autoref{table:Convergence Test} between higher-resolution trials. For our training dataset, and for all other synthetic observations in this study, we choose velocity-subsampling factors such that $\delta{v} = \SI{0.391}{\km/\s}$, which we find is computationally tractable. Crucially, however, we find that convergence with respect to radial resolution occurs only at very high resolution. For our test snapshot and observational parameters, we find approximate visual convergence above a radial resolution of $r = 4096$, corresponding to a radial cell size of $\delta{r} \leq \SI{0.110}{\parsec}$. 
    \par Due to the quadratic time-complexity scaling of the synthetic-observation process described in \autoref{subsec:Foreground and Background}, high-resolution data generation is computationally expensive. While such expense is necessary, however, in maximizing physicality of the training dataset for generalization to real observational data, as is the future objective of the IRIS project, maximal physicality is not strictly necessary in establishing a proof-of-concept of the supervised-reversion method on synthetic data. Since the effect of insufficient radial resolution is merely consistent degradation of observation quality and overestimation of optical depth, successful training of our reversion model on a training dataset generated at such insufficient resolution is still a valuable demonstration of the potential of the method, particularly when expanded in the future to more sophisticated training data. Adoption of a somewhat insufficient radial resolution in our training data is also consistent with the analysis in this study of a limited training dataset containing only a single simulation run in which all not physics are fully activated, as described in \autoref{subsec:AREPO Zoom Simulations}.
    \par For this study, we therefore choose a lower training-data resolution of $r = 512$, corresponding to $\delta{r} = \SI{0.881}{\parsec}$. We compare synthetic observation at this lower resolution to synthetic observation at our approximate convergent resolution of $r = 4096$ and $\delta{r} \leq \SI{0.110}{\parsec}$ in \autoref{fig:Low-Res vs. High-Res}. In the left panel, we show the top-down (latitude-mean) density of our test AREPO snapshot from \citeLipman, described in \autoref{subsec:AREPO Zoom Simulations}. In the center panels, we show the mean $\ell, b$ and $\ell, v$ projections of our synthetic observation of this snapshot at $r = 512$. In the right panels, we show the mean $\ell, b$ and $\ell, v$ projections of our synthetic observation at $r = 4096$. We find a noticeable but marginal visual difference. Quantitatively, using the error metrics defined in \autoref{subsec:Error Metrics for Synthetic Observation}, we find $\TSREcube(T_{512}, T_{4096}) \approx 48.2\%$ and $\TSRElv(T_{512}, T_{4096}) \approx 40.2\%$, with $\TAM(T_{512}) \approx \SI{0.29}{\K}$ and $\TAM(T_{4096}) \approx \SI{0.38}{\K}$, illustrating significant error. Reconstruction of training data at a higher radial resolution will therefore be important in future work focused on generalization to real observations.
    \par We adopt this lower radial resolution of $\delta{r} = \SI{0.881}{\parsec}$ in \autoref{fig:Synthetic Reversions}. In \autoref{fig:RADMC-3D vs IRIS}, we choose a substantially lower radial resolution of $\delta{r} = \SI{4.71}{\parsec}$ due to the limiting computational expense of RADMC-3D \citep{Dullemond2012}, which, as described in \autoref{subsec:RADMC-3D Configurations}, still serves to validate IRIS-SO under like observational conditions. For \autoref{fig:POLARIS vs IRIS}, we use a higher resolution of $\delta{r} = \SI{0.586}{\parsec}$, which better matches the velocity-interpolation method implemented in POLARIS \citep{Reissl2016, Brauer2017} as described in \autoref{subsec:POLARIS Configurations}. While this resolution is still lower than the approximate convergent resolution of $\delta{r} \leq \SI{0.110}{\parsec}$ that we find on the zoom-simulations from \citeLipman, we nonetheless find that it is sufficient for convergence on the lower-resolution simulations we observe in our side-by-side comparisons. (See \autoref{subsec:POLARIS Configurations} for an explanation of our use of lower-resolution simulations for these side-by-side comparisons due to the computational expense of POLARIS.) For all other synthetic observations in this publication, we choose a value of $\delta{r} = \SI{0.220}{\parsec}$ that we find provides approximate visual convergence, with a numerical deviation from $\delta{r} \leq \SI{0.110}{\parsec}$ given in \autoref{table:Convergence Test} as $\TSRElv(T_{2048}, T_{4096}) \approx 10.4\%$.

\subsection{RADMC-3D Configurations} \label{subsec:RADMC-3D Configurations}
    \par We now turn to verifying IRIS-SO against known synthetic-observation codes. For the most direct verification of the synthetic-observation process itself, we perform a side-by-side verification against the synthetic-observation code RADMC-3D \citep{Dullemond2012}. See \autoref{subsec:Side-by-Side Verification} and \autoref{fig:RADMC-3D vs IRIS} for our final results and error analysis. See also \autoref{table:Synthetic Observation Parameters} for a compilation of all observational parameters for this side-by-side. Among the ecosystem of synthetic-observation codes, RADMC-3D is notable not only for its established track record and widespread use, but for its breadth of features, which exceed the more limited options provided by IRIS-SO. Among these more advanced features are computation of dust heating and scattering via Monte-Carlo ray-tracing, polarization effects, and Zeeman splitting of spectral lines under the influence of magnetic fields.
    \par As detailed below, we turn most of these features off for the purpose of this verification, which focuses on reproducing the exact conditions observed by IRIS, which are in turn optimized for computational efficiency (see \autoref{table:IRIS-SO Modeling Effects} for a list of IRIS-SO features). For this verification, we provide RADMC-3D and IRIS the exact same physical tensor, which we compute from a snapshot sampled from a low-resolution AREPO simulation due to \citeLipman, different from the high-resolution simulation used in our IRIS training data. This alternate simulation follows the same general parameters as those described in \autoref{subsec:AREPO Zoom Simulations}, except in the absence of sink particles and lack of the zoom-in refinement scheme. We select this lower-resolution snapshot in order to provide consistency with our POLARIS-IRIS side-by-side, for which we are constrained to a lower resolution by computational limits (see \autoref{subsec:POLARIS Configurations} for more details). We further process this physical tensor for observation by RADMC-3D by applying the same $\ThirteenCO$ abundance function as described in \autoref{subsec:Spectral-Line Configuration} and the same $3 : 1$ ortho-$\HTwo$-to-para-$\HTwo$ ratio utilized by IRIS.
    \par Observation of the same physical tensor under like conditions first requires that we define our IRIS coordinate system in RADMC-3D. Unlike in IRIS, which, for computational efficiency, couples the physical-tensor coordinate system and grid with the ray-tracing computation during synthetic observation, RADMC-3D keeps these two computational considerations separate. In RADMC-3D, the observed field may be specified via a regular cartesian grid, an octree grid, or a spherical-coordinate grid. The primary mode of observation supported by RADMC-3D is then a parallel-plane projection of whichever coordinate system the user specifies, which entails tracing across grid boundaries, off grid alignment. In order to provide a more direct comparison with IRIS, however, we use RADMC-3D's local-observer mode, which allows us to observe in a spherical projection. Setting the focus of our spherical projection as the origin of our IRIS spherical-coordinate system, these settings exactly mirror our IRIS configurations, prompting RADMC-3D to compute line emission and absorption, thermal dust emission, and dust extinction via the same direct ray-tracing along fixed angular coordinates in each physical tensor.
    \par We further specify that the number of pixels observed by RADMC-3D is the same as the angular pixel-resolution of our physical tensor. For this comparison, we choose this resolution in IRIS to provide square pixels, as opposed to the rectangular pixel dimensions specified in \autoref{table:Dataset Parameters}, in order to provide compatibility with RADMC-3D. We are also limited in the radial resolution we are able to observe in RADMC-3D, due to computational expense. As described in \autoref{subsec:Resolution Convergence}, we therefore choose a radial resolution that we find is insufficient for convergence. Nonetheless, because RADMC-3D and IRIS are observing the exact same physical tensor, our comparison still serves to verify the IRIS synthetic-observation algorithm itself, even if observation at this resolution is not an accurate observation of the underlying AREPO snapshot. In order to verify the accuracy of synthetic observation against the ground truth of the simulation, we instead rely on our POLARIS verification (see \autoref{subsec:POLARIS Configurations}). 
    \par As in IRIS, we then configure RADMC-3D to observe both a line-plus-dust-continuum cube and a pure dust-continuum cube, and perform a post-transfer continuum subtraction (see \autoref{subsec:Continuum Subtraction}). For the line cube, we configure RADMC-3D to observe the exact velocity-channel centers as in the IRIS fine-velocity grid used in our optically-thick transfer (see \autoref{subsec:Transfer Solution: Optically Thick Transfer}). We then perform the same post-transfer velocity integration of the fine-velocity grid so that the velocity resolutions of our final synthetic observations match exactly, with equivalent velocity-channel averaging. For the dust cube, we configure RADMC-3D to observe only the final coarse velocity channels, as in IRIS. We further configure RADMC-3D to observe just thermal dust emission and dust extinction. We introduce a constant dust opacity at all wavelengths, which we set to the same value from \citet{Ossenkopf1994} that we use in IRIS, as described in \autoref{subsec:Dust Emission and Absorption}. While RADMC-3D provides the ability to compute dust temperatures due to radiative transfer via a Monte-Carlo ray-tracing algorithm, we configure RADMC-3D instead to use the same external dust temperatures as we employ in IRIS, provided from the AREPO chemical network (see \autoref{subsec:AREPO Zoom Simulations}). Under these conditions, the dust continuum observed by RADMC-3D is identical to that observed by IRIS.
    \par In RADMC-3D, we turn off all scattering and flux-conservation, and use first-order integration, which are the settings that are fastest and that most directly parallel IRIS. The first-order integration scheme used by RADMC-3D is equivalent to the formal integration scheme detailed in \autoref{subsec:Transfer Solution: Formal and Smooth Integration of Optically Thick and Selectively Thin Transfer}. Since we configure IRIS to use our preferred smooth integration (see \autoref{subsec:Transfer Solution: Formal and Smooth Integration of Optically Thick and Selectively Thin Transfer}), this aspect of our comparison is one that is unlike. We separately verify, however, that smooth and formal integration yield negligibly different results in \autoref{subsec:Comparison of Formal and Smooth Integration}, including via a dedicated side-by-side visualization in \autoref{fig:Formal vs Smooth}.
    \par For the RADMC-3D line cube, we configure RADMC-3D to use the same OT level balance used by IRIS (see \autoref{subsec:Level Populations: The Optically Thin Assumption}). Unlike in the synthetic observations of our training dataset, however, for this comparison, we use the original LAMDA $\ThirteenCO$ data-file \citep{Schoier2005} for both RADMC-3D and IRIS, without the added H collisions (see \autoref{subsec:Spectral-Line Configuration}). We then provide RADMC-3D only the gas density of $\HTwo$. Under these conditions, both RADMC-3D and IRIS consider only $\HTwo$ collisions in computing the level balance (\autoref{subsec:Level Populations: Mathematical Solution}). Since we are unable to specify H collision rates to POLARIS (see \autoref{subsec:POLARIS Configurations}), this simplification ensures our IRIS, RADMC-3D, and POLARIS observations are all directly comparable. In the same vein, we set the continuum background temperature in our side-by-side to $\overline{T} = \SI{2.75}{\K}$ (as opposed to our preferred value in the training dataset of $\overline{T} = \SI{2.73}{\K}$, see \autoref{table:Synthetic Observation Parameters}), since this value is hard-coded into the POLARIS OT level balance. For simplicity of comparison, however, we set the radiative-transfer background to $\SI{0}{\K}$. 
    \par In computing this level balance, RADMC-3D cannot compute temperatures that exceed the temperature ceiling specified in our LAMDA file. For this reason, we cap gas temperatures for RADMC-3D at $\SI{2995}{\K}$, just under the temperature ceiling in the LAMDA file of $\SI{3000}{\K}$. By contrast, IRIS pre-computes a grid of emission and absorption coefficients at temperatures ranging between $\SI{0}{\K}$ and the LAMDA ceiling of $\SI{3000}{\K}$, and then computes a linear interpolation of these coefficients above the temperature ceiling (see \autoref{subsec:Grid Precomputation}). While this aspect of our RADMC3D-IRIS side-by-side is not a like comparison, our final results demonstrate that these algorithmic differences yield error within an acceptable tolerance. To double-check our level balance specifically, we also performed a numerical comparison of the RADMC-3D OT level balance and the IRIS OT level balance over a large array of temperatures, gas densities, and $\HTwo$ abundances, and found results identical within floating-point precision. See \autoref{subsec:Side-by-Side Verification} and \autoref{fig:RADMC-3D vs IRIS} for results and analysis of our RADMC3D-IRIS comparison.

\subsection{POLARIS Configurations} \label{subsec:POLARIS Configurations}
    \par In addition to our RADMC-3D side-by-side, we also provide a separate verification against the synthetic-observation code POLARIS \citep{Reissl2016, Brauer2017}. See \autoref{subsec:Side-by-Side Verification} and  \autoref{fig:POLARIS vs IRIS} for our final results and error analysis. See also \autoref{table:Synthetic Observation Parameters} for a compilation of all observational parameters for this side-by-side. Unlike our RADMC-3D comparison, which serves to proof the IRIS synthetic-observation algorithm under the most like conditions, the aim of our POLARIS comparison is to investigate the error introduced by the preprocessing steps executed in the IRIS pipeline prior to synthetic observation. These steps include the nearest-neighbor interpolation of the AREPO Voronoi mesh utilized during physical-tensor construction, as well as the velocity blurring applied to each physical tensor to eliminate cell-edge artifacts (see \autoref{subsec:Physical Tensor Interpolation}).
    \par POLARIS is notable for its ability to natively handle Voronoi grids, of the kind used in AREPO (see \autoref{subsec:Physical Tensors}), which is why we select this particular synthetic-observation code for this task. Like RADMC-3D, POLARIS also provides a much wider breadth of features than those modeled by IRIS-SO (see \autoref{table:IRIS-SO Modeling Effects} for list of IRIS-SO features). Among these more advanced features are (as in RADMC-3D), computation of dust heating and scattering via Monte-Carlo ray-tracing, polarization effects, and Zeeman splitting of spectral lines under the influence of magnetic fields. As in our RADMC-3D side-by-side, we turn these features off for a more direct comparison. In these configurations, the POLARIS line radiative transfer, thermal dust emission, and dust extinction implement a similar direct ray-tracing algorithm as does IRIS and RADMC-3D, but over the Voronoi grid itself as opposed to over a physical-tensor coordinate grid.
    \par For this comparison, we observe the same simulation snapshot as in our RADMC-3D side-by-side, which is sampled from a low-resolution AREPO simulation due to \citeLipman, different from the high-resolution simulation used in our training dataset. This alternate simulation follows the same parameters that are described in \autoref{subsec:AREPO Zoom Simulations}, other than in the absence of sink particles and lack of the zoom-in refinement scheme. We choose this alternate simulation because, due to the extraordinary resolution within the CMZ region of the zoom-in simulations of \citeLipman, the use of POLARIS to synthetically observe them via the original Voronoi mesh is computationally infeasible. This reduced-resolution comparison still provides an end-to-end verification of both the physical-tensor interpolation step (\autoref{subsec:Physical Tensor Interpolation}) and the synthetic-observation pipeline.
    \par As described in \autoref{subsec:Resolution Convergence}, we configure IRIS with a radial resolution that we find is sufficient for convergence. A sufficient radial resolution is necessary for an accurate comparison, because POLARIS implements an adaptive solution to the radial-resolution problem that is independent of the resolution of the observed grid. This adaptive solution involves the cubic interpolation of velocity at each substep of the Runge-Kutta-Fehlberg method POLARIS employs in solving each ray \citep{Reissl2019}. The output of POLARIS is therefore a good estimate of a synthetic observation in the limit of infinite radial resolution, which serves to strengthen our use of POLARIS in verifying our entire simulation-processing pipeline. Note that this higher radial resolution renders our IRIS results in our POLARIS verification different, however, from those in our RADMC-3D verification (see \autoref{subsec:RADMC-3D Configurations}).    
    \par The preparation of our AREPO snapshot for synthetic observation by POLARIS differs substantially from the preparations detailed in \autoref{subsec:RADMC-3D Configurations} for our RADMC-3D side-by-side, since we are no longer observing the physical tensor directly. We begin by sampling all cells of the AREPO Voronoi grid within the spherical-coordinate bounds of our physical tensor. This step ensures that POLARIS is observing the same underlying gas field as IRIS. Over each of the sampled Voronoi cells, we then use the same methods described in \autoref{subsec:Tensor Variables and Gas Temperature} and \autoref{subsec:Spectral-Line Configuration} to compute gas density, kinetic temperature, and $\ThirteenCO$ abundance. We also use the same $T_\infty = \SI{5e4}{\K}$ temperature threshold described in \autoref{subsec:Observability Determination}. We additionally configure a dust field, as described in \autoref{subsec:POLARIS Dust Treatment}, to match our IRIS settings.
    \par To perform our synthetic observation in POLARIS, we run both a line radiative-transfer routine (\texttt{CMD\_LINE\_EMISSION})---which performs direct ray-tracing of the total intensity of the spectral line and dust continuum, without scattering---and a continuum-only routine (\texttt{CMD\_DUST\_EMISSION})---which computes a single dust-continuum intensity at the line center. We then subtract the continuum intensity from the total-intensity cube as in IRIS (see \autoref{subsec:Continuum Subtraction}). In computing the line radiative transfer, we configure POLARIS to use the same OT level balance as in IRIS (\autoref{subsec:Level Populations: The Optically Thin Assumption}, termed \textit{full escape probability} or \textit{FEP} in POLARIS). Since POLARIS only computes this balance under the assumption of a $\SI{2.75}{\K}$ background, we equivalently set the IRIS parameter $\overline{T} = \SI{2.75}{\K}$, although we leave the radiative transfer background temperature in both codes as $T_\text{CMB} = \SI{0}{\K}$. 
    \par As in our RADMC-3D comparison, we use the standard LAMDA \citep{Schoier2005} $\ThirteenCO$ file for both POLARIS and IRIS, and provide POLARIS the gas density of only $\HTwo$. POLARIS then internally sets the same $3 : 1$ ortho-$\HTwo$-to-para-$\HTwo$ ratio that we configure in our default IRIS settings. Under all these conditions, IRIS, POLARIS, and RADMC-3D compute the identical level balance, considering collisions only with $\HTwo$. These configurations contrast those in our training data, which incorporates collision rates for monotomic H due to \citet{Walker2015} via our augmented LAMDA file (see \autoref{subsec:Spectral-Line Configuration} and \autoref{table:Synthetic Observation Parameters}). We make this simplification since we are not able to configure POLARIS to compute H collisions.
    \par While the volume of our AREPO snapshot that is observed by POLARIS is identical to that observed by IRIS, a direct comparison via equivalent observational geometry is not feasible. POLARIS does not implement a spherical projection scheme over a regular angular grid in dimensions of longitude and latitude that is equivalent to the IRIS projection scheme (as defined by the physical-tensor grid; see \autoref{subsec:Physical Tensors}). Instead, we implement a plane detector, observing towards the galactic center in parallel projection. We then transform this parallel projection into angular coordinates via reprojection under a small-angle approximation, which yields negligible distortion within the limited angular extent of our observation ($\ell \in [\SI{-3}{\degree}, \SI{3}{\degree}], b \in [\SI{-1}{\degree}, \SI{1}{\degree}]$). Following observation via both POLARIS routines, subtraction, and angular reprojection, the output of our POLARIS pipeline is a single, continuum-subtracted PPV cube, which is directly comparable to the output of IRIS under the same configurations. The direct side-by-side comparison is visualized, in both $\ell, b$ and $\ell, v$ mean reduction, in \autoref{fig:POLARIS vs IRIS}.

\subsection{POLARIS Dust Treatment} \label{subsec:POLARIS Dust Treatment}
    \par POLARIS is also notable, however, in the degree of sophistication with which it treats dust, to include allowance for multi-species dust with varying grain geometry and orientation, which in turn contribute to diverse polarization effects \citep{Reissl2016, Brauer2017}. Each dust species must be specified via a configuration file that includes grain properties such as density as well as complex refractive indices over a spectrum of discrete wavelengths. The user must also specify a grain-size distribution. This native complexity in the POLARIS dust treatment is an additional unlike element of our comparison, since IRIS considers only a single dust species with a scaleless, constant dust opacity. 
    \par To best match these conditions in POLARIS, we produce a custom dust-configuration file with attributes that we calculate below to match our IRIS constant dust opacity, as described in \autoref{subsec:Dust Emission and Absorption}. Our derived dust parameters are
    \begin{align}
        \rho_\text{dust} &= .01\rho_\text{gas} \text{ ,} \\
        \rho_\text{grain} &= \SI{1}{\g\cm^{-3}} \text{ ,} \qquad \text{and} \\
        m &= 1.0 + i \, \SI{1.0822e-2}{} \text{ ,}
    \end{align}
    where $\rho_\text{dust}$ and $\rho_\text{gas}$ are the dust and gas mass-densities in the ISM, respectively, $\rho_\text{grain}$ is the material density of an individual grain, and $m$ is the grain's complex refractive index at the center wavelength $\lambda \approx \SI{1.36}{\mm}$ of the $\ThirteenCOTwoOne$ transition. This proportionality of $\rho_\text{dust}$ and $\rho_\text{gas}$ matches the assumed $100 : 1$ gas-to-dust ratio we use in deriving our IRIS opacity of $\kappa = \SI{1e-2}{\cm^2\g^{-1}}$ from \citet[][\autoref{subsec:Dust Emission and Absorption}]{Ossenkopf1994}. We further configure POLARIS with a standard distribution of dust number-density according to a power-law in grain radius,
    \begin{equation}
        n(r) \propto r^\text{-3.5} \text{ ,}
    \end{equation}
    with radii ranging from $r_\text{min} = \SI{5e-9}{\m}$ to $r_\text{max} = \SI{2.5e-7}{\m}$. 
    \par Since $r \ll \lambda$, then assuming idealized, spherical grains, the absorption cross-section of an individual dust grain is approximated by
    \begin{equation}
        C_\text{abs} \approx \frac{8\pi^2r^3}{\lambda}\Im{\frac{m^2 - 1}{m^2 + 2}}
    \end{equation}
    \citep{Bohren1983}. Letting $\mu(r)$ denote the mass of an individual dust grain, the average dust opacity (defined per unit dust mass) for the fixed wavelength $\lambda$ is then given by
    \begin{align}
        \kappa_\text{dust} &\approx \frac{\int_{r_\text{min}}^{r_\text{max}} C_\text{abs} \, dn(r)}{\int_{r_\text{min}}^{r_\text{max}} \mu \, dn(r)} \\
        &= \frac{\frac{8\pi^2}{\lambda}\Im{\frac{m^2 - 1}{m^2 + 2}} \int_{r_\text{min}}^{r_\text{max}} r^3 \, dn(r)}{\frac{4\pi\rho_\text{dust}}{3} \int_{r_\text{min}}^{r_\text{max}} r^3 \, dn(r)} \notag \\
        &= \frac{6\pi}{\lambda\rho_\text{dust}}\Im{\frac{m^2 - 1}{m^2 + 2}} \notag \\
        &\approx \frac{6\pi}{(\SI{1.36e-3}{\m})(\SI{1e3}{\kg\m^{-3}})} \notag \\
        &\hspace{1cm} \cdot \Im{\frac{(1.0 + i \, \SI{1.0822e-2}{})^2 - 1}{(1.0 + i \, \SI{1.0822e-2}{})^2 + 2}} \notag \\
        &\approx \SI{1e-1}{\m^2\kg^{-1}} \text{ .} \notag
    \end{align}
    Hence, the average dust opacity per unit gas density is then
    \begin{equation}
        \kappa \approx .01\kappa_\text{dust} \approx \SI{1e-2}{\cm^2\g^{-1}} \text{ ,}
    \end{equation}
    matching our IRIS value. Note also that this result is invariant to the grain-size distribution under this idealized model of dust-grain absorption cross section.

\subsection{Side-by-Side Verification} \label{subsec:Side-by-Side Verification}
    \begin{figure*}[t]
        \centering
        \includegraphics[width=1\linewidth]{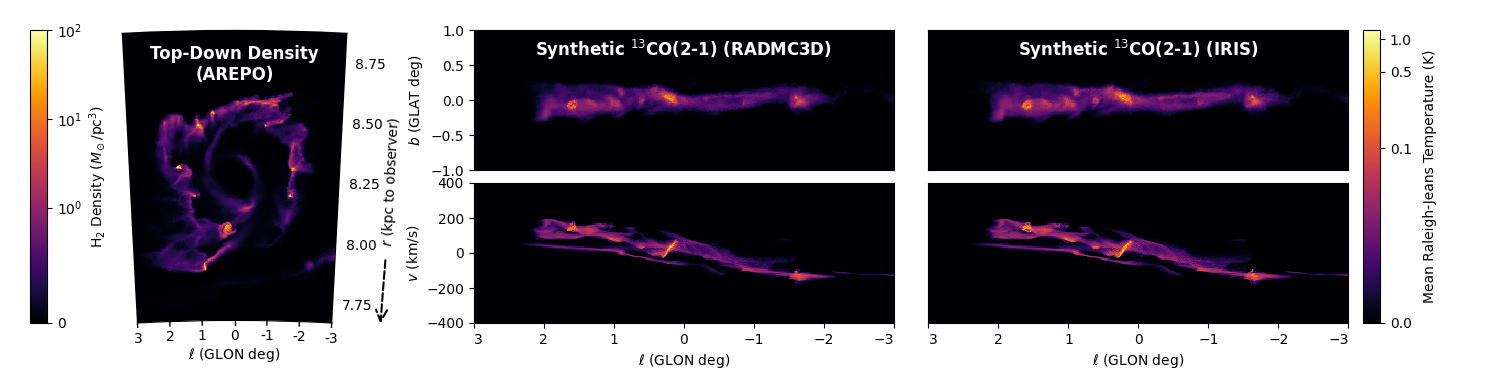}
        \caption{\textbf{RADMC3D-IRIS Side-by-Side:} A comparison of the RADMC-3D \citep{Dullemond2012} and IRIS synthetic-observation codes, in the $\ThirteenCOTwoOne$ spectral line. We define a set of test conditions, as described in \autoref{subsec:RADMC-3D Configurations}, such that RADMC-3D and IRIS-SO compute like observations of the same physical tensor (see \autoref{subsec:Physical Tensors} for physical-tensor definitions). The observed simulation is due to \citeLipman. All color bars are scaled with an $\arcsinh$ nonlinearity. See \autoref{subsec:Side-by-Side Verification} for a detailed discussion of the figure. Note that the observations are not comparable to those in \autoref{fig:POLARIS vs IRIS} due to differing radial resolutions (see \autoref{subsec:Resolution Convergence}). We see strong alignment consistent with our quantitative comparisons detailed in \autoref{subsec:Side-by-Side Verification}, proving the sound implementation of the IRIS synthetic-observation code.}
        \label{fig:RADMC-3D vs IRIS}
    \end{figure*}
    \begin{figure*}[t]
        \centering
        \includegraphics[width=1\linewidth]{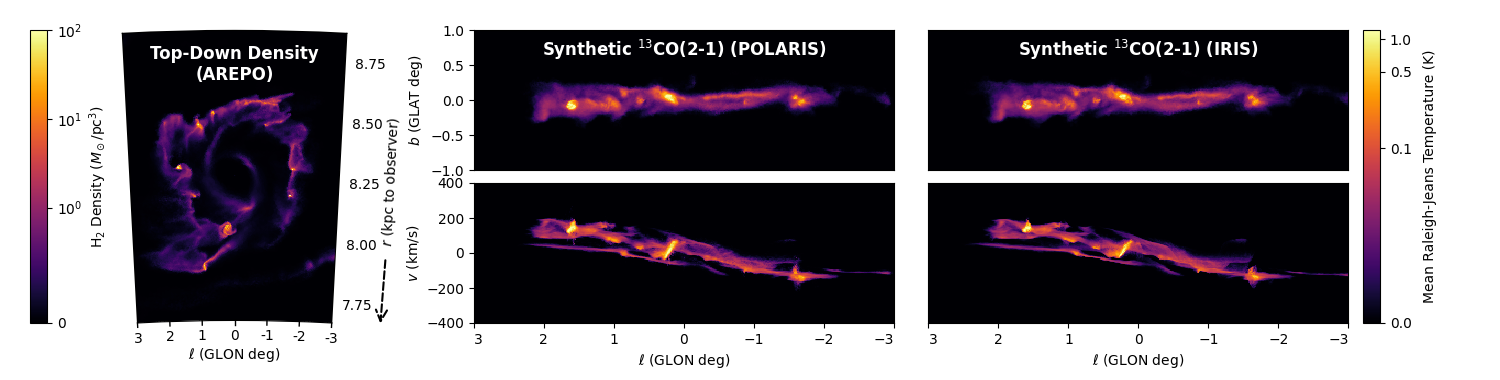}
        \caption{\textbf{POLARIS-IRIS Side-by-Side:} A comparison of the POLARIS and IRIS synthetic-observation codes, in the $\ThirteenCOTwoOne$ spectral line. We define a set of test conditions, as described in \autoref{subsec:POLARIS Configurations}, such that POLARIS and IRIS-SO compute like observations. Since POLARIS directly observes the AREPO Voronoi mesh as opposed to the physical tensor observed by IRIS, this comparison provides insight into the error introduced by our simulation-preprocessing pipeline. The observed simulation is due to \citeLipman. All color bars are scaled with an $\arcsinh$ nonlinearity. See \autoref{subsec:Side-by-Side Verification} for a detailed discussion of the figure. Note that the observations are not comparable to those in \autoref{fig:RADMC-3D vs IRIS} due to differing radial resolutions (see \autoref{subsec:Resolution Convergence}). We see excellent alignment consistent with the quantitative analysis we provide in \autoref{subsec:Side-by-Side Verification}, demonstrating the end-to-end accuracy of the entire simulation-preprocessing pipeline and synthetic-observation code.}
        \label{fig:POLARIS vs IRIS}
    \end{figure*}

    \par Our RADMC-3D and POLARIS verifications provide the benefit of both qualitative and quantitative comparison. For quantitative comparison, we begin with our RADMC-3D verification. Since our RADMC-3D and IRIS synthetic observations were computed under the same spherical projection, over the exact same observational resolution, we are able to perform a voxel-wise error analysis over the entire PPV cube. Using the error metrics defined in \autoref{subsec:Error Metrics for Synthetic Observation}, we find the whole-cube error to be $\TSREcube(\text{RADMC-3D}, \text{IRIS}) \approx 34.8\%$ and the error in our latitude-mean $\ell, v$ images (which are the actual data products provided as inputs to our reversion model, see \autoref{subsec:Supervised Reversion: Objective}) to be $\TSRElv(\text{RADMC-3D}, \text{IRIS}) \approx 33.7\%$. At the lower radial resolution adopted in this RADMC-3D comparison (see \autoref{subsec:Resolution Convergence}), we find the voxel-wise maxima in units of continuum-subtracted Raleigh-Jeans brightness temperature (see \autoref{subsec:Intensity Versus Temperature}) to be $\max(T_\text{IRIS}) \approx \SI{59.1}{\K}$ and $\max(T_\text{RADMC}) \approx \SI{83.9}{\K}$. For the Threshold-Active Mean we define in \autoref{subsec:Error Metrics for Synthetic Observation}, we find $\TAM(T_\text{IRIS}) \approx \SI{0.29}{\K}$ and $\TAM(T_\text{RADMC}) \approx \SI{0.26}{\K}$. These results indicate reasonable quantitative alignment.
    \par Quantitative comparison between POLARIS and IRIS is more challenging. As described in \autoref{subsec:Error Metrics for Synthetic Observation}, the geometric differences between the parallel projection employed by POLARIS and the spherical projection employed by IRIS preclude a voxel-wise error analysis, so we limit ourselves to the analysis of global statistics. At the higher radial resolution adopted in this POLARIS comparison (see \autoref{subsec:Resolution Convergence}), we find voxel-wise maxima of $\max(T_\text{IRIS}) \approx \SI{78.2}{\K}$ and $\max(T_\text{POLARIS}) \approx \SI{66.5}{\K}$. We also find $\TAM(T_\text{IRIS}) \approx \SI{0.54}{\K}$ and $\TAM(T_\text{POLARIS}) \approx \SI{.60}{\K}$. While not as comprehensive as a voxel-wise error analysis, these results still suggest reasonable quantitative alignment.
    \par For a qualitative comparison, we provide side-by-side visualizations of both RADMC-3D versus IRIS and POLARIS versus IRIS (see \autoref{fig:RADMC-3D vs IRIS} and \autoref{fig:POLARIS vs IRIS}). Each side-by-side figure shows, in the left panel, the top-down $\HTwo$ density of our test AREPO snapshot according to a latitude-mean projection over the full latitude extent of the physical tensor. The center and right panels each show the $\ell, b$ and $\ell, v$ projections of synthetically observed Raleigh-Jeans brightness temperature, with the IRIS observation in the right panels and the alternate observation in the center panels. Consistent with the quantitative alignment described above, we find that our RADMC-3D and POLARIS side-by-side visualizations each yield near-total alignment. 
    \par Note that the observed intensities in the RADMC-3D comparison are different from those in the POLARIS comparison, since, as described in \autoref{subsec:RADMC-3D Configurations}, we are constrained to a lower radial resolution in RADMC-3D by computational expense. For parity, we adopt this lower resolution in IRIS in the RADMC-3D comparison. POLARIS, by contrast, as described in \autoref{subsec:POLARIS Configurations}, employs a velocity-interpolation scheme that yields results closer to the limit of radial resolution, which we approach with a higher radial resolution. While our two side-by-side comparison figures thus yield incomparable results, each separately represents a like comparison of IRIS with another synthetic-observation code. Overall, we find that these results provide a strong, end-to-end verification of our entire preprocessing pipeline as described in \autoref{sec:Simulation Processing and Data Production} and the IRIS synthetic-observation code as described in \autoref{sec:Synthetic Observation Overview and Theory} and \autoref{sec:Synthetic Observation Implementation and Algorithms}.

\subsection{Optical Depth and Dust Effects} \label{subsec:Optical Depth and Dust Effects}
    \begin{figure*}[t]
        \centering
        \includegraphics[width=1\linewidth]{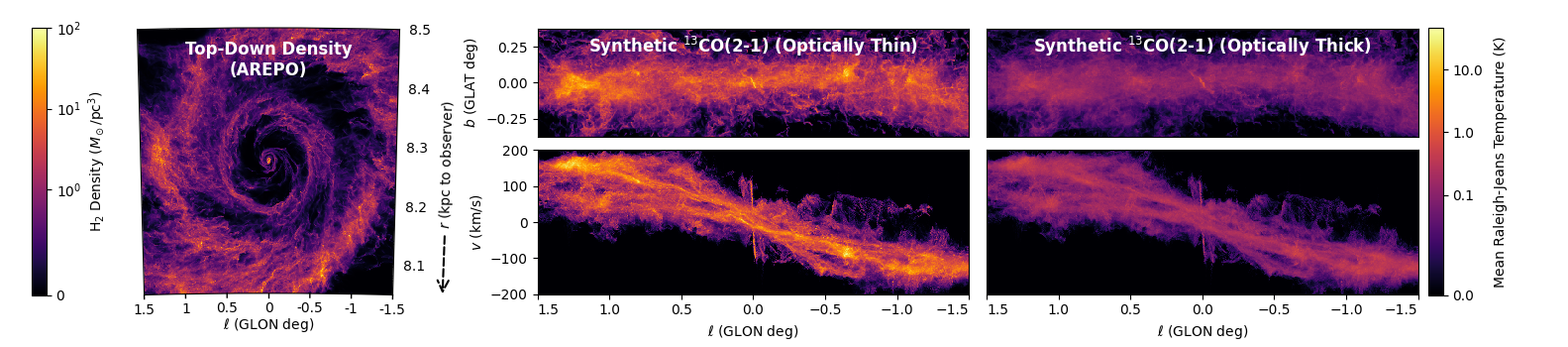}
        \caption{\textbf{Synthetic-Observation Optical-Depth Analysis (Line Only):} We compare IRIS-SO in optically thin transfer (\autoref{subsec:Transfer Solution: Optically Thin Transfer}, modeling spontaneous emission of the $\ThirteenCOTwoOne$ spectral line only), versus in optically thick transfer (\autoref{subsec:Transfer Solution: Optically Thick Transfer}, modeling spontaneous emission, stimulated emission, and absorption of the $\ThirteenCOTwoOne$ spectral line). In this figure, we do not include dust in our optically thick observation, instead isolating the optically thick behavior that is due to the spectral line alone. The simulation featured is due to \citeLipman. All color bars are scaled with an $\arcsinh$ nonlinearity. See \autoref{subsec:Optical Depth and Dust Effects} for a detailed figure discussion. We see a substantial degree of optically thick behavior in the line, indicating the necessity of modeling line absorption and stimulation in our training data.}
        \label{fig:Optically Thin vs Thick}
    \end{figure*}
    \begin{figure*}[t]
        \centering
        \includegraphics[width=1\linewidth]{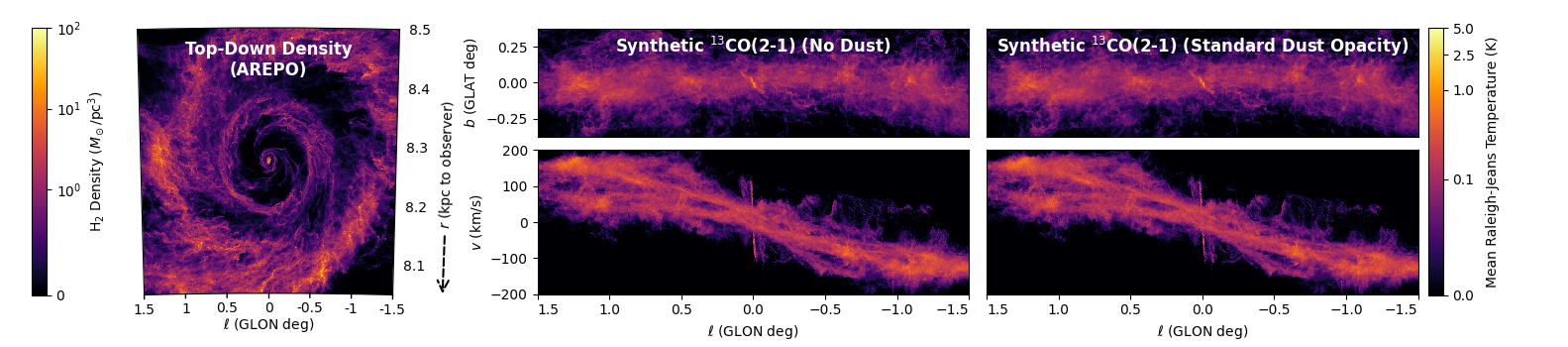}
        \caption{\textbf{Synthetic Observation With and Without Dust (Standard Opacity):} This figure follows the same format as \autoref{fig:Optically Thin vs Thick}, except that the center panels now show synthetic observation by IRIS-SO under optically thick transfer without dust, while the right panels show optically thick transfer with the addition of thermal dust emission and dust extinction, under our assumed dust opacity from \citet{Ossenkopf1994} of $\kappa = \SI{1e-2}{\cm^2\g^{-1}}$ (see \autoref{subsec:Dust Emission and Absorption}). See \autoref{subsec:Optical Depth and Dust Effects} for a detailed figure discussion. We see negligible difference between the dust and no-dust observations at this assumed opacity, indicating that dust has only a marginal effect on synthetic observation of the $\ThirteenCOTwoOne$ spectral line over our training-data simulations due to \citeLipman, as described in \autoref{subsec:AREPO Zoom Simulations}.}
        \label{fig:No Dust vs Dust}
    \end{figure*}
    \begin{figure*}[t]
        \centering
        \includegraphics[width=1\linewidth]{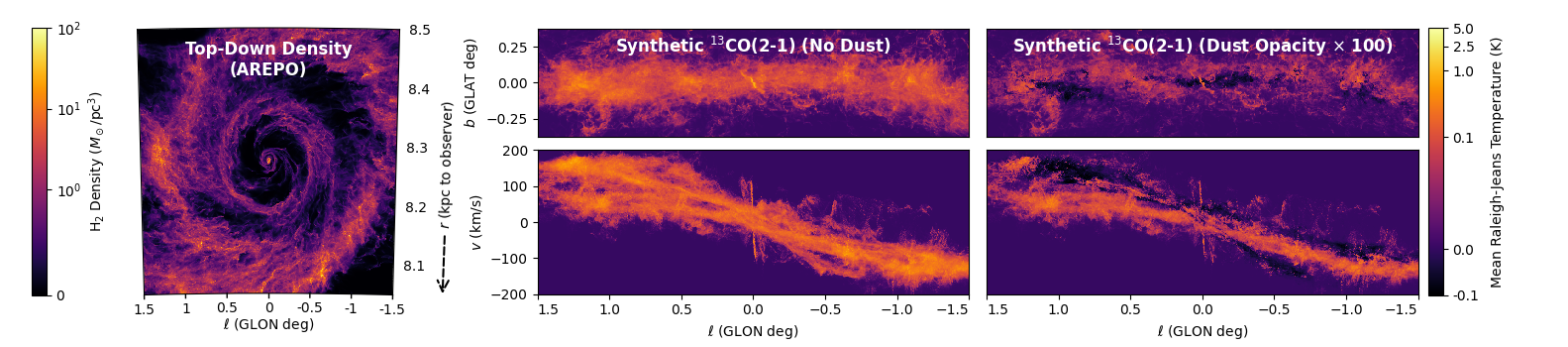}
        \caption{\textbf{Synthetic Observation With and Without Dust (Opacity Threshold for Visual Impact):} This figure follows the format of \autoref{fig:No Dust vs Dust} exactly, except that the right panels now show synthetic observation by IRIS-SO with dust at $\kappa = \SI{1}{\cm^2\g^{-1}}$, or 100 times our standard opacity. We now see prominent negative-intensity dust-extinction features in our continuum-subtracted observation. This figure illustrates that IRIS has the ability to model dust effects, and that these effects become prominent at sufficiently high densities of dust in the ISM, but that the necessary dust densities for a substantial visual impact significantly exceed realistic values.}
        \label{fig:No Dust vs Dust x100}
    \end{figure*}

    \par Having verified in \autoref{subsec:Side-by-Side Verification} that IRIS-SO correctly implements the synthetic-observation process in comparison to other known synthetic-observation codes, we now turn to validating some of the assumptions we have made in \autoref{sec:Synthetic Observation Overview and Theory} and \autoref{sec:Synthetic Observation Implementation and Algorithms}. Specifically, we verify that these assumptions contribute to a synthetic-observation process that faithfully models real observation. In the following subsections, we provide a series of comparisons and figures using synthetic observations of the zoom-in simulations of \citeLipman, as described in \autoref{subsec:AREPO Zoom Simulations}, that we incorporate into our training dataset (see \autoref{table:Dataset Parameters}). Since we are no longer bound to a lower-resolution simulation by the computational intractability of observing a high-resolution Voronoi mesh in POLARIS (see \autoref{subsec:POLARIS Configurations}), analyzing synthetic observations of the same simulations in our training dataset provides the best means of verifying that our synthetic-observation process is sufficiently realistic for the implementation of our machine-learning experiments (see \autoref{sec:Machine-Learning Methods and Reversion}).
    \par In the first three of these figures, we investigate the optical depth of our synthetic observations, which is affected both by absorption and stimulation of the $\ThirteenCOTwoOne$ spectral line itself and by dust extinction. In gauging the relevance of modeling these effects in the synthetic observations for our training dataset, we compare synthetic observation modeling emission only versus modeling emission and absorption, and we compare synthetic observation without dust against synthetic observation with dust at two different opacities. In enabling these comparisons, we refer back to the optically thin and optically thick transfer modes described in \autoref{subsec:Transfer Solution: Overview}. In optically thin transfer, only spontaneous emission of the line is computed, with no stimulated emission or line absorption, and no dust extinction. In optically thick transfer, line spontaneous emission, stimulated emission, and absorption are all computed. Dust extinction is also computed in optically thick transfer if a nonzero dust opacity is set.
    \par In \autoref{fig:Optically Thin vs Thick}, we compare a synthetic observation computed via optically thin transfer against a synthetic observation of the same physical tensor computed via optically thick transfer without dust, illustrating the optically thick behavior that is due to the $\ThirteenCOTwoOne$ spectral line itself. We also set the temperature $T_\text{CMB}$ of the radiative-transfer background to zero in both these observations, eliminating all continuum radiation, although we leave the temperature $\overline{T}$ of the continuum background used in the OT level balance at the default $\overline{T} = \SI{2.73}{\K}$ (see \autoref{subsec:Level Populations: The Optically Thin Assumption}). These configurations isolate the effect of line self-absorption in the radiative transfer, as opposed to continuum absorption by the line, which we describe in further detail in \autoref{subsec:Validity of the Optically Thin Level Balance}. In the left panel, we show the top-down (latitude-mean) density of our AREPO snapshot. In the center panels, we show the mean $\ell, b$ and $\ell, v$ projections of our optically thin synthetic observation of this snapshot. In the right panels, we show the mean $\ell, b$ and $\ell, v$ projections of our optically thick synthetic observation without dust. 
    \par We note, by the visual discrepancies between our optically thin and optically thick observations, a substantial amount of optically thick behavior of the spectral line at our assumed $\ThirteenCO$ abundance (see \autoref{subsec:Spectral-Line Configuration}). The quantitative error metrics defined in \autoref{subsec:Error Metrics for Synthetic Observation} illustrate substantial error, with $\TSREcube(\text{THIN}, \text{THICK}) \approx 54.8\%$ and $\TSRElv(\text{THIN}, \text{THICK}) \approx 116\%$. We also find substantial deviation in Raleigh-Jeans brightness-temperature statistics defined in the same section (see also \autoref{subsec:Intensity Versus Temperature}), with $\TAM(T_\text{THIN}) \approx \SI{3.10}{\K}$ and $\TAM(T_\text{THICK}) \approx \SI{0.48}{\K}$. We further note, as described in \autoref{subsec:Resolution Convergence}, that, while the insufficient radial resolution of our training dataset tends to yield an overestimation of optical depth, this comparison is conducted at a higher radial resolution at which this overestimation is minimal. Radial resolution thus cannot account for the large discrepancies we see in this comparison. These discrepancies therefore indicate the importance of modeling the effects of line absorption and stimulation in our training dataset, although maximal physicality will depend upon future reconstruction of the dataset at a higher resolution that accurately models optical depth.
    \par In \autoref{fig:No Dust vs Dust} and \autoref{fig:No Dust vs Dust x100}, which follow the exact same visual format, we compare the results of optically thick transfer with and without dust. In \autoref{fig:No Dust vs Dust}, the center panels show a synthetic observation without dust, and the right panels show one with dust at the opacity we assume for the IRIS training data ($\kappa = \SI{1e-2}{\cm^2\g^{-1}}$, see \autoref{subsec:Dust Emission and Absorption}). While quantitative analysis shows minute discrepancies, with $\TSREcube(\text{NO DUST}, \text{DUST}) \approx 1.0\%$ and $\TSRElv(\text{NO DUST}, \text{DUST}) \approx 0.8\%$, we also find $\TAM(T_\text{NO DUST}) \approx \SI{0.43}{\K} \approx \TAM(T_\text{DUST})$, and that these discrepancies are visually imperceptible. We conclude that modeling dust in our synthetic observations is not relevant in generating our training dataset, although we leave the dust settings turned on in our training-data production code. This is a positive result, since we model dust with substantially less sophistication than that with which we model spectral lines. We therefore conclude that these substantial simplifications in our dust modeling are not detrimental to the overall accuracy of our supervised-reversion scheme (see \autoref{subsec:Supervised Reversion: Objective}).
    \par In \autoref{fig:No Dust vs Dust x100}, we conduct a sanity test, searching for a rough threshold at which dust effects become visually significant in our synthetic observations. While the center panels still show the same optically thick transfer without dust, the right panels now show optically thick transfer with dust at 100 times the opacity assumed in \autoref{subsec:Dust Emission and Absorption} ($\kappa = \SI{1}{\cm^2\g^{-1}}$. At this opacity, we begin to see strong dust effects, and even prominent negative-temperature extinction features. Our quantitative metrics inflate to $\TSREcube(\text{NO DUST}, \text{DUST} \times 100) \approx 89.6\%$ and $\TSRElv(\text{NO DUST}, \text{DUST} \times 100) \approx 63.2\%$, and $\TAM(T_\text{NO DUST}) \approx \SI{0.43}{\K}$ versus $\TAM(T_{\text{DUST} \times 100}) \approx \SI{0.35}{\K}$. As described in \autoref{subsec:Dust Emission and Absorption}, however, this $\kappa$ value would only be consistent with a gas-to-dust mass ratio of roughly $1 : 1$, which is unrealistic in the CMZ ISM. We conclude that, while IRIS demonstrates the capacity to model the impact of dust on spectral lines, such modeling is only relevant to synthetic observation of $\ThirteenCOTwoOne$ in extremely dusty environments such as dense cores and protoplanetary disks.

\subsection{Validity of the Optically Thin Level Balance} \label{subsec:Validity of the Optically Thin Level Balance}
    \begin{figure*}[t]
        \centering
        \includegraphics[width=1\linewidth]{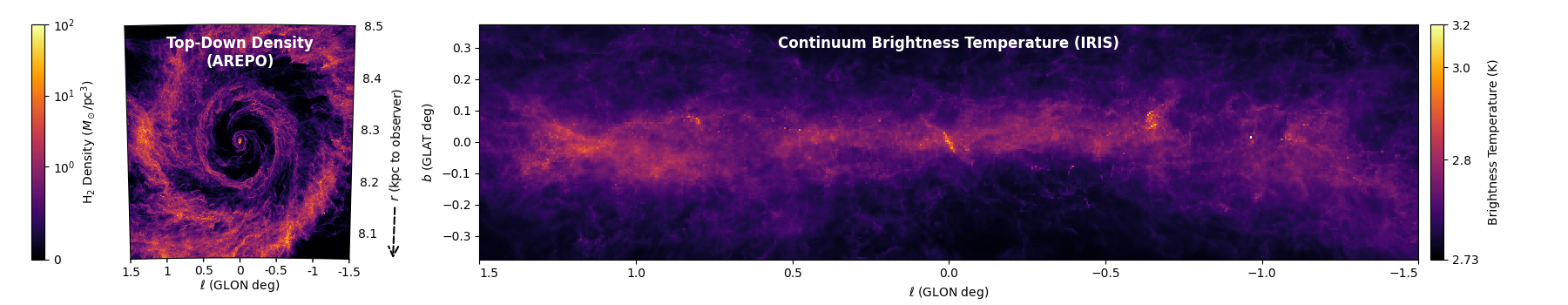}
        \caption{\textbf{Synthetic Observation of the Continuum Temperature:} We show the continuum baseline observed by IRIS-SO over our training-data simulations from \citeLipman, which we describe in \autoref{subsec:AREPO Zoom Simulations}. Here, we observe the continuum at the $\ThirteenCOTwoOne$ transition frequency, but without contribution of the $\ThirteenCOTwoOne$ spectral line. The components of this continuum baseline are the CMB as well as thermal dust emission and dust extinction at our standard opacity from \citet{Ossenkopf1994} of $\kappa = \SI{1e-2}{\cm^2\g^{-1}}$ (see \autoref{subsec:Dust Emission and Absorption}). In this figure, unlike the rest of our synthetic-observation figures, we plot the Planck brightness temperature as opposed to the Raleigh-Jeans brightness temperature, for ease of comparison with $T_\text{CMB}$, which is defined as a Planck brightness temperature. As described in \autoref{subsec:Validity of the Optically Thin Level Balance}, the small deviation in our observed continuum from the CMB baseline of $T_\text{CMB} = \SI{2.73}{\K}$ proves that the continuum is dominated by the CMB in our synthetic observations. This finding is consistent with our conclusions in \autoref{subsec:Optical Depth and Dust Effects} that dust is a minority effect under our observed conditions and assumed dust opacity.}
        \label{fig:Continuum Temperature}
    \end{figure*}
    \begin{figure*}[t]
        \centering
        \includegraphics[width=1\linewidth]{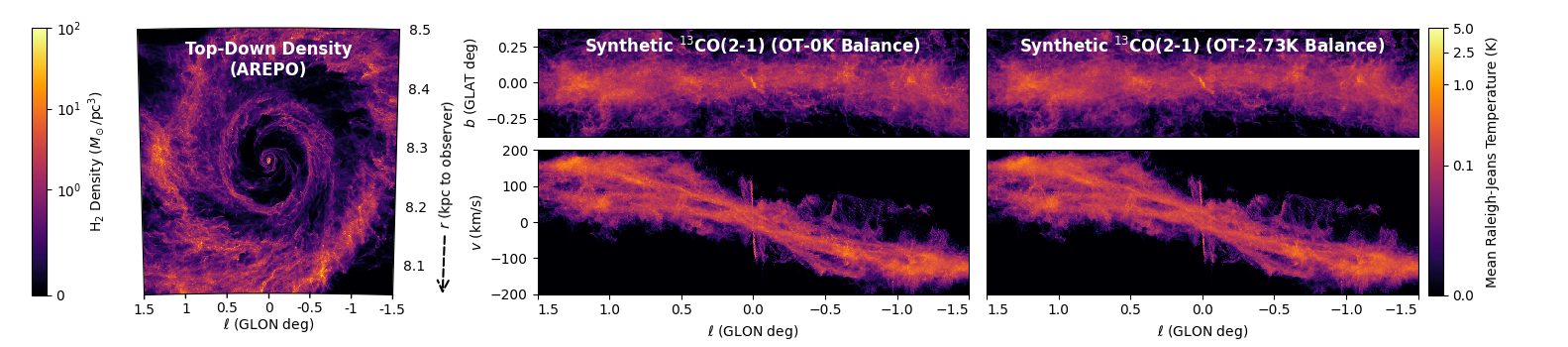}
        \caption{\textbf{Deviation Below Expectation of the Background Temperature in the OT Level Balance:} This figure follows the same format as \autoref{fig:Optically Thin vs Thick}, except that the center panels now show synthetic observation by IRIS-SO (in optically thick transfer, with dust at our standard opacity given in \autoref{subsec:Dust Emission and Absorption}) computed according to the OT level balance with $\overline{T} = \SI{0}{\K}$, while the right panels show the same synthetic observation computed according to the OT level balance with $\overline{T} = T_\text{CMB} =\SI{2.73}{\K}$ (see \autoref{subsec:Level Populations: The Optically Thin Assumption}). As described in \autoref{subsec:Validity of the Optically Thin Level Balance}, the negligible visual difference indicates the insensitivity of the OT level balance to deviations of $\overline{T}$ below expectation.}
        \label{fig:OT-OK vs OT-CMB Balance}
    \end{figure*}
    \begin{figure*}[t]
        \centering
        \includegraphics[width=1\linewidth]{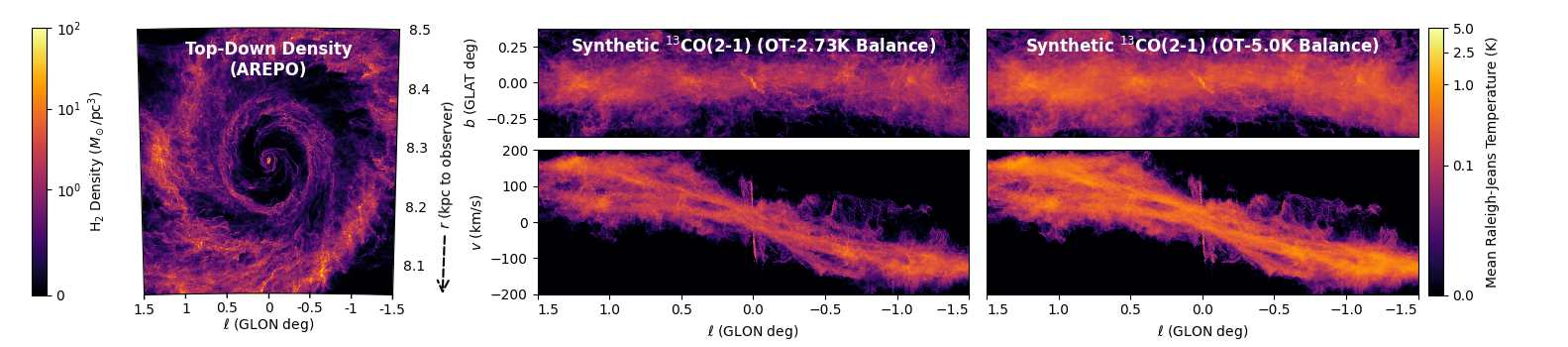}
        \caption{\textbf{Deviation Above Expectation of the Background Temperature in the OT Level Balance:} This figure follows the exact format of \autoref{fig:OT-OK vs OT-CMB Balance}, except that the center panels now show synthetic observation by IRIS-SO computed according to the OT level balance with $\overline{T} = T_\text{CMB} =\SI{2.73}{\K}$ (identical to the right panels in \autoref{fig:OT-OK vs OT-CMB Balance}), while the right panels show the same synthetic observation computed according to the OT level balance with $\overline{T} =\SI{5}{\K}$ (see \autoref{subsec:Level Populations: The Optically Thin Assumption}). As described in \autoref{subsec:Validity of the Optically Thin Level Balance}, the substantial visual difference indicates the sensitivity of the OT level balance to deviations of $\overline{T}$ above expectation.}
        \label{fig:OT-CMB vs OT-5K Balance}
    \end{figure*}
    \begin{figure*}[t]
        \centering
        \includegraphics[width=1\linewidth]{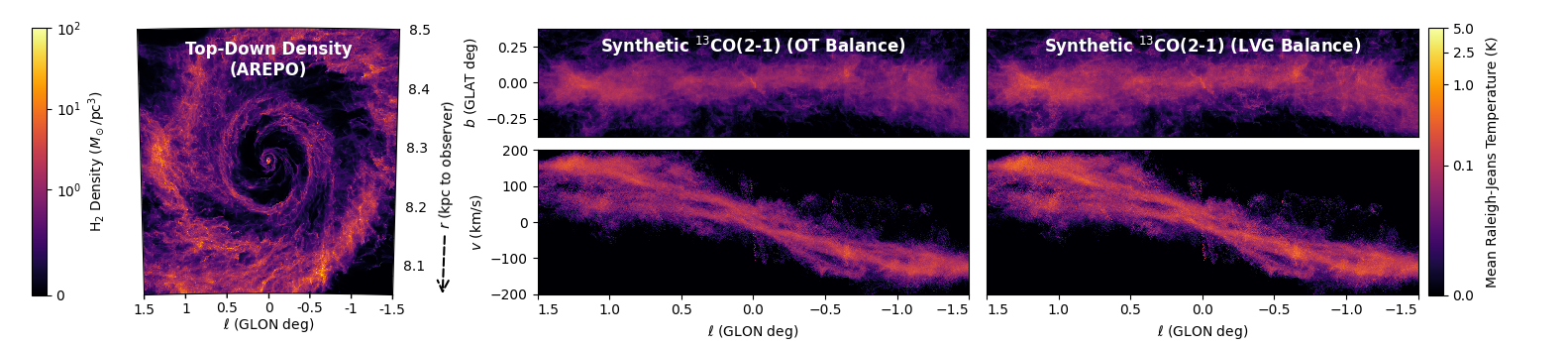}
        \caption{\textbf{Synthetic Observation via the OT vs. LVG Level Balance:} This figure follows the same format as \autoref{fig:Optically Thin vs Thick}, except that the center panels now show synthetic observation by RADMC-3D (with dust at our standard opacity given in \autoref{subsec:Dust Emission and Absorption}) computed according to the OT level balance (with $\overline{T} = T_\text{CMB} =\SI{2.73}{\K}$, see \autoref{subsec:Level Populations: The Optically Thin Assumption}), while the right panels show the same synthetic observation computed by RADMC-3D according to the LVG level balance (see \autoref{subsec:Level Populations: Computational Challenges}). As described in \autoref{subsec:Validity of the Optically Thin Level Balance}, we take the LVG balance, which is the most robust but computationally expensive level-balance approximation, as our ground truth. The negligible visual differences in synthetic observation between the OT and LVG balances then prove the validity of the OT approximation under our observed conditions of the $\ThirteenCOTwoOne$ spectral line over our training-data simulations from \citeLipman, as described in \autoref{subsec:AREPO Zoom Simulations}.}
        \label{fig:OT vs LVG Balance}
    \end{figure*}
    \begin{figure*}[t]
        \centering
        \includegraphics[width=1\linewidth]{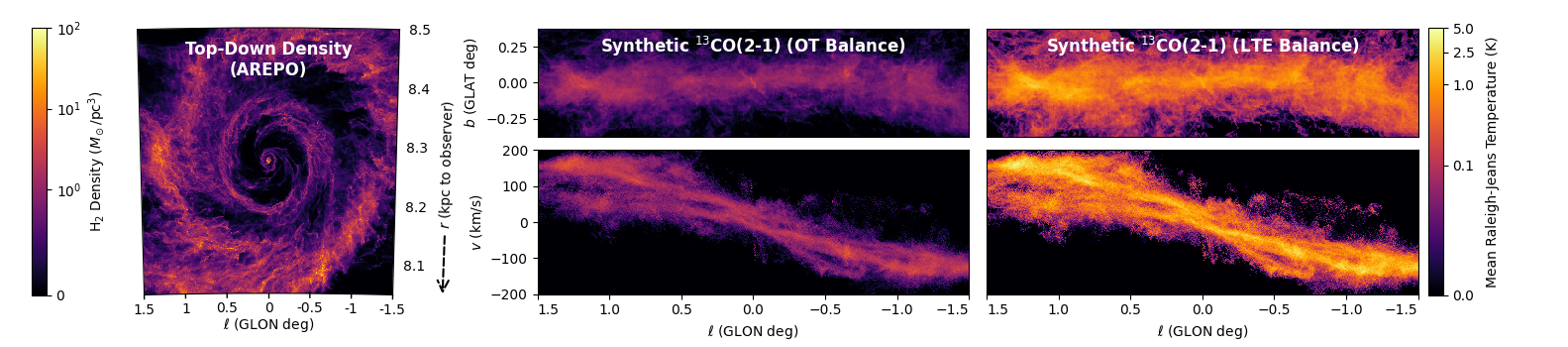}
        \caption{\textbf{Synthetic Observation via the OT vs. LTE Level Balance:} This figure follows the exact format as \autoref{fig:OT vs LVG Balance}. The center panels still show synthetic observation by RADMC-3D (with dust at our standard opacity given in \autoref{subsec:Dust Emission and Absorption}) computed according to the OT level balance (with $\overline{T} = T_\text{CMB} =\SI{2.73}{\K}$, see \autoref{subsec:Level Populations: The Optically Thin Assumption}), but the right panels now show the same synthetic observation computed by RADMC-3D according to the LTE level balance (see \autoref{subsec:Level Populations: Computational Challenges}). The previous figure \autoref{fig:OT vs LVG Balance} shows that, over our observed conditions of the $\ThirteenCOTwoOne$ line on our training-data simulations from \citeLipman, which we describe in \autoref{subsec:AREPO Zoom Simulations}, the OT level balance matches the results of the LVG balance, which we adopt as our ground truth. The substantial visual discrepancies in this figure between the OT and LTE balances therefore demonstrate the inadequacy of the LTE balance over these same observational conditions. See \autoref{subsec:Validity of the Optically Thin Level Balance} for more detailed discussion.}
        \label{fig:OT vs LTE Balance}
    \end{figure*}
    
    \par The substantial degree of optically thick behavior noted in \autoref{subsec:Optical Depth and Dust Effects} calls into question the validity of the optically thin (OT) level balance we assume in modeling all spectral-line effects (see \autoref{subsec:Level Populations: The Optically Thin Assumption}). As described in \autoref{subsec:Level Populations: Computational Challenges}, computing an exact solution to the level balance at every point in our observed field is an intractable problem, since this balance is a function of the radiation field itself, and so this tractability issue must be resolved via some approximation. The OT approximation we employ in IRIS assumes that the level balance is not substantially affected by absorption or stimulation of the tracer molecule at any energy level. In \autoref{fig:Optically Thin vs Thick}, however, we see that line absorption is substantial throughout our synthetic observation of the $\ThirteenCOTwoOne$ line. We therefore devote the entirety of this subsection, to include a series of five figures, towards probing the limits of the OT assumption and definitively demonstrating that this assumption is still appropriate under the conditions modeled in our training dataset. 
    \par First, note that, as discussed in \autoref{subsec:Level Populations: The Optically Thin Assumption}, the presence of optically thick behavior along observed lines of sight is not automatically inconsistent with use of the OT level balance. The validity of the OT approximation depends upon optical depth as a local property of each point in the observed gas, in the omnidirectional average, which incorporates primarily out-of-disk rays. We may reasonably assume this local, omnidirectional behavior is substantially more thin than the total optical depth over the entire span of only the most dense lines of sight that are confined within the galactic disk. This reasoning provides a strong initial justification in suspecting that the OT level balance may still be appropriate, even under the optically thick conditions we observe in \autoref{fig:Optically Thin vs Thick}.
    \par It is of particular relevance, moreover, to identify the dominant components of the total radiation field. In IRIS, our total radiation field is assumed to be fed by three separate emission sources---the CMB, thermal dust, and the spectral line. Investigating the strength of the thermal-dust and spectral-line emission in comparison to the CMB is relevant because, as discussed in \autoref{subsec:Level Populations: The Optically Thin Assumption}, our OT balance still provides the ability to model the affect to the level populations of absorption of, and stimulation by, a continuum background with a constant blackbody temperature $\overline{T}$. As a best estimate in IRIS, we set $\overline{T} = T_\text{CMB} = \SI{2.73}{\K}$, but it remains to show that this simplified assumption is reasonable by showing that the CMB is the dominant component of the total radiation field, in the omnidirectional mean, within our synthetic observations.
    \par In \autoref{fig:Continuum Temperature}, we provide a continuum-temperature image generated by IRIS in synthetically observing the zoom-in simulations of \citeLipman that we use in generating our training dataset (see \autoref{subsec:AREPO Zoom Simulations}). In the left panel, we show the top-down (latitude-mean) $\HTwo$ density of our observed snapshot. In the right panel, we show a mean $\ell, b$ temperature image of the continuum. No line intensity is computed in this image, so it represents the ideal continuum baseline. By contrast to the rest of the figures in this publication, the temperature we plot in this figure is the true brightness temperature via Planck's law, as opposed to the Raleigh-Jeans brightness temperature (see \autoref{subsec:Intensity Versus Temperature}), which we prefer here for ease of comparison with $T_\text{CMB}$, which is similarly defined as a true brightness temperature. 
    \par Notably, we find that the continuum temperatures in this image range only between a minimum of $\sim T_\text{CMB} = \SI{2.73}{\K}$ and a maximum of $\sim \SI{3.2}{\K}$. Hence, the continuum is dominated by the CMB in our synthetic observations, with only a marginal contribution from thermal dust emission at our assumed opacity (see \autoref{subsec:Dust Emission and Absorption}), which is consistent with our findings in \autoref{subsec:Optical Depth and Dust Effects} that dust extinction of the spectral line is small in our synthetic observations. We may then be tempted to compare this continuum-temperature image directly with our continuum-subtracted spectral-line observations of this same snapshot (e.g. \autoref{fig:OT-OK vs OT-CMB Balance}). Noting that these observations yield temperatures not exceeding around $\sim \SI{1}{\K}$ in most observed regions, we may initially conclude that the spectral line is also a minority component of the total radiation field in our synthetic observations, and that this total radiation field is dominated entirely by the CMB. In this case, it would be reasonable to assume that any effect to the level balance by the radiation field can be reasonably approximated by modeling the effect of the CMB, which we already incorporate into our OT level balance by setting $\overline{T} = T_\text{CMB}$.
    \par But this conclusion is not as immediate, since the continuum-subtracted temperature we plot in our synthetic line observations, which is a Raleigh-Jeans temperature, cannot be directly compared against the Planck brightness temperature in our continuum map. Per the definitions described in \autoref{subsec:Intensity Versus Temperature}, our observed Raleigh-Jeans temperature cap of $\sim \SI{1.0}{\K}$ in our continuum-subtracted line observations is equivalent to a Planck brightness temperature of $\sim \SI{4.3}{\K}$, which does appear to dominate the continuum contribution along our observed lines of sight. We therefore next wish to stress-test our OT level balance under different $\overline{T}$ values, to determine how much deviations in these values affect our final synthetic observations.
    \par In \autoref{fig:OT-OK vs OT-CMB Balance}, we first test the difference between $\overline{T} = \SI{0}{\K}$ and $\overline{T} = T_\text{CMB} = \SI{2.73}{\K}$. For this figure, we follow the format of our previous side-by-side comparisons in \autoref{subsec:Side-by-Side Verification} and \autoref{subsec:Optical Depth and Dust Effects}, using the same snapshot from \citeLipman as in our continuum-temperature image. In the left panel, we show the top-down (latitude-mean) density. In the center panels, we show mean $\ell, b$ and $\ell, v$ projections of a synthetic observation (optically thick transfer with dust) computed via the OT level balance with $\overline{T} = \SI{0}{\K}$. In the right panels, we show the mean $\ell, b$ and $\ell, v$ projections of the same synthetic observation computed via the OT level balance with $\overline{T} = \SI{2.73}{\K}$. Visually, we see negligible difference. According to the quantitative error metrics defined in \autoref{subsec:Error Metrics for Synthetic Observation}, we find $\TSREcube(\SI{0}{\K}, \SI{2.73}{\K}) \approx 16.0\%$ and $\TSRElv(\SI{0}{\K}, \SI{2.73}{\K}) \approx 8.6\%$, and Raleigh-Jeans temperature statistics of $\TAM(T_{\SI{0}{\K}}) \approx \SI{0.45}{\K}$ versus $\TAM(T_{\SI{2.73}{\K}}) \approx \SI{0.43}{\K}$. This result provides encouraging initial evidence that deviation of the total radiation field within our expected bounds yields only minimal effect in our final synthetic observation.
    \par In \autoref{fig:OT-CMB vs OT-5K Balance}, however, having tested the effect to the level balance of deviation of the radiation field below expectation, we now test deviation of the radiation field above expectation, comparing $\overline{T} = \SI{2.73}{\K}$ against $\overline{T} = \SI{5}{\K}$. This figure follows the exact same format as \autoref{fig:OT-OK vs OT-CMB Balance}, except that $\overline{T} = \SI{2.73}{\K}$ is now featured in the center panels, and $\overline{T} = \SI{5}{\K}$ is featured in the right panels. In this case, we now see a marked visual discrepancy. Our quantitative metrics are also substantially elevated, with $\TSREcube(\SI{2.73}{\K}, \SI{5}{\K}) \approx 94.1\%$ and $\TSRElv(\SI{2.73}{\K}, \SI{5}{\K}) \approx 80.9\%$, and Raleigh-Jeans temperature statistics of $\TAM(T_{\SI{2.73}{\K}}) \approx \SI{0.43}{\K}$ versus $\TAM(T_{\SI{5}{\K}}) \approx \SI{0.71}{\K}$. We therefore find that above-expectation deviations in the radiation field, by contrast to below-expectation deviations, impact the OT level balance substantially. (Note that since we find $\TAM(T_{\SI{0}{\K}}) > \TAM(T_{\SI{2.73}{\K}}) < \TAM(T_{\SI{5}{\K}})$, indicating a $\TAM$ minimum for $\overline{T}$ in the interval $[\SI{0}{\K}, \SI{5}{\K}]$, we further verified that below-expectation deviations impact our synthetic observation minimally by testing $\overline{T}$ values between $\SI{0}{\K}$ and $\SI{2.73}{\K}$, with similar results.) Since Planck background temperatures of $\overline{T} = \SI{5}{\K}$ are within the expected range of the total observed radiation, this susceptibility of the level balance to such reasonable deviations in the radiation field is another point of concern for the OT approximation.
    \par The definitive test is therefore to compare the results of our synthetic observations under level balances other than the OT approximation. While IRIS does not currently support level approximations other than the OT balance up to a configurable background temperature, RADMC-3D \citep{Dullemond2012} allows us to observe the same simulation snapshot under each of the OT, LTE, and LVG approximations described in \autoref{subsec:Level Populations: Computational Challenges}. As a ground truth, we compare against the LVG approximation, which is the most robust and general of these approximations, albeit also the most computationally intensive by far. In \autoref{fig:OT vs LVG Balance}, we once again follow the same figure format on the same simulation snapshot, except that center panels now show the OT balance (with $\overline{T} = \SI{2.73}{\K}$) as observed by RADMC-3D, and the right panels now show the LVG balance as observed by RADMC-3D. 
    \par We see that the visual results of the OT and LVG balances are negligibly different. Our quantitative metrics illuminate some discrepancy, with $\TSREcube(\text{OT}, \text{LVG}) \approx 31.0\%$ and $\TSRElv(\text{OT}, \text{LVG}) \approx 33.3\%$, and $\TAM(T_\text{OT}) \approx \SI{0.36}{\K}$ versus $\TAM(T_\text{LVG}) \approx \SI{0.41}{\K}$. But we conclude by the visual analysis that this error is within tolerance, and thus that the OT level balance is a suitable approximation under the conditions of our $\ThirteenCOTwoOne$ observations. And indeed, this finding that the OT and LVG balances yield approximately equivalent results in our synthetic observations is not inconsistent with our previous finding that a deviation from $\SI{2.73}{\K}$ to $\SI{5}{\K}$ in the mean radiation field---a deviation which we observe in the total intensity of our synthetic observations---yields a substantial impact to observed intensity. This is because, as previously described, the mean radiation field specified according to $\overline{T}$ is the omnidirectional average intensity. While we encounter intensities in substantial excess of the CMB along dense lines of sight, we infer, by the minimal discrepancy of the LVG balance, that the omnidirectional average radiation field is still approximately the CMB, since the majority of lines of sight in this average are outside the galactic plane. We thus conclude that $\overline{T} = T_\text{CMB}$ is also a good choice in the OT balance.
    \par Finally, we again use RADMC-3D to compare the OT and LTE balances. Since the LTE balance is substantially more computationally efficient than the OT balance by elimination of the level balance system (see \autoref{subsec:Level Populations: Mathematical Solution}), it is worth testing whether we can improve the efficiency of IRIS even further by using the LTE as opposed to OT balance. We note, of course, as describe in \autoref{subsec:Grid Precomputation}, that this additional efficiency would only be incurred at start-up, rather than during synthetic observation itself, due to the implementation of grid precomputation for fast runtime interpolation of emission and absorption coefficients. This point is moot, however, because \autoref{fig:OT vs LTE Balance} clearly shows that the LTE level balance is not an appropriate approximation under our conditions. 
    \par Following the same figure format, the center panels in \autoref{fig:OT vs LTE Balance} again show the OT balance (with $\overline{T} = T_\text{CMB}$) as observed by RADMC-3D. The right panels, however, now show the LTE balance, as observed by RADMC-3D. We see substantial visual deviation, clearly indicating the failure of the LTE assumption under these conditions. Under our quantitative metrics, we find $\TSREcube(\text{OT}, \text{LTE}) \approx 156\%$ and $\TSRElv(\text{OT}, \text{LTE}) \approx 168\%$, and $\TAM(T_\text{OT}) \approx \SI{0.36}{\K}$ versus $\TAM(T_\text{LTE}) \approx \SI{2.61}{\K}$. Overall, we conclude that the OT balance is the most appropriate assumption under our conditions, particularly effective as a first-order method for fast synthetic observation in generating the $\sim 100\text{k}$ synthetic observations incorporated into our training dataset, given the lack of fast alternatives for higher accuracy. 
    
\subsection{Comparison of Formal and Smooth Integration} \label{subsec:Comparison of Formal and Smooth Integration}
    \begin{figure*}[t]
        \centering
        \includegraphics[width=1\linewidth]{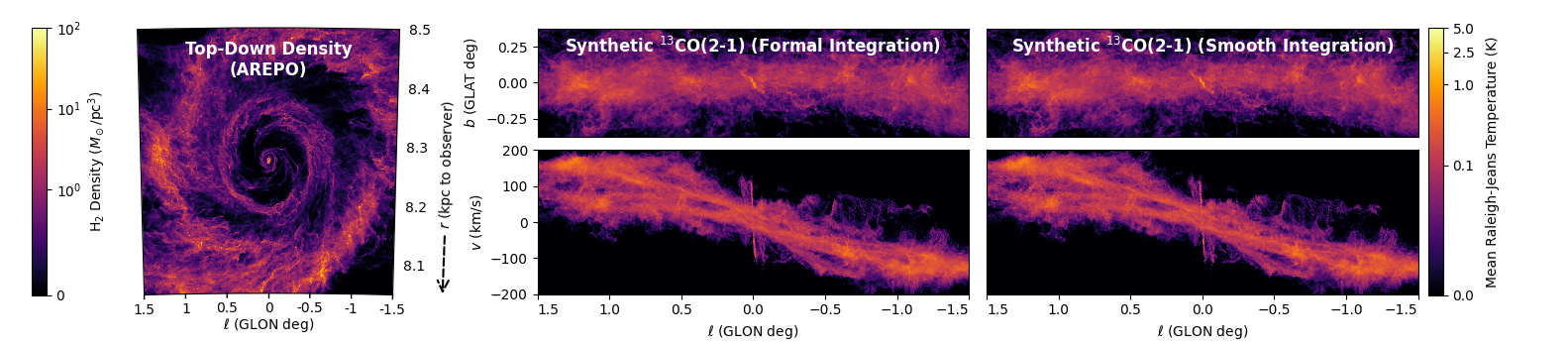}
        \caption{\textbf{Synthetic Observation in Formal vs. Smooth Integration:} This figure follows the same format as \autoref{fig:Optically Thin vs Thick}, except that the center panels now show synthetic observation by IRIS-SO computed via formal integration of optically thick transfer (with dust, at our standard opacity, as in \autoref{subsec:Dust Emission and Absorption}), while the right panels show the same optically thick transfer (with dust) in smooth integration. See \autoref{subsec:Transfer Solution: Formal and Smooth Integration of Optically Thick and Selectively Thin Transfer} for a description of these integration modes. The visually identical results in this comparison prove that our preferred smooth integration, which yields on average $\sim 1.5 \times$ speedups via the elimination of expensive transcendental operations, does so at a negligible cost to accuracy.}
        \label{fig:Formal vs Smooth}
    \end{figure*}
    
    \par We next compare the difference of our formal versus smooth integration schemes. As described in \autoref{subsec:Transfer Solution: Formal and Smooth Integration of Optically Thick and Selectively Thin Transfer}, these integration schemes provide different solutions to the same radiative-transfer equations. Formal integration provides an exponential step computation that is exact assuming stepwise-constant emission and absorption coefficients. Smooth integration, by contrast (in optically thick and selectively thin transfer modes), treats the radiative-transfer equation as a general stiff ODE (see \autoref{subsec:Transfer Solution: Formal and Smooth Integration of Optically Thick and Selectively Thin Transfer}) via an implicit step computation with an algebraic solution that is a low-order polynomial. While smooth integration is no longer exact in the stepwise-constant case, we prefer it since we find that it provides $\sim 1.5\times$ speedups by elimination of computationally expensive transcendental operations. It only remains to demonstrate that these speedups are not achieved at the cost of unacceptable error.
    \par In \autoref{fig:Formal vs Smooth}, we again follow the template of our other side-by-side figures, observing the same AREPO test snapshot due to \citeLipman as in these other figures as well (see \autoref{subsec:AREPO Zoom Simulations}). In the left panel, we show the top-down (latitude-mean) $\HTwo$ density of our observed snapshot. In the center panels, we show the mean $\ell, b$ and $\ell, v$ projections of a synthetic observation (with optically thick transfer and dust, according to the opacity given in \autoref{subsec:Dust Emission and Absorption}) computed via formal integration. In the right panels, we see the mean $\ell, b$ and $\ell, v$ projections of the same synthetic observation computed via smooth integration. Visually, we find the differences between these observations to be negligible, providing a strong proof of our smooth integration algorithm. Quantitative analysis illuminates a marginal difference. Using the error metrics defined in \autoref{subsec:Error Metrics for Synthetic Observation}, we find $\TSREcube(\text{FORMAL}, \text{SMOOTH}) \approx 2.5\%$ and $\TSRElv(\text{FORMAL}, \text{SMOOTH}) \approx 2.6\%$, with Raleigh-Jeans brightness-temperature statistics of $\TAM(T_\text{FORMAL}) \approx \SI{0.42}{\K}$ versus $\TAM(T_\text{SMOOTH}) \approx \SI{0.43}{\K}$. We consider this error as within tolerance, however, justifying our use of smooth integration in all other synthetic observations for our training dataset and for our publication figures.

\subsection{Synthetic Observation Versus Density-Tracing} \label{subsec:Synthetic Observation Versus Density-Tracing}
    \begin{figure*}[t]
        \centering
        \includegraphics[width=1\linewidth]{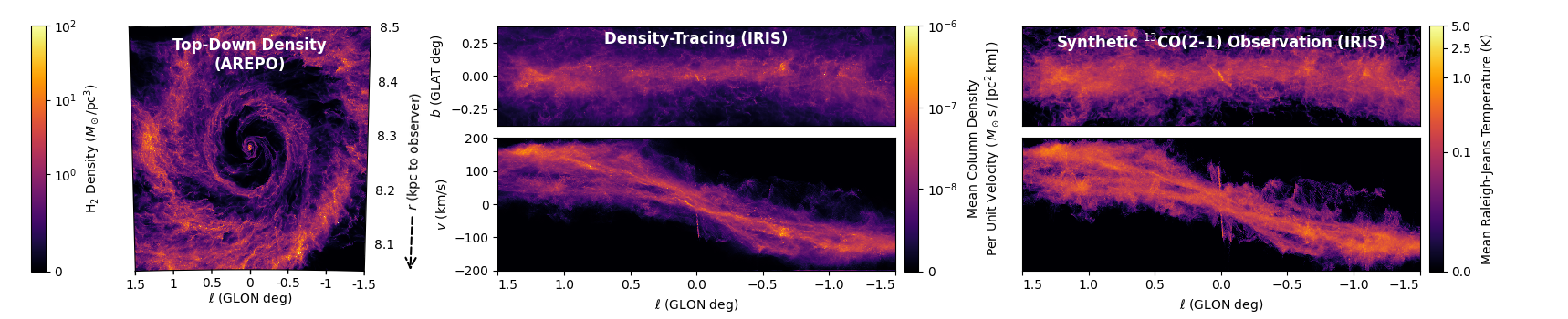}
        \caption{\textbf{Density-Tracing vs Synthetic Observation Comparison:} We compare the results of density-tracing (\autoref{subsec:Synthetic Observation Versus Density-Tracing}) versus synthetic observation by IRIS-SO. The simulation featured is due to \citeLipman, as described in \autoref{subsec:AREPO Zoom Simulations}, which we feature in our training dataset. All color bars are scaled with an $\arcsinh$ nonlinearity. As described in \autoref{subsec:Synthetic Observation Versus Density-Tracing}, we use this comparison to probe the ideal limit of density information encoded in a synthetic or real observation. We see a substantial amount of qualitative alignment, providing a strong rationale in attempting to reconstruct (top-down) density information from observations.}
        \label{fig:Simple vs Synth}
    \end{figure*}

    \par As a final point of analysis, we compare the characteristics of our synthetic observations against those of a density-tracing projection (implemented in IRIS-SO under the name \textit{simple observation}). As explained below, density-tracing provides an insight into the theoretical limit of density information that can be recovered from a real or synthetic observation. In \autoref{fig:Simple vs Synth}, we show the top-down density of an AREPO snapshot in the left panel, mean $\ell, b$ and $\ell, v$ projections of the corresponding density-tracing observation in the center panels, and mean $\ell, b$ and $\ell, v$ projections of the synthetic observation in the right panels. We show the same zoom-in simulations from \citeLipman that we use for our actual training-data construction (see \autoref{subsec:AREPO Zoom Simulations} for discussion of these simulations). 
    \par We compute a density-tracing observation as a projection of column density per unit velocity over dimensions of longitude, latitude, and velocity. This process yields a PPV cube that is comparable to a true synthetic observation, while not representing any real observational process. Density-tracing is a meaningful diagnostic tool because the coefficient of spontaneous emission of a spectral-line transition is proportional to number density of the transition's upper energy level (see \autoref{eqn:Line Emission}). In LTE, at fixed temperature and fixed abundance of the tracer and collision partners, this number density is proportional to the total gas density via Maxwell-Boltzmann statistics. This relationship is expressed via the equation
    \begin{equation}
        n_u = Z^{-1}g_ie^{-\varepsilon_i/[kT]}N
    \end{equation}
    \citep{Rybicki1986}, where $N$ is the total tracer number density, $n_u$ is the upper-level number density, $T$ is the gas temperature, $Z$ is the tracer partition function, $g_i$ and $\varepsilon_i$ are the respective statistical weights and energies of each level, and $k$ is the Boltzmann constant (see also \autoref{subsec:Level Populations: Computational Challenges}).
    \par In a true optically thin regime, absent absorption and stimulated emission, the line-of-sight derivative of intensity is the spontaneous emission coefficient: $dI_\nu \approx j_\nu \, ds$. Therefore, in this ideal regime, the intensity per unit frequency (or unit velocity) of the integrated ray, which is invariant with respect to distance of the observer, is proportional to the column density per unit frequency (or unit velocity) of the ray. This column density is precisely a density-tracing observation. We therefore provide density-tracing observations in both \autoref{fig:Sims Overview} and \autoref{fig:Simple vs Synth} as a visualization of the ideal information content of a PPV cube in the absence of the many confounding factors of the real observational process. 
    \par In general, \autoref{fig:Simple vs Synth} shows similar characteristics between our synthetic observations and density-tracing. This finding provides a strong justification for a central element of the supervised-reversion hypothesis---that density information can be recovered by our reversion model through the analysis of real and synthetic observations (see \autoref{subsec:Supervised Reversion: Objective}). In theory, optically thick behavior and dust extinction distort an observation in comparison to pure density-tracing via masking effects. These masking effects may then degrade the structural information encoded in the observation, although we note that such degradation is not an entirely foregone conclusion, since masking also provides an alternative means of encoding positional information along the line of sight. (See also \citet{Lipman2025} and \citet{Lipman2026} for related uses of such masking effects in inferring positional information regarding CMZ structure.) If masking does represent information degradation, however, then such degradation poses an inference challenge our reversion model must learn to mitigate via the application of physical trends learned from our training dataset.
    
\subsection{Speed Testing} \label{subsec:Speed Testing}
    \begin{figure*}[t]
        \centering
        \includegraphics[width=1\linewidth, trim={0cm 2cm 0cm 0cm}, clip]{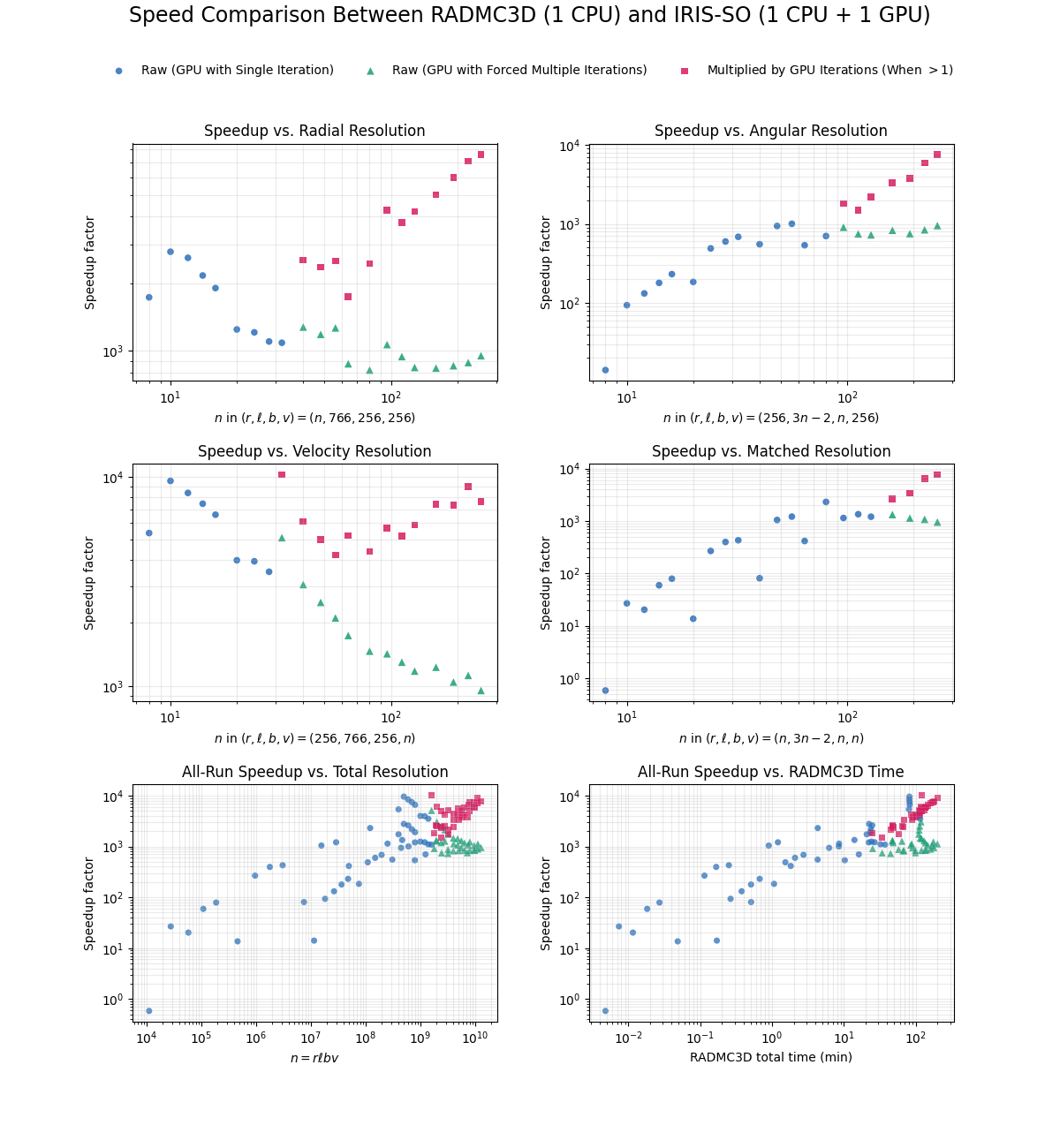}
        \caption{\textbf{IRIS-SO Speed Testing:} We summarize our speed comparisons between IRIS-SO and RADMC-3D, which we describe in \autoref{subsec:Speed Testing}. Each RADMC-3D observation is computed on a single AMD EPYC 7713 CPU, whereas each IRIS observation is computed on one AMD EPYC 7713 CPU plus one 40GB NVIDIA A100 GPU. We ran a total of 81 separate speed tests, running two trials of each test, exploring the effect of varying resolution on the total speedup factor (defined in \autoref{subsec:Speed Testing}). In the top four panels, we show trial-averaged speedup factor versus resolution. The resolutions varied are radial/line-of-sight ($r$, top-left), latitude and longitude ($\ell, b$, top-right), velocity ($v$, middle-left), and all-dimension matched resolution (middle-right). In the bottom-left, we plot the speedup factor versus total resolution $r{\ell}bv$ for all trials. In the bottom-right, we plot the speedup factor versus the RADMC-3D time for all trials. At certain resolutions, we record speedups in excess of $10{,}000\times$. IRIS achieves these speedups by parallelizing the observation on GPU, across each separately observed ray. At high resolutions, however, the 40GB GPU memory prevents us from computing the entire observation in parallel, forcing us to break the sky plane into separate chunks and iterate one chunk at a time. Since this iteration artificially deflates the speedup factor, we use separate ticks to plot low-resolution trials with no forced iteration, high-resolution trials with iteration, and the speedup factor multiplied by number of iterations (if $> 1$), which represents an ideal upper bound for speedup in the absence of a GPU memory cap.}
        \label{fig:Speed Test}
    \end{figure*}
    
    \par In addition to verifying the accuracy of the IRIS synthetic-observation algorithm, our RADMC-3D verifications provide the opportunity to conduct a battery of speed comparisons under like conditions. See \autoref{fig:Speed Test} for a summary of our results. For each speed comparison, we compare RADMC-3D running on a single AMD EPYC 7713 CPU versus IRIS-SO running on one AMD EPYC 7713 CPU plus one 40GB NVIDIA A100 GPU. We configure RADMC-3D exactly as in \autoref{subsec:RADMC-3D Configurations}, observing the exact same simulation due to \citeLipman. In local observer mode using first-order integration with scattering and flux-conservation turned off, RADMC-3D is configured for maximal speed while maintaining like conditions with IRIS. Under these conditions, we probe the total speedup provided by IRIS under varying resolution of the radial, angular, and velocity dimensions. 
    \par As a like comparison, we start our timer with the observed physical tensor existing in memory in our Python environment, since most astrophysics use-cases, including this one, are Python-based. In timing RADMC-3D, we thus include the disk-write time of the configuration files provided to RADMC-3D, the runtime of the RADMC-3D executable (implemented in FORTRAN), and the disk-write time of the output images for both our line and continuum observations. We do note that our disk-write times are subjective to the particular networked drive setup we used on the UConn Storrs HPC, although this setup was generally fast, and contributed only a minority of the total RADMC-3D time. We do not, however, time the Python-based, image-file read-in and post-processing steps, to include velocity integration. Since we do time these post-processing steps in IRIS, we are providing a lower speedup estimate than if we also timed IRIS only up to the velocity integration step.
    \par Under these conditions, we define the total speedup factor as the ratio of total RADMC-3D time and total IRIS time. At our tested resolutions, we find speedups in excess of $10{,}000\times$. One complication that arises is that, under these resolutions, we are forced, on our 40GB A100 GPUs, to chunk our synthetic observations along the sky plane and iterate the synthetic-observation process one chunk at a time, which increases IRIS time and reduces the speedup factor. In practice, this time increase is not an exact scaling by the number of observed chunks. The chunk-multiplied speedup factor, however, provides an upper bound on the additional speedups possible with a higher GPU memory capacity at which we can conduct fully parallelized (single-iteration) synthetic observations at high resolution. While we find chunk-multiplied speedups are similarly capped around $10{,}000\times$, we reach chunk-multiplied speedups in the thousands and up to the $10{,}000\times$ cap at a broader range of resolutions.
    \par These speed results provide a strong justification for the custom development of IRIS-SO for this research study. Nearing our training-data resolution, we find that RADMC-3D times are measured in hours. This long runtime makes generation of the $\sim 100\text{k}$ synthetic observations required for our training dataset infeasible using RADMC-3D or other traditional, CPU-based synthetic-observation codes of comparable performance, with which compute time for our training-data construction would be measured in core-years. For IRIS, however, each synthetic observation is computed in only $\sim 5$ seconds, enabling construction of our full dataset as detailed in \autoref{table:Dataset Parameters}.
    
\section{Machine-Learning Methods and Reversion} \label{sec:Machine-Learning Methods and Reversion}
    \par The ultimate intent of the IRIS project is to leverage machine intelligence to reverse the observational process, yielding information about the hidden line-of-sight dimension of the CMZ. To this end, we explored not just one but a number of widely varying methods. We ultimately had success with one of these methods---the supervised-reversion approach---which we detail in this paper as our primary contribution. We also take a moment, however, to highlight the wider space of possible methods, including those alternate methods that we did not pursue or with which we did not have success. Some of these alternate methods, we believe, may still be of independent and future interest to other researchers, and may provide the basis of some of the possibilities for future work we detail in \autoref{sec:Summary and Conclusion}. We therefore discuss more details regarding the numerous variations within these method classes over the following section.
    \par The two primary classes of methods we explored over the course of this research effort were the \textit{supervised-reversion method} and the \textit{neural-fields method}. In the supervised-reversion method, a neural network is trained as a function that maps observed objects of some kind to their original, or \textit{reverted}, physical representations prior to observation. In the neural-fields method, the space of possible physical representations of a single observed object is itself treated as a differentiable tensor, or \textit{neural field} \citep{Park2019, Mescheder2019, Mildenhall2021, Xie2022}, over which we conduct a gradient descent process in order to arrive at the best-fit representation. While we ultimately pivoted away from a neural-fields approach after substantial exploration, we note the recent success with such a method in the related work of \citet{Levis2025}, which investigated the reverse-imaging of protoplanetary disks as opposed to the CMZ, and which we believe may invite reexamination of the neural-fields approach in future CMZ-related research. 

\subsection{Supervised Reversion: Objective} \label{subsec:Supervised Reversion: Objective}
    \par The primary concept explored by this study is that of the supervised-reversion model. The terminology \textit{supervised}, in this context, is invoked in the standard sense of training on a dataset of paired inputs and outputs \citep{Goodfellow2016}. In this scheme, a dataset of physical tensors is constructed from a bank of simulation snapshots, which are each synthetically observed. A neural network is then trained as a function that is a quasi-inverse of the synthetic-observation process. More specifically, let $P$ denote the space of physical tensors. That is, each physical tensor $p \in P$ is a map from a discretized spherical-coordinate space in dimensions $r, \ell, b$ to the product space of each physical-tensor channel---radial velocity, gas mass-density, gas temperature, $\HTwo$ abundance, CO abundance, and dust temperature. Then let $P'$ be the space of top-down density images, i.e. scalar density maps over $r, \ell$ space, with the latitude-mean reduction
    \begin{align}
        t \, : \, P &\to P' \\
        p &\mapsto \Big[(r, \ell) \mapsto \meanb \, p_\rho(r, \ell, b)\Big] \text{ ,} \notag
    \end{align}
    where $\meanb$ is the mean over all latitude space, and $p_\rho$ is the density coordinate projection (density channel) of a physical tensor. Next, let $S$ denote the space of observed spectral (PPV) cubes. That is, scalar maps from the discretized space $\ell, b, v$ to the space of either radiative intensity or equivalent brightness/Raleigh-Jeans temperature (see \autoref{subsec:Intensity Versus Temperature}). Define $S'$ to be the space of $\ell, v$ observations with the reduction
    \begin{align}
        u \, : \, S &\to S' \\
        s &\mapsto \Big[(\ell, v) \mapsto \meanb \, s(\ell, b, v)\Big] \text{ .} \notag
    \end{align}
    Then, letting $o \, : \, P \to S$ denote the synthetic-observation process, and choosing some training data $D(P) \subset P$, we seek to learn a reversion map $r' \, : \, S' \to P'$ such that
    \begin{equation}
        t\mid_{D(P)} = r' \circ u \circ o\mid_{D(P)} \text{ ,}
    \end{equation}
    where $\mid_{D(P)}$ denotes the domain restriction to the data $D(P)$---i.e., in standard algebraic convention, the diagram commutes:
    \begin{equation}
        \begin{tikzcd}
            D(P) \arrow{r}{o} \arrow{d}{t} & S \arrow{d}{u} \\
            P' & S' \arrow[swap]{l}{r'}
        \end{tikzcd}
    \end{equation}
    We note that the concept of reconstructing 3D/top-down density structure based on analysis of $\ell, v$ observations is similar to the method presented in \citep{Sormani2015b}, although this method relies on human-identified features rather than a machine-learning approach.

\subsection{Supervised Reversion: Training Theory} \label{subsec:Supervised Reversion: Training Theory}
    \par To learn the reversion function $r'$, we train a neural network on a dataset of pairs $(t(p), (u \circ o)(p))$ where $p \in D(P)$, via a supervised loss function
    \begin{equation}
        \mathcal{L}_{P'} \, : \, P' \times P' \to \R \text{ .}
    \end{equation}
    We then intend that the quasi-inverse property of the maps $o\mid_{D(P)}$ and $r'$ extends to some broadened domain $D'(P) \supset D(P)$ with the real observation $s_\text{real}$ \citep[in our case, SEDIGISM $\ThirteenCOTwoOne$,][]{Schuller2021} contained in the image $o[D'(P)]$. Of course, a true (two-sided) inverse $o^{-1} \, : \, S \to P$ of the full observational map $o \, : \, P \to S$ is certainly impossible. In the optically thick limit, foreground masking destroys all information regarding the background of an observation, meaning that $o$ is in general highly non-injective, and in separate modes than the non-injectivity of the latitude-averaging map $t$. Therefore, definition of an ideal map $r'$ requires a choice between degenerate options. The purpose of the supervised training scheme is to allow this choice to be informed by the physics encoded into the simulations themselves, which manifest many complex and emergent properties not obvious from a first-principles standpoint.
    \par For this proof-of-concept, we specifically aim to revert synthetic $\ThirteenCOTwoOne$ observations, with an ultimate intent of generalizing to the SEDIGISM $\ThirteenCOTwoOne$ data (although our primary focus in this study remains on proving the feasibility of supervised reversion on synthetic data alone). To perform these reversions, the network is not provided any prior information or ``hint''. Rather, the network learns, from scratch, how to conduct a reversion based solely upon information encoded in the training dataset. We may speculate that the network learns something about the observation process and its relationship to CO density, something about the relationship between CO density and $\HTwo$ density, and something about how to fill in missing details in a top-down perspective by completing featural motifs such as common orbital patterns. Nonetheless, due to the black-box nature of machine learning, we can never know for sure how the network operates or the precise nature of what it has learned, only whether it has learned enough to perform satisfactorily on a test dataset.

\subsection{Supervised Reversion: Higher Dimensionality} \label{subsec:Supervised Reversion: Higher Dimensionality}
    \par In our earliest explorations of the supervised-reversion method, we sought instead to find a true one-sided inverse $r \, : \, S \to P$ of the data-restricted observational map $o|_{D(P)} \, : \, D(P) \to S$, mapping full PPV cubes to full physical tensors such that
    \begin{equation}
        1 = \begin{tikzcd} D(P) \arrow{r}{o} & S \arrow{r}{r} & P \end{tikzcd} \text{ ,}
    \end{equation}
    where $1 \, : \, D(P) \to D(P)$ is the identity function. We aimed to find such a one-sided inverse by training a neural network on a dataset of pairs $(p, o(p))$ where $p \in D(P)$, via a supervised loss function
    \begin{equation}
        \mathcal{L}_{P} \, : \, P \times P \to \R \text{ .}
    \end{equation}
    Letting $s_\text{real} \in S$ be a real observation, the ideal limit of the training process is that
    \begin{equation}
        \begin{gathered}
            (r \circ o \circ r)(s_\text{real}) = r(s_\text{real}) \text{ .} \\[.5em]
            \begin{tikzpicture}
                \coordinate (LeftColumn) at (0,0);
                \coordinate (RightColumn) at (2.5,0);

                \draw[thick] (LeftColumn) ellipse [x radius=.8cm, y radius=1.95cm];
                \draw[thick] (RightColumn) ellipse [x radius=.8cm, y radius=1.95cm];
                \node at (0,1.45) {$R$};
                \node at (2.5,1.45) {$S$};

                \node (LeftLower) at (0,-1.25) {$\bullet$};
                \node (RightUpper) at (2.5,0.65) {$\bullet$};
                \node[anchor=west] at ([yshift=-10pt]RightUpper.west) {$s_{\mathrm{real}}$};
                \node (RightLower) at (2.5,-1.25) {$\bullet$};

                \draw[->] (RightUpper) -- node[pos=.45,left] {$r$} (LeftLower);
                \draw[->, bend left=15] (LeftLower) to node[above] {$o$} (RightLower);
                \draw[->, bend left=15] (RightLower) to node[below] {$r$} (LeftLower);
            \end{tikzpicture}
        \end{gathered}
    \end{equation}
    In particular, however, this more robust scheme also allows us to check whether
    \begin{equation}
        \begin{gathered}
            (o \circ r)(s_\text{real}) = s_\text{real} \text{ .} \\[.5em]
            \begin{tikzpicture}
                \coordinate (LeftColumn) at (0,.1);
                \coordinate (RightColumn) at (2.5,.1);

                \draw[thick] (LeftColumn) ellipse [x radius=.8cm, y radius=1.4cm];
                \draw[thick] (RightColumn) ellipse [x radius=.8cm, y radius=1.4cm];
                \node at (0,.8) {$R$};
                \node at (2.5,.8) {$S$};

                \node (Left) at (0,0) {$\bullet$};
                \node (Right) at (2.5,0) {$\bullet$};
                \node[anchor=west] at ([yshift=-10pt]Right.west) {$s_{\mathrm{real}}$};

                \draw[->, bend left=15] (Left) to node[above] {$o$} (Right);
                \draw[->, bend left=15] (Right) to node[below] {$r$} (Left);
            \end{tikzpicture}
        \end{gathered}
    \end{equation}
    \par And indeed, we may also incorporate this check into the training process as a secondary supervised loss \begin{equation}
        \mathcal{L}_{S} \, : \, S \times S \to \R
    \end{equation}
    evaluated only over some sampling distribution on $S$ that is narrow around $s_\text{real}$. Here, narrow may be via a simple metric on $S$ itself or via some perceptual metric defined in the latent space of some convolutional neural network over $S$. Unfortunately, we found that learning a neural network from a single-channel 3D space to a multi-channel 3D space was dimensionally prohibitive. Far fewer standard heuristics have been established for 3D-to-3D transcoder-type neural networks than for 2D-to-2D image transcoders. Moreover, the creation of a 3D-to-3D training dataset requires orders of magnitude more storage space than a 2D-to-2D training dataset, which also translates to orders of magnitude more training time due to increased disk-to-memory load latency. For these reasons, we pivoted away from the true 3D-to-3D approach for this initial proof-of-concept, but note it here as a possibility for future expansions of this research. We discuss such future research possibilities in greater detail in \autoref{sec:Summary and Conclusion}.

\subsection{Neural Fields: General Approach} \label{subsec:Neural Fields: General Approach}
    \par The second class of methods we considered early on as an alternative to the supervised-reversion approach is that of the neural-fields method. Neural fields, which have wide applicability in image reconstruction, are an established class of machine-learning approaches based on optimization of a differentiable (or \textit{neural}) input rather than a differentiable (\textit{neural}) function \citep{Park2019, Mescheder2019, Mildenhall2021, Xie2022}. In this scheme, rather than training a neural-network function as a quasi-inverse or true one-sided inverse to the observational process, we treat the input of the observational map $o$ as a trainable neural field. We then run a gradient-descent optimization over the space of physical tensors $p \in P$ by minimizing both $\mathcal{L}_P^{(k)}(p)$, for some collection
    \begin{equation}
        \mathcal{L}_P^{(1)}, \dots , \mathcal{L}_P^{(n)} \, : \, P \to \R
    \end{equation}
    of physics-informed loss functions, as well as $\mathcal{L}_S[o(p), s_\text{real}]$ for some observational loss function
    \begin{equation}
        \mathcal{L}_S \, : \, S \times S \to \R \text{ .}
    \end{equation}
    This approach has the advantage of optimizing for a less general object. Rather than attempting to find a general mapping that inverts the observational process over a broad set of possible observations, we seek only to invert the real observation itself.
    \par Since the space of possible physical tensors satisfying this inversion is, again, highly degenerate, we must constrain the optimization problem via some set of physics constraints. In this case, however, rather than relying on the simulations as physical constraints, we rely on physics-informed loss functions that evaluate the plausibility of a physical tensor. In order to compute the gradient of the observational loss function with respect to the neural-field/physical-tensor parameters, we require a differentiable synthetic-observation process implemented in the PyTorch backend \citep{Paszke2019} so that gradients can be backpropagated through the synthetic-observation process itself to the neural field. This was the original reason we built IRIS-SO as a differentiable module (see \autoref{sec:Synthetic Observation Implementation and Algorithms} for more details on IRIS-SO differentiability).
    \par Ultimately, we did not pursue this general neural-fields approach since the specification of appropriate physics-informed loss functions is a challenging task that fails to take advantage of the rich information available from our sophisticated simulation database. Additionally, even if such loss functions can be appropriately specified, it is not clear that the total loss function
    \begin{equation}
        \mathcal{L} = \mathcal{L}_P^{(1)} + \dots + \mathcal{L}_P^{(n)} + \mathcal{L}_S
    \end{equation}
    will yield a manifold over physical-tensor space $P$ that is convenient for gradient descent. 
    \par We note similarity, however, between this method and that employed successfully by \citet{Levis2025} in the reverse-imaging of protoplanetary disks. While the IRIS project was conceived and developed independently from and concurrently with \citet{Levis2025}, we take both this independent convergence of methods and the success demonstrated in \citet{Levis2025} as an indication of potential future promise for the neural-fields concept within the context of CMZ research. For this reason, we note this approach here in order to stimulate a deeper discussion regarding future avenues of research.

\subsection{Neural Fields: Latent-Constrained} \label{subsec:Neural Fields: Latent-Constrained}
    \par A variant of the neural-fields method that we explored in more detail is that of a constrained neural-fields method. In this method, we apply a similar general approach to that of the neural-fields concept, but seek to optimize a neural field over a constrained space rather than the full space $P$ of physical tensors. The aim of this approach is to leverage the simulations as a constraining prior on the neural-field optimization by creating a \textit{perfect} optimization space $Z$ with a continuous and differentiable embedding $d \, : \, Z \to P$ such that:
    \begin{enumerate}[(i)]
        \item $Z$ is lower-dimensional than $P$, reducing the number of trainable parameters in the neural field;
        \item all points in the simulation training data have a representation in the constrained space, i.e. $d(Z) \supseteq D(P)$;
        \item all points in the constrained space look like points in the training data, i.e. there is a uniform bound $\varepsilon > 0$ such that for all $z \in Z$, there exists some $p \in D(P)$ and some $\varepsilon$-sized ball $B(p, \varepsilon)$, under a suitable perceptual metric, satisfying $d(z) \in B(p, \varepsilon)$; and
        \item the manifold of the $\mathcal{L}_S$ loss over $Z$ is convenient for gradient descent.
    \end{enumerate}
    In this method, no physics-informed losses are applied. The only loss function is $\mathcal{L}_S$, and the optimization process is imbued with physical context via the constraint of the neural field from $P$ to $Z$.
    \par We devoted substantial effort trying to solve this \textit{perfect encoding} problem. Originally, we aimed to produce $Z$ as the latent space of an autoencoder
    \begin{equation}
        \begin{tikzcd} P \arrow{r}{e} & Z \arrow{r}{d} & P \end{tikzcd}
    \end{equation}
    trained in an unsupervised manner over the training dataset $P(D)$ of physical tensors \citep[see][for details on autoencoding and unsupervised training]{Hinton2006, Goodfellow2016}. We quickly determined, however, that this method was unsatisfactory in that it failed the condition (iii). Similarly, producing $Z$ by training $d$ as a purely generative model via a GAN strategy \citep[see][for details on GANs]{Goodfellow2014, Goodfellow2016} suffers mode collapse and quickly fails (ii). We made additional attempts to combine these approaches by simultaneously training $d \circ e$ as an autoencoder and $d$ as a GAN, but with the effect of simply failing both (ii) and (iii) rather than succeeding at both.
    \par Ultimately, we conjecture that the perfect encoding problem is generally unsolvable by virtue of topological constraints. In particular, any map that satisfies (i), (ii), and (iii) will very likely be highly discontinuous, complicating condition (iv). For these reasons, we abandoned the constrained neural-fields approach. We have been prompted to reevaluate the potential of neural-fields-based approaches, however, in light of the success of \citet{Levis2025}. We therefore include this description of our unsuccessful experiments with the constrained-neural-fields method in order to stimulate a deeper discussion regarding future research possibilities.

\subsection{Implementation of Reversion: Architecture} \label{subsec:Implementation of Reversion: Architecture}
    \begin{figure*}[t]
        \centering
        \includegraphics[width=.9\linewidth, trim={0cm 0cm 0cm 2cm}, clip]{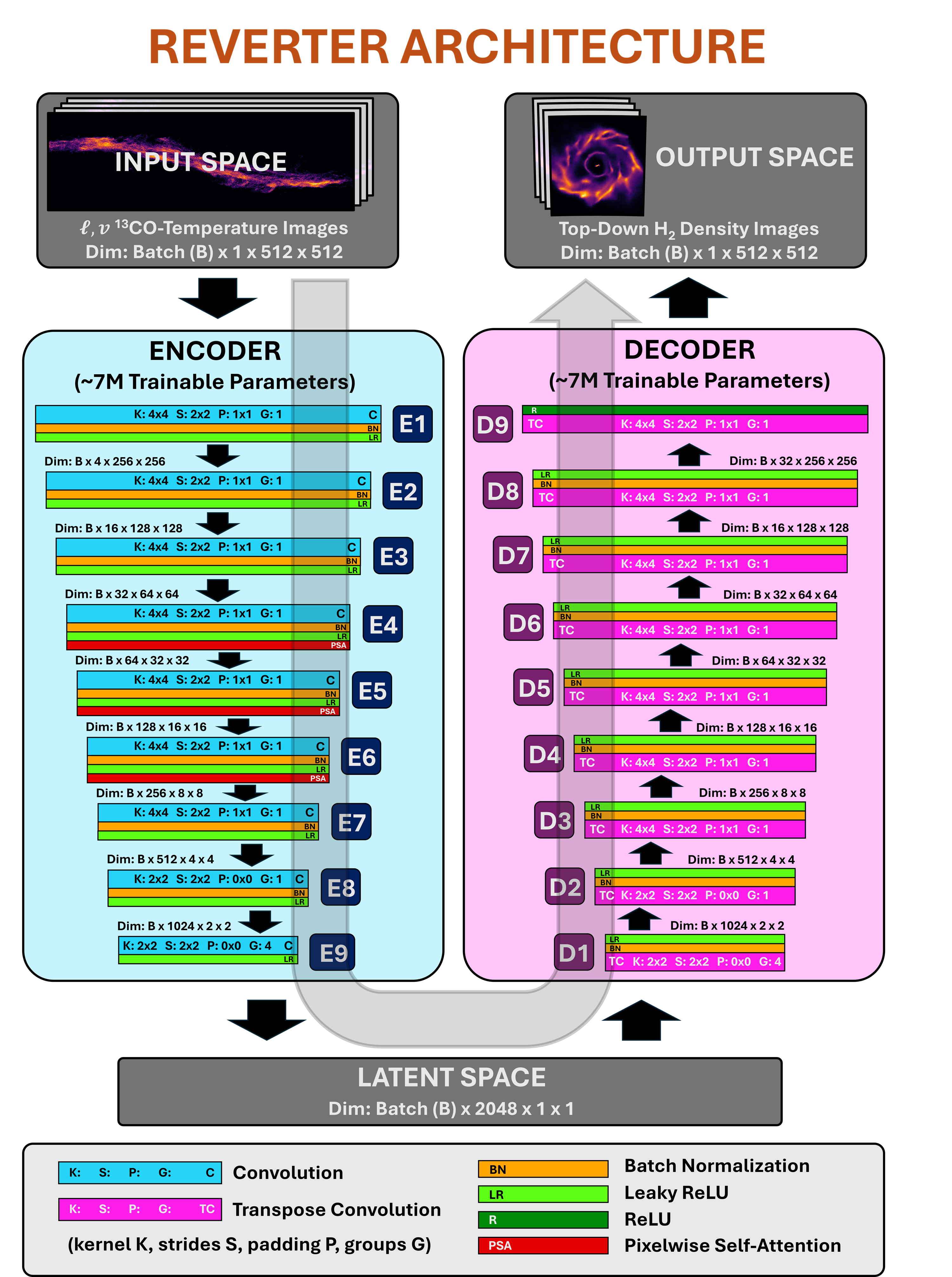}
        \caption{\textbf{Reverter Architecture:} A schematic diagram illustrating the architecture of our custom reversion model, or \textit{reverter}, as described in \autoref{subsec:Implementation of Reversion: Architecture}. The reverter is a deep convolutional neural network (CNN) with pixelwise self-attention. The architecture follows a classic encoder-decoder heuristic, mapping from the observed image space to the top-down image space via a fully featural latent space out of which all image-dimensionality has been reduced. Other than in the implementation of self-attention layers in the encoder and the hard ReLU output activation of the decoder, the encoder and decoder are mirror images. The encoder and decoder each contain $\sim 7\text{M}$ trainable parameters, for a model total of $\sim 14\text{M}$ trainable parameters. See also \autoref{subsec:Supervised Reversion: Objective} for a precise description of the objective the reverter is designed to learn.}
        \label{fig:Architecture}
    \end{figure*}

    \par We tested many architectural variations of the IRIS reversion model for practical implementation of the supervised-reversion method (\autoref{subsec:Supervised Reversion: Objective}). A complete overview of our final network architecture is given in \autoref{fig:Architecture}. In designing this architecture, we employed heuristics common within the field of computer vision. These heuristics apply well to this use-case because, to an artificial neural network, a transformation from an $\ell, v$ reduced observation to a top-down density image is just a standard single-channel image-to-image transformation. The only difference is that, in our case, the pixel values represent quantities in physical units as opposed to grayscale intensities.
    \par The role of our research was thus not in innovating new machine-learning techniques, but in porting wisdom from the well-researched domain of computer vision to the domain of astrophysics, in which these established techniques are less widely applied. See, in particular, image segmentation as an image-to-image problem \citep{Long2015, Badrinarayanan2017}. The code for the IRIS reversion model, or \textit{reverter}, which is implemented exclusively in PyTorch \citep{Paszke2019}, is all openly accessible on the IRIS GitHub. We have opted, however, not to openly release trained model weights at this time, since we do not yet find definitive convergence of the trained model on real observational data, and since research into the application of the trained model to such real observations is ongoing. (See \autoref{sec:Results and Discussion} for further discussion of training results).
    \par The IRIS reverter is a convolutional neural network \citep[CNN,][]{LeCun2002, Krizhevsky2012, Goodfellow2016} with pixelwise self-attention \citep{Wang2018, Zhang2019} that operates via a simple encoder-decoder heuristic common within computer vision research \citep{Long2015, Badrinarayanan2017}. The network may be viewed as a map
    \begin{equation}
        \begin{tikzcd} r' \, : \, S' \arrow{r}{e} & Z \arrow{r}{d} & P' \end{tikzcd}
    \end{equation}
    where the encoder $e$ maps an element in the latitude-reduced observed space $S'$  to a purely featural latent space $Z$, and the decoder $d$ maps an element in $Z$ to the space $P'$ of top-down density images. It is typical to treat the image channel count as an additional dimension preceding the image dimensions themselves. Thus, $S'$ has dimensions $1 \times 512 \times 512$, $Z$ has dimensions $2048 \times 1 \times 1$, and $P'$ has dimensions $1 \times 512 \times 512$.
    \par In the latent space $Z$, the image dimensions are entirely removed, replaced by a purely featural dimension. The size of the latent space is also compressed from the size of the input space by a factor of $512 \cdot 512 / 2048 = 128$. This compression encourages the network to learn a more efficient, abstract representation of the input object and discard unnecessary information. The output resolution is chosen to be equivalent to the input resolution, although the effective resolution of the output space may be lower (see \autoref{subsec:Synthetic Reversion} and \autoref{fig:Synthetic Reversions}). An entirely dimensionless featural space is used because the input and output spaces of the network are dislike images spaces. Removing all dimensionality ensures that no visual features of the input image are retained in literal form in the output image. For the same reason, we apply no skip-connections between the encoder and decoder of the kind frequently used in image segmentation \citep[a mapping between like image spaces,][]{Ronneberger2015}.
    \par The encoder and decoder are direct mirrors of each other, with a one-for-one replacement of convolutions in the encoder with transpose convolutions in the decoder---except for the pixelwise attention layers, which are only applied in the encoder. The assumption in this specific asymmetry is that the encoder performs the hard work of ``understanding'' the input by mapping it into an abstract featural space that illuminates its true structure in the most efficient possible representation. Attention in the encoder augments this understanding process by allowing local features to be considered within a broader global context. Once organized into the featural space, the decoder's work is merely that of an ``illustrator'' that produces an output image based upon the coherent set of instructions encoded into the featural space, for which global attention is assumed unnecessary. Nonetheless, our experiments indicate that the addition of attention layers provides only a marginal performance enhancement from the pure CNN design.
    \par The encoder and decoder each consist of nine layer blocks. Each block is a stack of either a convolution \citep{LeCun2002} or transpose convolution \citep{Long2015}, a batch normalization \citep[if applied,][]{Ioffe2015}, a leaky ReLU \citep[or ReLU for the output layer,][]{Krizhevsky2012, Maas2013}, and pixelwise attention \citep[if applied,][]{Wang2018, Zhang2019}. Downsampling in the encoder is performed by strided convolutions rather than by fixed pooling, and is performed by every convolution rather than by dedicated downsampling layers. This heuristic provides the most direct possible path for the input observation and allows the one-for-one replacement of downsampling convolutions in the encoder with upsampling transpose convolutions in the decoder. Attention layers consist of a layer norm \citep{Ba2016}, followed by a multi-head attention applied pixelwise, followed by another layer norm. To maximize effect while minimizing the added computational overhead, attention is used only in blocks 4, 5, and 6 of the encoder. The final decoder layer is not normalized, and performs a hard ReLU to ensure that a negative density prediction is not made.
    \par Convolutions 1--7 of the encoder use a kernel size of $4 \times 4$, strides of $2 \times 2$, and padding of $1 \times 1$. Convolutions 8 and 9 use kernel sizes of $2 \times 2$ with no padding, with convolution 8 using strides of $2 \times 2$ and convolution 9 performing the final strideless convolution. Convolution 9 is divided into four groups in order to cap parameter count. Transpose convolution 1 of the decoder is a strideless, unpadded transpose convolution with kernel size $2 \times 2$. Transpose convolution 2 is unpadded with kernel size $2 \times 2$ and strides $2 \times 2$. Transpose convolutions 3--9 use kernel size $4 \times 4$, strides $2 \times 2$, and padding $1 \times 1$. Transpose convolution 9 is also divided into four groups in order to cap parameter count. The entire model contains about $\sim 14\text{M}$ trainable parameters.

\subsection{Implementation of Reversion: Choice of Loss Function} \label{subsec:Implementation of Reversion: Choice of Loss Function}
    \par In any machine-learning application, half the learning capacity lies not in the model architecture but in an effective training scheme, which begins with the choice of an appropriate loss function. In the supervised-reversion method (\autoref{subsec:Supervised Reversion: Objective}), the model must be trained to make the inferred top-down density images look like the ground-truth, top-down density images. The loss function is the mathematical item quantifying this comparison, according to which model parameter gradients are computed and the optimizer is stepped. In computer vision, the most common image-image loss function is mean squared error (MSE). For our application, however, we find that MSE fails to produce model convergence. We infer that the reason for this lack of convergence is because MSE does not appropriately scale the units of density for effective comparison.
    \par To resolve this issue, we apply the principle that if the human eye can \textit{see} the data, then the training optimizer can \textit{see} the data also. In other words, whatever nonlinearity yields the best visualization of the model outputs should also be applied to the image inputs of the loss function. Based upon this principle, we arrived at an $\arcsinh$-scaled MSE-variant we refer to in this publication as \textit{Mean Arc-Hyperbolic-Sinusoidal Square-Error (MASE)}. We define this loss function as
    \begin{align}
        &\MASE \, : \, P' \times P' \to \R \\
        &\MASE \, (p_1', p_2') = \mean\left(\arcsinh\left[\left(\frac{p_1' - p_2'}{\gamma}\right)^2\right]\right) \text{ ,} \notag
    \end{align}
    where a density normalization constant
    \begin{equation}
        \gamma = \SI{1.0e-19}{\kg/\m^3}
    \end{equation}
    is applied to ensure the loss function is units-invariant. A units-invariant loss function allows comparison of loss metrics across training runs conducted on a variety of datasets with differing units. As described in \autoref{subsec:Computational Considerations}, the units of each separate dataset are set in order to approximate unit standard deviation in both the input and output spaces $S'$ and $P'$, which promotes better training of the reversion model.

\subsection{Implementation of Reversion: Training Hyperparameters, Overfitting, and Regularization} \label{subsec:Implementation of Reversion: Training Hyperparameters, Overfitting, and Regularization}
    \par During training, we divided our training data into an 80\%--20\% training-validation split. Average loss scores were accumulated over each epoch on both the training and validation data, but gradients were not computed during validation and the optimizer was only stepped during training. Comparing the training and validation loss metrics provides a first-order approximation of how overfit the model is onto the training data.
    \par Of course, this approximation is extremely first-order. Because there are 64 physical tensors per snapshot (see \autoref{subsec:Physical Tensor Interpolation} and \autoref{table:Dataset Parameters}), then even with random distance perturbations (see \autoref{subsec:Physical Tensor Perturbations}), the total dataset may be somewhat saturated. Since the validation subset is randomly selected, it is still highly correlated with the training subset. The true degree of overfitting, then, as pertains to the larger observational space containing both real observations and even synthetic observations from other simulation runs, may be systematically underestimated by a coarse training-validation comparison. We return to this discussion of overfitting in \autoref{subsec:Loss Convergence} and \autoref{subsec:Discussion}.
    \par During training, we also made random additions of both noise and foreground features (see \autoref{subsec:Noise Addition} and \autoref{subsec:Foreground and Background}) to the latitude-reduced observations in order to train the reversion model to intelligently ignore these features when inferring top-down density images. Since both these additions are made randomly at each training step without correlation to the supervised data pairs themselves, we find they are also natural sources of regularization, which prevent overfitting and reduce the gap between training and validation loss. We see systematic training/validation gap reductions for terminal loss on training runs where noise, foreground, or both of these features are added, in comparison with training runs without such additions.
    \par We ran several independent training runs over the IRIS training dataset under a variety of hyperparameters. From these experiments, we settled upon a hyperparameter configuration that we found to produce best results in terms of both terminal training/validation loss and generalization to the real SEDIGISM data \citep{Schuller2021}. On evaluating the quality of the SEDIGISM generalization, we used a qualitative, by-eye judgment based on the expectation established in the literature \citep[e.g. by][]{Walker2025, Lipman2025, Lipman2026} of what a true top-down density image of the Milky-Way CMZ should look like (see \autoref{sec:Introduction} and \autoref{fig:CMZ Overview}). Primarily, we simply assessed to what degree the SEDIGISM reversion resembled an elliptical orbit or a coherent, open trajectory versus an unphysical noise distribution. We avoided applying speculative biases, instead allowing the reversion method itself to guide results as pertain to specific orbital parameters and cloud-scale geometry. Once we had settled on an optimal hyperparameter configuration, we ran the same training setup several times over the same training data, in order to compare the consistency of results on both the synthetic data and the real SEDIGISM data (see \autoref{fig:True Reversions}).
    \par Our ideal training hyperparameters are as follows: We train the reversion model over the entire training dataset for 32 epochs. We use the Adam optimization algorithm with a scheduled learning rate. We hold the learning rate constant at $\SI{1e-3}{}$ for the first 16 epochs, decreasing on epoch 17 to $\SI{5e-4}{}$, on epoch 18 to $\SI{2.5e-4}{}$, on epoch 19 to $\SI{1.25e-4}{}$, and holding constant at that rate for the remainder of model training. 
    \par While the IRIS code provides configurability for multi-node, multi-GPU training, we trained almost exclusively on a single 40GB NVIDIA A100 GPU with one CPU manager process and 16 CPU workers that work in parallel to stream data from the disk into memory. We find that this streaming latency is the primary bottleneck during training, so our setup benefits more in terms of compute time from the incorporation of additional CPU workers than from the incorporation of additional GPUs. Overall, our training pipeline requires around 3--6 hours to complete on the UConn Storrs HPC, although the exact training time is highly variable, as it depends primarily on the fluctuating load latency from a networked drive into memory.
    \par We experimented with both small and large batch sizes. We find that a single 40GB GPU accommodates a batch size of 768, which is already in excess of the batch size found to maximize training performance. In fact, we find that using a small batch size of 8, in combination with an effective batch size of 128, produced by accumulating gradients over 16 batches, yields another natural and critical source of regularization in reducing overfitting and improving validation loss scores. In theory, training in this setup with an effective batch size of 128 via batches of 8 and gradient updates every 16 batches versus with an actual batch size of 128 and gradient updates every batch are identical, except in how these setups differ with respect to the running statistics incorporated in each batch-normalization computation of the reverter. In principle, the stock batch-normalization layers incorporated in the IRIS reverter could be replaced by a custom layer that separates batch normalization into internal groups of 8 in order to mimic this effect. This alteration would allow single batch sizes of 128 and reduce total GPU usage time. We did not perform this optimization for this publication, however, because we find that GPU throughput is not the primary compute bottleneck. 

\section{Results and Discussion} \label{sec:Results and Discussion}
    \par We now discuss the results of our reversion experiments, illustrated through a series of figures. The primary aim of these experiments is to establish a proof-of-concept of the supervised-reversion method (\autoref{subsec:Supervised Reversion: Objective}). The most critical objective is thus demonstrating the ability of the model to train and perform in the simplest and most readily controlled environment of synthetic data. In \autoref{subsec:Loss Convergence} and \autoref{subsec:Synthetic Reversion}, we discuss the statistical results of the training process itself, as measured in loss metrics (see \autoref{subsec:Implementation of Reversion: Choice of Loss Function}), and then test the application of the trained model to synthetic data the model has not seen during training. We then probe some of the failure modes of the trained model (\autoref{subsec:Failure Modes}). Generally, we establish excellent success on synthetic data. 
    \par In generalizing our trained model to the real SEDIGISM data \citep{Schuller2021}, we are limited in that our training dataset for this publication (see \autoref{table:Dataset Parameters} for details), while containing over 100k top-down density images and synthetic $\ell, v$ images (see \autoref{subsec:Supervised Reversion: Objective} for definitions), is still minimal. Specifically, our training dataset was constructed from snapshots of a single run of the AREPO simulations due to \citeLipman, evolved over a still relatively short timescale of $\sim \SI{7}{\mega\year}$. Moreover, due to our current constraints in availability of simulation snapshots, we incorporated snapshots from early in the simulation run during which not all physics was fully turned on (see \autoref{subsec:AREPO Zoom Simulations}). Therefore, we consider experiments in generalization of our trained model to the real SEDIGISM data \citep{Schuller2021} still to be somewhat premature.
    \par Nonetheless, towards investigating the potential of our method, and its ability to generalize to real observations given sufficient training data, we do apply instances of our trained model to the SEDIGISM data in \autoref{subsec:Reversion of SEDIGISM Data}, and present top-down views of the Milky Way's CMZ generated by these trained models. We find that these top-down predictions are somewhat plausible, yielding some potentially notable evidence of an $x_2$ orbit that is larger than typically theorized, although these predictions contain some artifacts from the training process and do not converge to the same prediction across separate training runs. Rather than expounding in too much detail on scientific conclusions regarding CMZ structure, which remain tenuous, we primarily interpret this plausible but inconsistent generalization as evidence that, while real observational data is currently out-of-domain with respect to our training data, our method will likely yield much more consistent and meaningful scientific predictions when expanded to a much larger and more varied training dataset such that real observations become in-domain. We roughly conjecture that a data constructed from 1000 or more independent simulation runs may be ideal. We provide additional context and discussion in \autoref{subsec:Discussion}.

\subsection{Loss Convergence} \label{subsec:Loss Convergence}
    \begin{figure}[t]
        \centering
        \includegraphics[width=1\linewidth]{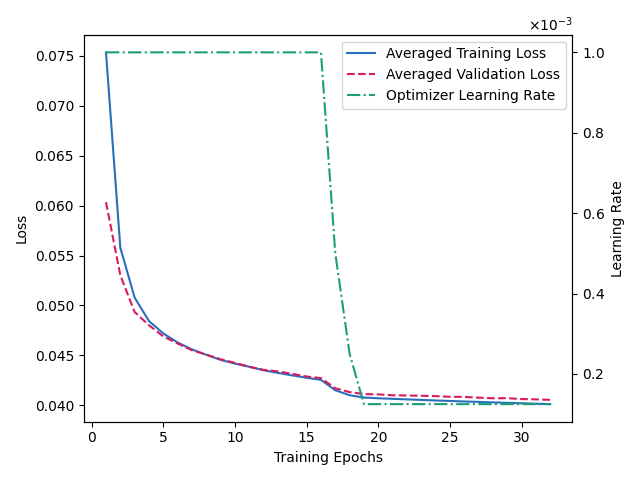}
        \caption{\textbf{Training Loss Convergence:} We plot the average training and validation loss values (on the left-side vertical axis) for our reversion model versus our training schedule as measured in training epochs. We average both metrics across one full epoch at a time and across six separate training runs. We also plot the optimizer learning rate (on the right-side vertical axis) versus our training schedule. We see that both training and validation loss converge, over the course of the training schedule, to a similar limiting value, indicating an absence of gross overfitting. See \autoref{subsec:Loss Convergence} for detailed analysis of our loss performance during training.}
        \label{fig:Loss Trajectory}
    \end{figure}

    \par In \autoref{fig:Loss Trajectory}, we illustrate the reliable convergence dynamics of both training and validation loss in relation to the schedule of training epochs and learning rates. More specifically, we average the training and validation loss values (see \autoref{subsec:Implementation of Reversion: Choice of Loss Function}) over each training epoch (one iteration over the entire training dataset) and across six different training runs, where each training run is initialized identically. As discussed in \autoref{subsec:Implementation of Reversion: Training Hyperparameters, Overfitting, and Regularization}, the validation data differs from the training data in that the validation data is used for statistics only, rather than training of the model. We then plot these cross-run averages for each epoch against the training schedule of 32 total epochs, overlaying the stepped decreasing learning rate (see \autoref{subsec:Implementation of Reversion: Training Hyperparameters, Overfitting, and Regularization}). 
    \par The convergence in both training and validation loss towards the same limiting value, which we find consistently repeatable, indicates that the model is reaching maximal performance on the given training dataset, and that further training will only contribute to overfitting. The near-identical performance of the model on training and validation data rules out gross overfitting onto the training data itself, although we provide a more nuanced discussion on the potential for overfitting in \autoref{subsec:Discussion}. Note also the temporary acceleration of loss convergence following reduction of the learning rate, which we introduce beginning after epoch 16, once training approaches diminishing returns at our initial, higher learning rate.

\subsection{Synthetic Reversion} \label{subsec:Synthetic Reversion}
    \begin{figure*}[t]
        \centering
        \includegraphics[width=.82\linewidth]{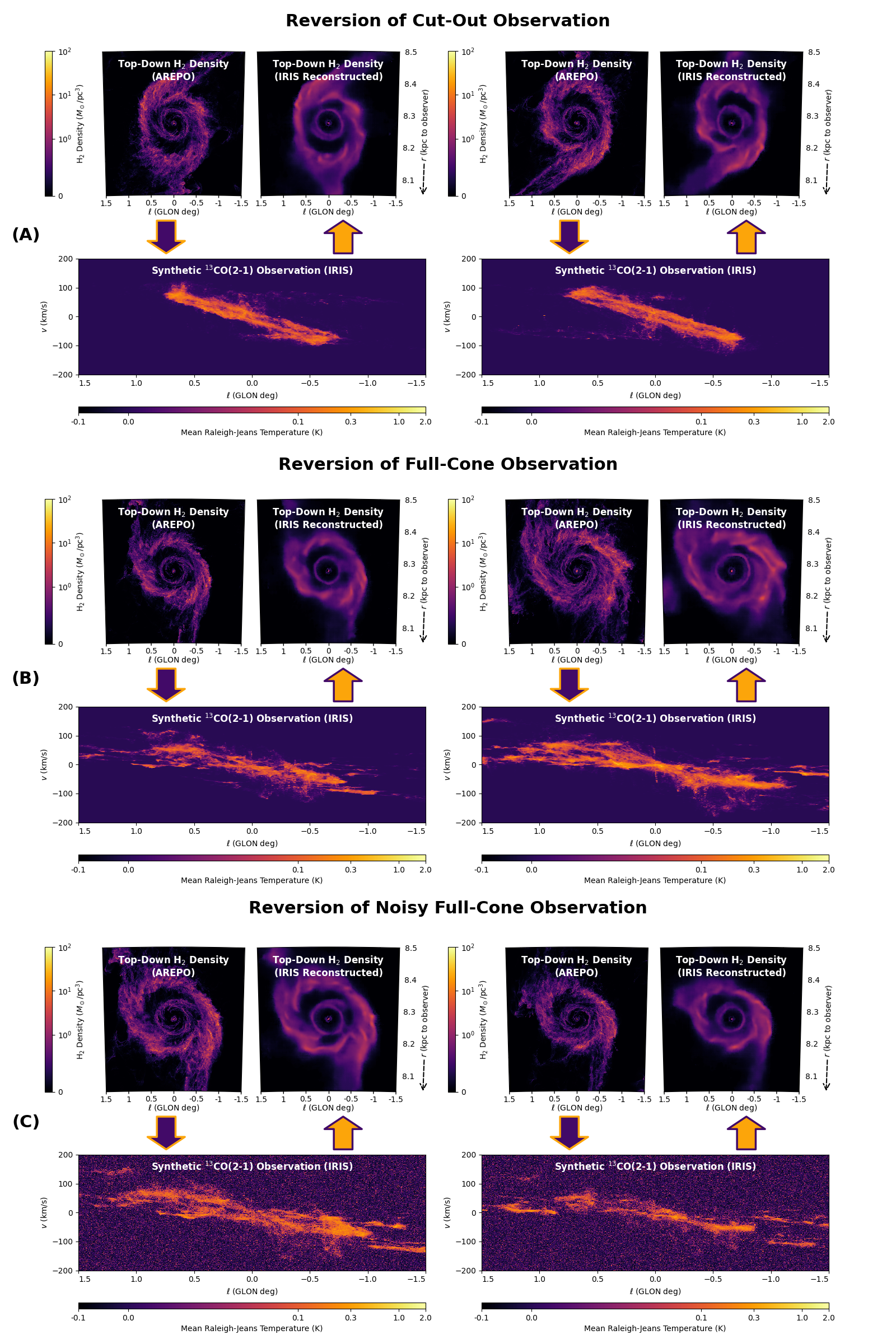}
        \caption{\textbf{Synthetic Reversions:} A visualization of successful reversion of synthetic data, as described in \autoref{subsec:Synthetic Reversion}. Each of the six panels shows reversion of a physical tensor (\autoref{subsec:Physical Tensors}) generated from our training simulation due to \citeLipman but not included in our training dataset. We applied random distance perturbations with scaling constants sampled from the interval $[0.4, 0.6)$ during physical-tensor construction, yielding a CMZ roughly half its original size in simulation (see \autoref{subsec:Physical Tensor Perturbations}). All color bars are scaled with an $\arcsinh$ nonlinearity. In all cases, we see strong alignment between the true top-down image and our top-down reversion, indicating model success.}
        \label{fig:Synthetic Reversions}
    \end{figure*}
    
    \par In \autoref{fig:Synthetic Reversions}, we illustrate the visual results of the trained model when applied to synthetic data generated from the training simulation, but not sampled from the training dataset itself. In other words, the model has never \textit{seen}, during training, the synthetic $\ell, v$ images pictured in this figure. In each of the six displayed panels, the true top-down density image of the original AREPO snapshot is displayed in the top left, the synthetically-observed $\ell, v$ image underneath, and the reverted top-down density image inferred by our trained model in the top right. (See \autoref{subsec:Supervised Reversion: Objective} for discussion regarding definitions of these data objects.) The two panels displayed in Row A feature standard synthetic observations of the CMZ region only, the panels in Row B feature full-cone synthetic observations that incorporate the foreground and background (see \autoref{subsec:Foreground and Background}), and the panels in Row C display similar full-cone observations but with the addition of simulated antenna noise (see \autoref{subsec:Noise Addition}).
    \par In each of the six synthetic reversions featured in \autoref{fig:Synthetic Reversions}, we see strong alignment between the true and reverted top-down density images. In general, the reversions show more blurring than the true images, leading to a reduction in the small-scale density maxima similar in effect to the application of a smoothing convolution. As a result, the effective resolution of the reversions appears to be lower than either the fine-feature scale of the true images or the pixel resolution of each image itself. Nonetheless, we see strong correlation between true image and reversion in their large-scale features. These results indicate that the convergent performance indicated by \autoref{fig:Loss Trajectory} is meaningful and effective, at least on synthetic data very close to the training domain. The examples of successful reversions on full-cone and noisy full-cone synthetic observations also demonstrate the efficacy of the addition of noise and foreground features at training time as described in \autoref{subsec:Noise Addition}, \autoref{subsec:Foreground and Background}, and \autoref{subsec:Implementation of Reversion: Training Hyperparameters, Overfitting, and Regularization}.

\subsection{Failure Modes} \label{subsec:Failure Modes}
    \begin{figure*}[t]
        \centering
        \includegraphics[width=1\linewidth]{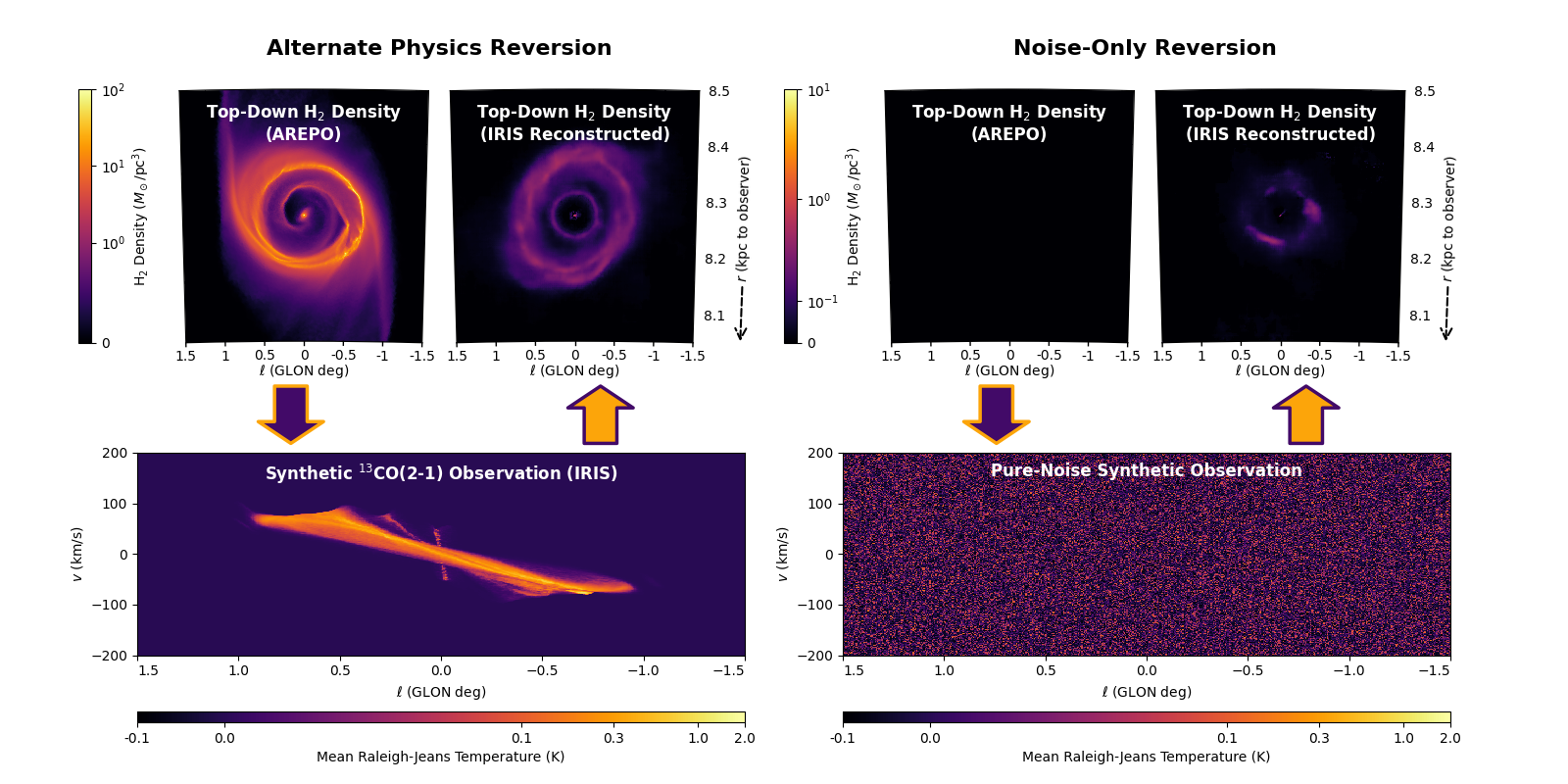}
        \caption{\textbf{Failure Modes:} A visualization of some of the failure modes of our trained reversion model, as described in \autoref{subsec:Failure Modes}. In the left panel, we perform a reversion on an AREPO simulation with very different physics than the simulation due to \citeLipman that we use for our training dataset (\autoref{subsec:AREPO Zoom Simulations}). The simulation we choose is a modified recreation of \citet{Hatchfield2021}. We see that the trained model fails to produce an accurate reversion. This failure mode is partially expected, since we intend for the model to reduce the degeneracy of the reversion problem by learning physics specific to the simulation(s) on which is it trained. In the right panel, we show the reversion of pure noise. We see that the model predicts the presence of an inner ring, which was always present in our training data, causing the model to always predict its presence, regardless of whether evidence for its existence exists in the observation. We expect this failure mode to be alleviated naturally by further expansion of our training dataset and enhancement of training-data variation, such that universal artifacts of this kind cease to exist. All color bars are scaled with an $\arcsinh$ nonlinearity.}
        \label{fig:Failure Modes}
    \end{figure*}

    \par In \autoref{fig:Failure Modes}, we probe the limitations and failure modes of one of our trained models both by testing this model on synthetic data generated from a simulation with substantially different characteristics/physics and by testing the model on pure noise. The figure follows the same format as \autoref{fig:Synthetic Reversions}, with each panel showing the true top-down density image from AREPO in the top left, the synthetically observed $\ell, v$ image underneath, and the reverted top-down density distribution inferred by our trained model in the top-right. (See \autoref{subsec:Supervised Reversion: Objective} for discussion regarding definitions of these data objects.)
    \par In the left panel, we show one such alternate-physics reversion. The simulation we choose for this test is a recreation of \citet{Hatchfield2021} with the substitution of the bar potential from \citet{Hunter2024}, and includes MHD but no star physics (i.e. no sink or star particles and no stellar feedback) and no gas self-gravity. As such, the physics of this simulation is substantially different than that of our training simulations due to \citeLipman detailed in \autoref{subsec:AREPO Zoom Simulations}. As expected, the figure shows that our model begins to fail on this out-of-domain data, indicating some combination of an over-preference towards the single simulation run from which the training data was constructed and importance of the accuracy of simulation physics in reducing the degeneracy of the image-reversion problem. 
    \par Even under ideal performance, this failure mode of the reversion model is not necessarily a failure mode of the supervised-reversion method. We intend for the model to reduce the degeneracy inherent to reversion in part by learning physics specific to the simulation(s) on which is it trained, or, more accurately, by recognizing the complex, emergent patterns produced by that physics. An effective model can then always be expected to fail on simulations with sufficiently distinct physics. To better understand the bounds of this failure mode and generalizability of the model to alternate physics, and to differentiate appropriate versus inappropriate model failure, we would then ideally like to test our trained model on simulations representing a variety of physics of varying similarity with our training simulation, as well other simulation runs based upon the exact same physics. Such tests would help to understand whether the model merely fails when it is applied to a very different physics, or if it fails even on simulations with the same physics but unfamiliar configurations. Unfortunately, however, we do not yet have the breadth of simulations available to stress-test the generalizability of the model in this way. We leave such explorations to future work.
    \par We note, however, the possibility of two separate failure modes in any such incorrect generalization. Since observation is a highly non-injective mapping, i.e. many hypothetical CMZ states may produce identical observations, the model must reduce this degeneracy in the reversion problem by application of physics learned from the training data. The model can then fail either by incorrectly resolving this degeneracy, predicting a top-down density image that is consistent with the observation but incorrect, or by predicting a top-down density image that is entirely inconsistent with the observation. Incorrect resolution of degeneracy is a more subtle failure mode, which further complicates the challenge of determining the reliability of the model predictions on real observational data for which a top-down density image is not known a priori. For future research, maximizing physicality and breadth of the training dataset is thus critical in ensuring that the true CMZ is similar to examples in the training data on which the model's correct reversion has been verified. Additionally, independent top-down models of the CMZ such as those provided in \citet{Walker2025}, \citet{Lipman2025}, and \citet{Lipman2026} can help to provide a touchpoint for the most large-scale features.
    \par In the right-hand panel of \autoref{fig:Failure Modes}, we further probe model limitations and failure modes by examining a reversion of pure noise. We find for this pure-noise reversion that the model predicts no outer $x_2$ orbit, but still predicts a consistent inner orbit, which was a relatively constant feature throughout the single simulation run on which our training dataset was generated. We conjecture that since the model always saw this specific feature during training, it did not learn to apply any critical thinking based on each specific observation as to whether the feature should actually be present, instead opting to automatically paint it into the reverted top-down image. We do note that this hallucinated inner orbit is only very faint, with a predicted density about an order of magnitude lower than the primary features in our successful reversions. But the presence of any consistent hallucination nonetheless emphasizes the importance of maximal training-data variation within the bounds of the possibility space determined by our simulation physics.
    \par We note that our crude distance perturbations (see \autoref{subsec:Physical Tensor Perturbations}) did not deviate this specific inner-ring feature sufficiently to prevent its average shape from being adopted as a constant prediction by the model. We did conduct some cursory experiments in which we inserted pure-noise samples randomly into our training data, by which we attempted to teach the model to provide a clean, zero-density prediction on this pure noise. We found this solution to be ineffective, however, only contributing to increased overfitting. A similar possible solution that we did not test would be to enforce the same zero-density prediction on pure noise as an additional penalty on the loss function. We interpret the reason for our lack of success in the first approach to be that this method forces the model to learn two very different behaviors (predicting a consistent inner orbit versus predicting a zero-field) on two very different input domains (the regular dataset versus pure noise). Without an explicit gating mechanism incorporated into the model architecture, the model is inconsistently incentivized to either learn such a gating mechanism or simply choose which behavior is cheaper to ignore. 
    \par Such inconsistency in training objective likely reduces training effectiveness, which we posit explains the failing of our attempted solution and would produce a similar failing in the loss-penalty variant. While the effectiveness of these solutions might be improved by incorporation into the model architecture of such an explicit gating mechanism that allows the model to learn two separate behaviors and choose when to apply each behavior, such a modification would be counterproductive. Our ideal trained reversion model is not one that that exhibits two separate behaviors in different circumstances, but one that exhibits a single consistent behavior across all circumstances, which nonetheless provides accurate predictions on both our regular training data and pure noise. Towards this end, we suggest that a more robust solution would be expansion of the training dataset to include a continuum of data ranging from an empty observed field to a standard system of orbits, with more sparse orbital structures in between. We leave such experiments to future work.

\subsection{Reversion of SEDIGISM Data} \label{subsec:Reversion of SEDIGISM Data}
    \begin{figure*}[t]
        \centering
        \includegraphics[width=.75\linewidth, trim={.8cm 2cm .8cm 2.5cm}, clip]{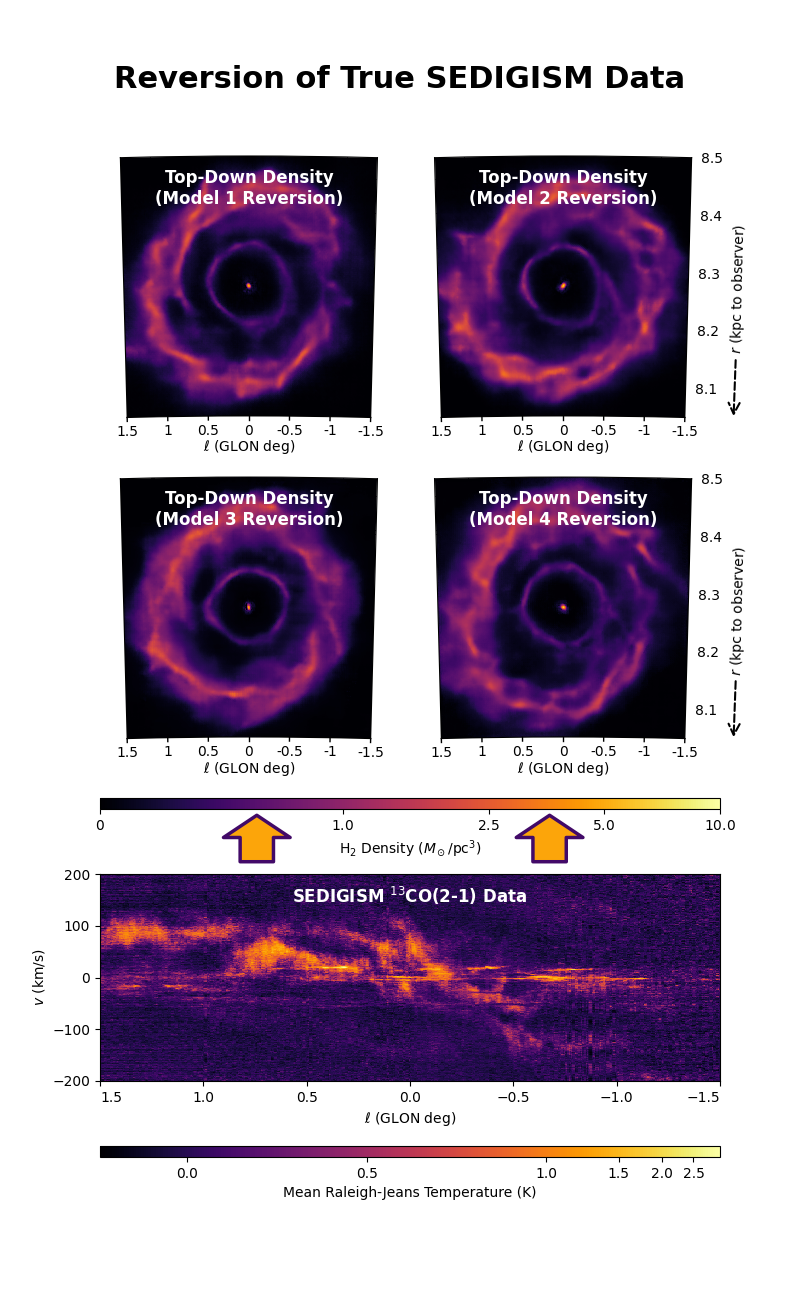}
        \caption{\textbf{SEDIGISM Reversions:} A visualization of some reversions of the SEDIGISM $\ThirteenCOTwoOne$ spectral-line survey of the CMZ \citep{Schuller2021}. As described in \autoref{subsec:Reversion of SEDIGISM Data}, we trained four separate, randomly initialized instances of our reversion model under identical conditions. We then applied all four trained models to the SEDIGISM data separately, generating four separate predictions of the top-down CMZ structure. We see that each reversion is somewhat plausible, roughly matching the elliptical eccentricity and rotation hypothesized by other studies \citep[e.g.][]{Walker2025, Lipman2025, Lipman2026}, albeit with the inner-ring artifact explored in \autoref{subsec:Failure Modes} and \autoref{fig:Failure Modes}. Across all four training runs, however, we see the model does not converge to a fully consistent, shared prediction. We conjecture that the lack of such definitive convergence is due to the minimality of our training dataset, which is generated from only a single simulation run evolved over a relatively short timescale (\autoref{subsec:AREPO Zoom Simulations}), and can be attained by expanding to a vastly larger training dataset generated from many independent simulation runs. All color bars are scaled with an $\arcsinh$ nonlinearity.}
        \label{fig:True Reversions}
    \end{figure*}

    \par In \autoref{fig:True Reversions}, we compare the application of four of our trained models to the real SEDIGISM $\ThirteenCOTwoOne$ data \citep{Schuller2021}. Each of these four models are instances of the same neural-network architecture (\autoref{subsec:Implementation of Reversion: Architecture}), randomly initialized under the same conditions, and subjected to the same training regimen and hyperparameters (\autoref{subsec:Implementation of Reversion: Training Hyperparameters, Overfitting, and Regularization}). All differences in convergent form and behavior emerged from the combined stochasticity of initialization and gradient descent. In each of the four featured panels in this figure, we show the top-down density image inferred from the SEDIGISM $\ell, v$ image by a given trained model instance. (See \autoref{subsec:Supervised Reversion: Objective} for discussion regarding definitions of these data objects.) The SEDIGISM data is shown in the bottom panel.
    \par We find that each model generalizes somewhat plausibly on the SEDIGISM data, but that the models do not fully converge to a consistent, shared prediction. All of the models do predict a consistent elliptical outer $x_2$ orbit of roughly the same size, with a semi-major axis of $\sim \SI{170}{\parsec}$ and a semi-minor axis of $\sim \SI{160}{\parsec}$. Moreover, these orbits are plausible in that they roughly match the elliptical eccentricity and rotation hypothesized by other works \citep{Walker2025, Lipman2025, Lipman2026}. We temper our characterization of these results as fully plausible, however, in that we find no observational evidence for the inner ring that we have established is burned into these reversions by our training data (see \autoref{subsec:Failure Modes}).
    \par We note, in particular, that the dimensions of these outer orbits are larger than those widely hypothesized in the literature \citep[e.g.][see \autoref{fig:CMZ Overview}]{Walker2025, Lipman2025, Lipman2026}. This observation is noteworthy in that, while inconsistent in fine features, each of these ellipses are consistent in size, whereas random scaling perturbations were applied to our training data to provide our reversion model with examples of CMZs ranging broadly in size (see \autoref{subsec:Physical Tensor Perturbations}). And moreover, our trained model accurately reverts synthetic observations across this entire size range. Nonetheless, we caution against drawing premature scientific conclusions from these preliminary results, since:
    \begin{enumerate}[(i)]
        \item we have only trained this model on snapshots from a single simulation run (see \autoref{table:Dataset Parameters});
        \item not all of these snapshots were sampled from periods of the simulation in which all physics were fully turned on (which weakens the total physicality of our training dataset, see \autoref{subsec:AREPO Zoom Simulations});
        \item the radial resolution of our training dataset is somewhat below the requirement for convergent synthetic observation (see \autoref{subsec:Resolution Convergence}); and
        \item the reversions of real observations are not yet fully convergent.
    \end{enumerate}
    Rather, the important conclusion we draw from these results, which we discuss in greater detail in the following subsection (\autoref{subsec:Discussion}), is the technical proof-of-concept of this novel method. We consider these somewhat plausible and semi-convergent SEDIGISM reversions as promising evidence of the potential of IRIS in providing more definitive scientific insight regarding true CMZ structure when expanded to a broader and more information-rich training dataset in future research.

\subsection{Discussion} \label{subsec:Discussion}
    \par We generally interpret our results as a proof-of-concept for our supervised-reversion method in the simplest case of a single-simulation dataset. But we further interpret these results to mean that, as expected, a single simulation run is not sufficiently constraining to yield consistent generalizations to real observational data. We interpret the plausibility of our semi-consistent real-data generalizations, up to the implausibility of the inner ring (see \autoref{subsec:Reversion of SEDIGISM Data}), as an indication of a strong likelihood of ultimate scientific success, should this method be expanded to a much broader dataset incorporating enough independent simulation runs to saturate the possibility space determined by the simulation physics. We coarsely conjecture that a dataset constructed from at least 1000 independent simulation runs and at least $\sim 1\text{M}$ top-down density images and synthetic observations would be optimal, although it is difficult to hypothesize a minimal necessary breadth.
    \par Relevant to the topic of overfitting we discussed in \autoref{subsec:Implementation of Reversion: Training Hyperparameters, Overfitting, and Regularization} and \autoref{subsec:Loss Convergence}, we note as well that it is difficult to determine to what degree the model is ``memorizing'' versus ``reasoning''. More specifically, it is difficult to know to what extent our model's success is by memorization of non-generalizing features and patterns in our limited, single-simulation training data, to what extent our model is learning a weak logic that compresses the memory required for adequate performance on the training dataset itself without generalizing beyond the training data, or to what extent the model is learning a robust process of reasoning that generalizes outside the training dataset.
    \par A standard tool we use in machine learning to answer this question is comparing training and validation loss. As discussed in \autoref{subsec:Loss Convergence}, and demonstrated in \autoref{fig:Loss Trajectory}, we find virtually identical performance in terminal training and validation loss. This alignment in loss values provides initial evidence that our model generalizes at least to the validation data on which it has never been trained. Complicating the conclusions we may draw from our validation-loss performance, however, is the high degree of internal and cross correlation of our training and validation datasets.
    \par Since all physical tensors and synthetic observations are generated from the same simulation run, evolved over a relatively short timescale (see \autoref{subsec:AREPO Zoom Simulations}), with each snapshot observed from 64 different angles in the galactic plane (see \autoref{subsec:Physical Tensor Interpolation}), we suspect that, even with the addition of random size perturbations (see \autoref{subsec:Physical Tensor Perturbations}), our training dataset may saturate the space of data we are able to produce. In other words, our validation data and training data may be highly correlated, limiting the utility of validation loss as an overfitting metric. Given that we are data-constrained in this proof-of-concept study, however, we find there is little more we can do to reserve some useful segment of the data such that our training and validation data can be made substantially independent.
    \par We are therefore left to consider alternate means of evaluating model overfitting, reasoning capacity, and generalizability. In considering the likelihood of our model performing by virtue of pure memorization, we could compute the total size of the training dataset ($.8 \cdot 210 \text{GiB} \approx 170 \text{GiB}$), and reason that given $\sim 14$M trainable parameters, such pure memorization would require an information density of roughly $\sim 15$kB per parameter, which, while plausible in an ideal, continuous parameter space, stretches plausibility in terms of the actual mechanics of single-precision parameters. But this naive computation would be misguided since, as articulated above, the datapoints in our training dataset are highly correlated.
    \par Instead of using the actual disk size of the training data to perform this information-density computation, we would prefer to first estimate the true information entropy of the dataset. While some practical methods exist for estimating this information size, such as principal-component analysis (PCA) or maximally compressive, lossless autoencodings \citep{Goodfellow2016}, this is a complex task we do not attempt. It is therefore somewhat difficult to rigorously rule out the possibility that this proof-of-concept reversion model has succeeded by at least a partial memorization strategy as opposed to a more generalizing, reasoning-based approach. Instead, we suggest these results perhaps not be viewed as proof positive of the potential of the method, but as a meaningful lack of falsification for such potential.
    \par In machine learning in general, such questions are usually not relevant, since most machine-learning problems involve the automation of verifiable predictions, or generalization to in-domain data on which a prediction may not be verifiable but for which we can trust reliability of a prediction by the verifiability of predictions on similar datapoints. The question is relevant in our case, however, specifically because we are data-constrained, and we are attempting to generalize to real data from the SEDIGISM survey \citep{Schuller2021} that is very likely, as evidenced by the lack of full convergence of our models on this real data across separate training runs (see \autoref{subsec:Reversion of SEDIGISM Data}), outside the domain defined by our training dataset. Presumably, such out-of-domain generalization is only possible if the model converges to a strong, reasoning-based approach.
    \par Rather than considering the unconfirmed reasoning and generalizing capacity of our model as a hard block, however, we interpret it merely as evidence of the further need to expand the breadth and physicality of our training data. In the hypothetical limit in which we have saturated the possibility space with highly physical training data such that the true CMZ state becomes, if not represented by an exact training datapoint, more properly in-domain, then this machine-learning application becomes more standard. Evaluation of the degree to which our model memorizes becomes, in turn, less relevant. We therefore consider a strong priority in future expansion of this research, towards the aim of producing meaningful scientific predictions regarding CMZ structure, to be expansion of the training dataset. We present this study as a promising initial exploration and minimal proof-of-concept of our novel method---the first step of a much broader research program.

\section{Summary and Conclusion} \label{sec:Summary and Conclusion}
    \par Our Milky Way Galaxy's CMZ is a highly complex region, the 3D structure of which is of profound importance in understanding the galaxy's energy cycles, placing the galaxy in a wider universal context, and constraining possible dark-matter annihilation signatures \citep[e.g.][]{Henshaw2023, Battersby2025a, Battersby2025b, Walker2025, Lipman2025}. In attempting to understand this elusive CMZ structure, we are aided by an ever-expanding wealth of multi-wavelength observations of escalating resolution and astrophysical simulations of increasing complexity and realism. But we argue that present methods of distilling the information present within the sum of all this data fail to leverage the complex correlations reflected across this data's full, multi-modal breadth. Where typical approaches focus on simple comparisons such as those performed ``by eye'' between individual simulation snapshots and observational images, we posit that an untapped wealth of information exists accessible only via inter-comparison of the entirety of available information.
    \par Towards achieving such massively data-informed scientific insight, we pioneered the IRIS project. Aiming to uncover the CMZ's 3D structure, we focused on producing such inter-comparisons via an automated, machine-learning approach. After exploring a variety of such approaches, including multiple variants of the supervised-reversion approach and the neural-fields approach (\autoref{sec:Machine-Learning Methods and Reversion}), we narrowed our focus to a minimal proof-of-concept study, which we present in this work. In this publication, by machine-learning on galactic simulations, we train an artificial neural network to map $\ell, v$ observations in a single spectral line to corresponding top-down density images for our galaxy's CMZ (see \autoref{subsec:Supervised Reversion: Objective} for definitions). We call this learned mapping \textit{reversion}.
    \par In realizing our IRIS proof-of-concept for simulation-informed observational reversion, we developed several novel and bespoke elements. We built a pipeline for processing snapshots of high-resolution AREPO galaxy simulations due to \citeLipman into a training dataset of physical tensors (\autoref{sec:Simulation Processing and Data Production}). We built a GPU-accelerated synthetic-observation pipeline from scratch, customized to the needs of our use-case and providing orders of magnitude of speedup (up to $10{,}000\times$) over CPU-based counterparts, for the purpose of mapping simulation snapshots/physical tensors to their observed counterparts in the massive volumes required. We designed a neural network to learn the specified task using a transcoder architecture implemented as a CNN with pixelwise self-attention. We then generated a massive dataset from a single simulation run and performed an exhaustive battery of training experiments with our model on this dataset.
    \par Our final IRIS reversion model demonstrates success on synthetic data within the general domain of the training dataset. On a simple training dataset generated from a single AREPO simulation run \citepLipman, sampled at a variety of timestamps, from a variety of observational perspectives, and subjected to random size perturbations (\autoref{subsec:Physical Tensor Interpolation}, \autoref{subsec:Physical Tensor Perturbations}), our model successfully learns to recover top-down density images of the simulated CMZ from $\ell, v$ observations on both training and validation data (\autoref{subsec:Synthetic Reversion}, \autoref{fig:Synthetic Reversions}). While discerning the extent to which the model is succeeding by virtue of ``memorizing'' versus ``reasoning'' poses some degree of challenge, we consider this question ultimately moot in relation to the future directions of this research. In such future work, we hope eventually to approach the limit at which we have expanded variation of the training dataset and improved training-data physicality to the extent that the real observational data to which we wish to generalize is effectively in-domain with respect to the training data, at which point we expect this distinction between memorizing and reasoning becomes less germane. 
    \par We note failure modes of the model, including unreliability in generalizing to synthetic data generated from simulations with substantially different physics \citep{Hatchfield2021, Hunter2024} and the replication of an inner-ring feature that was consistently present within the training data when applied to pure noise (\autoref{subsec:Failure Modes}, \autoref{fig:Failure Modes}). We interpret lack of generalizability outside the physics encoded into the training data, however, as a natural element of a system that relies on physics-based inference in reducing inherent degeneracy of the reversion problem. We expect data-derived hallucinations such as this inner-ring artifact to resolve naturally upon enhancement of training-data variation to the extent that no such constant features remain in the training dataset.
    \par We then apply separately trained instances of our IRIS model to the real SEDIGISM $\ThirteenCOTwoOne$ data \citep{Schuller2021} in order to generate top-down density predictions of the Milky Way's CMZ. Other than the inner ring that is ``burned'' into our model predictions by our training data, we find these predictions to be plausible. Specifically, in elliptical eccentricity and rotation, we find these predictions to compare favorably with recent models \citep{Walker2025, Lipman2025, Lipman2026}. Nonetheless, we note that these independent training runs do not converge to a consistent, shared prediction, which we interpret as evidence that training-data variation is as of yet insufficient to ensure such convergent generalization to real observations. We conjecture this shortcoming will also resolve with future expansion of the training dataset and enhancement of training-data variation and physicality.
    \par Despite this lack of total consistency in generalization to the SEDIGISM data, one area of consistency that we find potentially notable is in the predicted size of the primary $x_2$ orbit. This consistency may be of interest since, while the random size perturbations applied to the training data (\autoref{subsec:Physical Tensor Perturbations}) ensured that each model instance saw CMZs of a wide range of sizes during training, and while the trained model performs accurate reversions across this entire size range on synthetic data, the consistently predicted orbital size (semi-major axis of $\sim \SI{170}{\parsec}$ and a semi-minor axis of $\sim \SI{160}{\parsec}$) is larger than that predicted by typical models \citep{Walker2025, Lipman2025, Lipman2026}. Nonetheless, due to the narrow breadth of the training dataset, generated from only a single simulation run over a short timescale of $\sim \SI{7}{\mega\year}$ during which not all simulation physics were fully turned on (see \autoref{subsec:AREPO Zoom Simulations}), which is synthetically observed at a somewhat insufficient radial resolution (see \autoref{subsec:Resolution Convergence}), we believe that drawing, from this study, any definitive scientific conclusions regarding CMZ structure is still premature. Rather, we represent our main result as a strong proof-of-concept of the supervised-reversion method, indicating a promising potential in yielding meaningful scientific insight regarding true CMZ structure upon expansion of the training data.
    \par In parallel with the release of this publication, we have released all the IRIS code as an open-source repository on GitHub, including the synthetic-observation module, IRIS-SO (\url{https://github.com/bldubois/IRIS}). We believe IRIS-SO may be of independent interest to a variety of astrophysics researchers owing to the massive computational speedups afforded by its GPU-acceleration over CPU-based counterparts (up to $10{,}000\times$, see \autoref{subsec:Speed Testing}), its flexible differentiability modes, and its ease of use and modification within the PyTorch ecosystem \citep{Paszke2019}. IRIS-SO provides substantially more general functionality than is utilized in this publication. We leave open the possibility of a future independent release of the IRIS-SO code as a \textit{user-facing} Python package. In the case of such release, we intend that this publication will also serve as a primary reference. In addition to this publication and the IRIS code release itself, we have also published a comprehensive developer's documentation page on the IRIS GitHub describing all elements of the code in detail. We hope that this documentation provides a solid foundation for other researchers with whom we enthusiastically invite future collaboration.
    \par In presenting a general proof of the imagery-reversion concept via machine learning on simulations, we believe this publication may invite future research regarding the use of machine learning to compare observational and simulation-based data in astrophysical contexts even beyond the CMZ. Within the CMZ, however, we hope this publication is likewise only the beginning of a much larger program of research. Our vision is that variations of this method may provide not just large-scale structure of the CMZ in a top-down view, but true 3D, cloud-scale structure, dynamics, and characteristics to include density, temperature, and velocity. Towards achieving this end goal, we outline future work, on which we invite collaboration.
    \par First, we aim to advance the complexity to the underlying model to produce a greater variety of output variables based upon a greater variety of observational inputs. Such inputs could potentially incorporate not only multiple independent spectral-line observations but other types of observations such as dust-continuum maps and/or observations in other regions of the EM spectrum such as the x-ray band. Second, we aim to eventually adapt from a 2D-to-2D predictive framework to a 3D-to-3D framework. More specifically, we envision the use of 3D CNNs to transform entire PPV cubes to 3D spatial tensors. In the ideal scenario, reversion of full PPV cubes to full physical tensors would allow synthetic observation of the reversions of true CMZ observations, and then comparison of these synthetic observations with the original observational data. Such comparisons would provide another means of evaluating the plausibility of real-data reversions.
    \par Third, in parallel to these efforts, we aim to allow for modeling additional spectral tracers by either adapting the on-the-fly chemical network incorporated into the simulations themselves or application of post-processing chemical networks on simulation snapshots under certain equilibrium assumptions. These efforts will eventually allow us to revert real observations in a variety of other spectral lines, such as HNCO for comparison with ACES \citep{Longmore2026, Ginsburg2026, Lu2026, Hsieh2026, Walker2026}. Fourth, particularly in light of the success recently demonstrated by \citet{Levis2025} on a similar problem involving the reverse-imaging of protoplanetary disks, we anticipate that a renewed look at neural-fields approaches in the CMZ may be fruitful (see \autoref{subsec:Neural Fields: General Approach} and \autoref{subsec:Neural Fields: Latent-Constrained} for more discussion).
    \par Finally, in the long-term, substantial forward progress will require simulation-derived training datasets that are not only larger in size but broader in variation. We hope eventually to generate vastly more data with variation that fully saturates the physical possibility space by running our simulations many more times. We suggest a rough goal of a 1M-point dataset generated from 1000 independent simulation runs. To achieve this many simulation runs, acceleration of the simulations themselves will be necessary, possibly through GPU optimization and/or neural-network compression of computationally expensive simulation components. Since our current AREPO simulations are CPU-based, we conjecture substantial speedup and achievement of our total simulation goal may eventually be feasible, although such speedups might require the introduction of simplifying assumptions and/or the exploration of alternative hydrodynamic frameworks. We anticipate the possible benefit of developing, parallel to the primary suite of robust CMZ simulations due to \citet{Tress2020}, \citet{Sormani2020}, \citet{Tress2024}, \citet{Tress2025}, and \citeLipman, a separate strain of lightweight simulations that are, at the expense of maximum physicality and accuracy, speed-optimized specifically for application to machine-learning methods.
    \par In summary, in seeking to decipher the 3D structure of our Milky Way Galaxy's CMZ, we demonstrate proof-of-concept of the novel method of imagery reversion informed by machine-learning on galactic simulations. This method represents a new approach to the inter-comparison of large observational and simulation-based datasets. We test the method on both synthetic and real observational data, illustrating the method's strong potential and providing an early glimpse into new top-down views of our own CMZ. Implementation of this method in our IRIS code-base required development of many novel elements, such as our fully-differentiable and GPU-accelerated synthetic-observation code IRIS-SO, which may be of use to other researchers. We see this work as the start of a larger program and encourage interested researchers to connect and collaborate with us. In closure, we share our hope that the reward of this ongoing work will be an understanding of the Milky-Way CMZ, grounded in both real observational data and state-of-the-art simulations, as has never before been seen.

\phantomsection
\addcontentsline{toc}{section}{Acknowledgments}
\begin{acknowledgments}
    \par B.L.\ DuBois gratefully acknowledges funding from the National Science Foundation under a Veteran's Research Supplement, Award No. 2414862, to Award No. 2206510, as well funding from CAREER 2145689. C.\ Battersby  gratefully  acknowledges  funding  from  the National  Science  Foundation  under  Award  Nos. 2108938, 2206510, 2414862, and CAREER 2145689, as well as funding from the National Aeronautics and Space Administration through the Astrophysics Data Analysis Program under Award ``3-D MC: Mapping Circumnuclear Molecular Clouds from X-ray to Radio,” Grant No. 80NSSC22K1125 and participation in the PRIMA project under Grant No. 80NSSC25K7944. A.\ Ginsburg acknowledges support from the NSF under grants AAG 2206511 and CAREER 2142300. V.F.\ Ksoll acknowledges financial support by the Carl-Zeiss-Stiftung. M.C.\ Sormani acknowledges financial support from the European Research Council under the ERC Starting Grant ``GalFlow'' (grant 101116226) and from Fondazione Cariplo under the grant ERC attrattivit\'a no.\ 2023-3014. Z.\ Feng acknowledge financial support from the European Research Council under the ERC Starting Grant “GalFlow” (grant 101116226). The computational work for this project was conducted using resources provided by the Storrs High-Performance Computing (HPC) cluster (\url{https://hpc.uconn.edu/}). We extend our gratitude to the UConn Storrs HPC and its team for their resources and support, which aided in achieving these results.
\end{acknowledgments}

\phantomsection
\addcontentsline{toc}{section}{References}
\bibliography{refs}{}
\bibliographystyle{aasjournalv7}

\end{document}